\documentclass{aa}  

\usepackage{graphicx}
\usepackage{txfonts}
\usepackage{natbib}
\usepackage{arydshln}
\usepackage{xcolor}
\usepackage[colorlinks, citecolor=blue, linkcolor=blue, urlcolor=blue]{hyperref}

\newcommand{\gaia}{\textit{Gaia}~}
\newcommand{\Cov}{\mathrm{Cov}}
\usepackage{color}

\begin{document}

   \title{Purveyors of fine halos. II. Chemodynamical association of halo stars with Milky Way globular clusters\thanks{The full Tables \ref{Table: Extratidal target information} and \ref{Table: Associations to CN-strong stars} are only available in electronic form at the CDS via anonymous ftp to cdsarc.u-strasbg.fr (130.79.128.5) or via \url{http://cdsweb.u-strasbg.fr/cgi-bin/qcat?J/A+A/.}}}

   \author{Michael Hanke\inst{1}
          \and
          Andreas Koch\inst{1}
          \and
          Zden\v{e}k Prudil\inst{1}
          \and
          Eva K. Grebel\inst{1}
          \and
          Ulrich Bastian\inst{1}}

   \institute{Astronomisches  Rechen-Institut,  Zentrum  für  Astronomie  der  Universität  Heidelberg,  Mönchhofstr. 12-14,  D-69120  Heidelberg, Germany\\
              email: \href{mailto:mhanke@ari.uni-heidelberg.de}{mhanke@ari.uni-heidelberg.de}
              }

   \date{Received 2 March 2020 / Accepted 30 March 2020}
 
\abstract{A long-lasting open question in the field of Galactic archeology refers to the size of the contribution from former globular cluster (GC) stars to the formation of the  stellar halo of the Milky Way. We contribute to answering this important question by establishing observational links between the present-day halo field star population and GCs. To this end, we combined astrometric information such as space motions and parallaxes from the second data release of the \gaia mission (\gaia DR2) with spectroscopic radial velocities and metallicities ([Fe/H]) from the Sloan Digital Sky Survey (SDSS-IV, DR14) to end up with a seven-dimensional chemodynamical information space for more than $3\cdot10^5$ stars. Moreover, from our previous study, we incorporated the sample of halo giant stars with a distinct chemical signature (strong CN bandheads) that resembles the light-elements anomaly otherwise only seen in the second generation of globular cluster stellar populations. Using three different tagging techniques -- among which is the exploration of conservative integrals of motion -- we are able to establish unique associations between 151 extratidal stars in the neighborhood of eight GCs, which coincide with earlier findings of stellar envelopes beyond the tidal radius and even beyond (out to several tens of tidal radii). In addition, we trace the possible origin of about 62\% of the sample of CN-strong giants to their potential host clusters. We find a connection between several of the involved GCs   and the Gaia-Enceladus and Sequoia merger events. By establishing kinematic and chemical connections between 17 CN-strong stars and their surrounding fields, we identify co-moving groups of stars at the same [Fe/H] with a possible cluster origin. Some of these associations contain RR Lyrae variables, which allows meaningful distance inferences to be made. From these, we find strong evidence that four CN-strong stars and their associates are connected to the Sagittarius stream whilst their tightly confined [Fe/H] may hint to a birth site in M\,54, the massive cluster in Sagittarius' core remnant. Finally, by employing the counts of CN-strong and bona-fide CN-normal giants from our novel sample, we provide tentative estimates for the fraction of first-generation cluster stars among all stars lost to the halo. In the immediate cluster vicinity, this value amounts to $50.0\pm16.7\%$ while the associations in the halo field rather imply $80.2_{-5.2}^{+4.9}\%$. We speculate that -- if proven real by spectroscopic follow-up -- the disparity between these numbers could indicate a major contribution of low-mass clusters to the overall number of stars escaped to the halo or could alternatively suggest strong mass loss from the first generation during early cluster dissolution.}

\titlerunning{Chemodynamics of globular cluster escapees in the Galactic halo}
\maketitle

\section{Introduction}\label{Sec: Introduction}
The build-up of the stellar halo of the Milky Way (MW) has been a matter of extensive research for more than five decades. Existing formation scenarios distinguish between ``in-situ''  star formation --  where stars form within the host galaxy -- and ``ex-situ'' channels, where stars are born in satellite systems and only later accrete onto the massive galaxy. The proposed ratio of these two scenarios spans from purely in-situ formation \citep[e.g.,][]{Eggen62} to mixtures of both channels, with the degree of ex-situ contributions varying with Galactocentric distance \citep[e.g.,][]{Searle78, DeLucia08, Zolotov09, Pillepich15, Cooper15}.

There is a wealth of observational evidence of ongoing accretion in the form of persistent stellar streams, which support the ex-situ scenarios. These streams range from events as massive as the accretion of the Sagittarius (Sgr) dwarf spheroidal galaxy \citep{Ibata94, Belokurov06} to 
observations of more elusive stellar streams and envelopes attributed to globular clusters (GCs) that are in the process of being tidally disrupted \citep[e.g.,][]{Odenkirchen01,Grillmair06,Jordi10,Kuzma18}.

The advent of the \gaia space mission \citep{Gaia16} has revolutionized the field of Galactic archeology. Its latest data release 2 \citep[DR2,][]{GaiaDR2_18} provides positions, parallaxes, and proper motions for approximately one billion stars \citep{Lindegren18}. Moreover, for about seven million targets it offers the full 3D space-motion vector by adding radial velocities \citep{Cropper18}. \gaia DR2 led to the discovery of numerous kinematical substructures comprised of the remnants of massive merger events, such as the incorporation of the Gaia-Enceladus \citep{Helmi18b, Belokurov18} and Sequoia galaxies \citep{Myeong19} into the inner Milky Way halo and thick disk \citep[see, e.g.,][for the discussion of potential further merger events]{Forbes20}, but also many new indications of past and ongoing events of tidal disruption of GCs \citep{Malhan18, Ibata19, Borsato19}. 

Whether or not a sizable fraction of the halo originated from such disrupted clusters is a long-standing question that remains to be elucidated. One promising approach lies in chemical tagging of halo field stars that show unique chemical signatures in their light-element abundances that are otherwise only found in GCs. This abundance anomaly is attributed to high-temperature proton burning in the CNO-cycle and its Ne-Na chain. These give rise to the archetypal C-N and O-Na abundance anticorrelations found in nearly every GC studied to date \citep[e.g.,][]{Cohen78, Carretta09}. The less-frequently observed spreads in Mg, Al, Si, and potentially even K and Zn \citep[e.g.,][]{Gratton01, Hanke17, Pancino17, Gratton19} are attributed to burning at even higher temperatures \citep[$\gtrsim70\cdot10^6$~K, ][]{Arnould99}. Here, we follow the nomenclature of \citet{Bastian18} and distinguish a second, chemically ``enriched'' GC population (2P), which  manifests in enhanced N and Na abundances at the expense of depleted C and O. Models suggest that this second generation was enriched by the aforementioned nucleosynthesis in massive stars of the first, chemically ``normal'’ (i.e., C- and O-enhanced while being Na- and N-depleted) population (1P). While the latter is chemically indistinguishable from the general field star population, discoveries of chemical fingerprints in the halo that are reminiscent of 2P stars are in strong favor of a GC origin.

In turn, a number of theoretical studies investigated aspects of GC formation and evolution by focusing on the amount of mass loss that eventually leads to a direct inference of the fraction of GC stars that contributed to the present-day observed stellar halo \citep{Dercole08, Bastian13, Baumgardt17, ReinaCampos19}. Such models ultimately have to be informed by observations of stars from both 1P and 2P, the latter of which can be unambiguously tied to GCs. In this respect, \citet{Martell10} and \citet{Martell11} used low-resolution spectra of the Sloan Extension for Galactic Understanding and Exploration \citep[SEGUE-1 and SEGUE-2, ][]{Yanny09, Eisenstein11} as part of the Sloan Digital Sky Survey \citep[SDSS, ][]{York00} to identify 2P candidates in the halo from a combination of CN and CH bandstrengths in red giant branch (RGB) stars. From their finding that 2.5\% of their halo sample is CN-strong, these latter authors concluded that between 17 and 50\% of the halo may have originated from GCs. Similar arguments hold for an identification of GC-like stars, enabled by the infrared APOGEE survey \citep{Majewski16}, through their Mg and Al patterns as a product of hotter proton-burning cycles. Respective studies have been successfully carried out for the Galactic halo \citep{Martell16, FernandezTrincado19b, FernandezTrincado19}, and others also detected N-enriched stars in the Milky Way bulge \citep{Schiavon17}, hinting at a similarly (in-)effective formation channel of this old Galactic component.

Using the recent DR14 of SDSS-IV \citep{Abolfathi18} we doubled the number of known CN-strong stars by \citet{Martell10} and \citet{Martell11} to 118 \citep[][hereafter Paper I]{Koch19}. From these, we estimated a fraction of $2.6\pm0.2\%$  2P stars among all considered halo field giants, which led to a halo contribution from disrupted GCs of $11\pm1\%$.  

In the present study, we aim to explicitly tie halo field stars -- both of enriched and unenriched nature -- to a potential GC origin and to observationally test the fraction of bona fide 2P stars among the entire population of GC escapees. To this end, we combined radial velocities ($v_r$) and metallicities ([Fe/H]) from SDSS/SEGUE with \gaia DR2 astrometry to expand the previous chemical data used in Paper I into a seven-dimensional, chemodynamical information space (i.e., a three-component space vector, a three-component motion vector, and metallicty, [Fe/H]). This was achieved using three methods: First, we searched the immediate vicinity around GCs for extratidal stars that share the space motion and [Fe/H] of the clusters. Next, we used the conservation of action-angle coordinates to identify potential former host clusters for the sample of CN-strong stars from Paper I among the GCs that still exist in the MW. The third method is a modification of the first, and aims at finding stars that share the same portion of phase space as the CN-strong stars, which may indicate a common GC progenitor.

This paper is organized as follows: In Sect. \ref{Sec: Data}, we introduce the data sets employed throughout this work and discuss the involved statistical uncertainties and systematic errors. Next, in Sect. \ref{Sec: Methods}, the three employed tagging techniques are outlined, which is followed by the associated findings and a discussion thereof in Sect. \ref{Sec: Results}. Finally, we summarize our main conclusions and provide an outlook for further studies in Sect. \ref{Sec: Summary and Conclusions}.

\section{Data}\label{Sec: Data}
For the present analysis, we employed the latest data release of SDSS \citep[DR14,][]{Abolfathi18}, which contains optical stellar spectra at low resolution ($R\sim2000$) that were obtained throughout the two phases of the SEGUE \citep{Yanny09, Eisenstein11} survey and during the Extended Baryon Oscillation Spectroscopic Survey \citep[eBOSS,][]{Dawson16}. Of particular interest are the large samples of stellar metallicities and heliocentric radial velocities deduced using the SEGUE stellar parameter pipeline \citep[SSPP,][]{AllendePrieto08, Lee08a, Lee08b, Lee11, Smolinski11}. 

A large fraction of this study builds upon the compilation of CN-strong giants that were identified in Paper~I. There, SDSS spectra of 4649 bona fide halo red giants in the metallicity regime $-1.8\mathrm{~dex} < \mathrm{[Fe/H]}<-1.3\mathrm{~dex}$ were examined for a combination of spectral indices involving carbonaceous molecules that are indicative of an unusually strong CN enrichment that is atypical for the general halo population.

In order to make use of the full 6D phase-space information, we complemented SDSS data with proper motions ($\mu_{\alpha}\cos{\delta}$ and $\mu_\delta$) and parallaxes ($\varpi$) from \gaia DR2  \citep{GaiaDR2_18, Lindegren18}. In addition, the mean phase-space vectors of clusters were retrieved from the published collection by \citet{Vasiliev19}, which itself is based on the compilation of coordinates and distances in the Harris catalog of GC parameters \citep[][2010 edition]{Harris96} and on line-of-sight velocities from \citet{Baumgardt19}. Other properties used throughout this paper are cluster core radii ($r_\mathrm{c}$), King-model central concentrations ($c$), and half-light radii ($r_\mathrm{h}$), which were taken from the Harris catalog.

\subsection{Precision and accuracy assessments}\label{Subsec: Precision and accuracy assessment}
It is important to bear in mind that incorporating both realistic precisions and accuracies is essential for obtaining realistic probabilities in our Bayesian framework (Sect. \ref{Sec: Methods}). This latter is used to compare quantities from different sources and is therefore affected by potential systematic discrepancies. Below, we discuss the two main sources of error that were singled out throughout our analysis, namely inaccurate metallicities with additionally underestimated precisions, and underestimated errors for \gaia proper motions.

\subsubsection{SSPP parameters}
In light of the low-resolution ($R\sim2000$) nature of the SEGUE/eBOSS surveys and possible systematic trends among the parameter scales involved in the SSPP, caution must be taken when using the quoted internal uncertainties on the adopted parameters. A realistic assessment of the error budget should incorporate both systematic error contributions such as wavelength calibration errors or calibration inaccuracies of the SSPP parameter scales, as well as random errors caused by the finite signal-to-noise ratios ($S/N$s) of the underlying spectra. Substantial metallicity- and temperature-dependent departures of the SSPP stellar parameter scales with respect to high-resolution reference samples have already been pointed out by \citeauthor{Smolinski11} (\citeyear{Smolinski11}; see, in particular, their Figure A2). The authors further compared SSPP results for [Fe/H] and $v_r$ for several GCs with literature values and claim overall ``good internal and external consistency'' despite the evidence they provide for a wealth of residual substructure. However, an independent analysis of SEGUE spectra using the flux-ratio-based parameterization method ATHOS \citep{Hanke18}  shows very similar residual trends with respect to SSPP. This observation contradicts the favored explanation by \citet{Smolinski11}, who speculate that deviations may originate from inhomogeneities among the high-resolution studies that were used as reference.

\begin{figure}
    \centering
    \resizebox{\hsize}{!}{\includegraphics{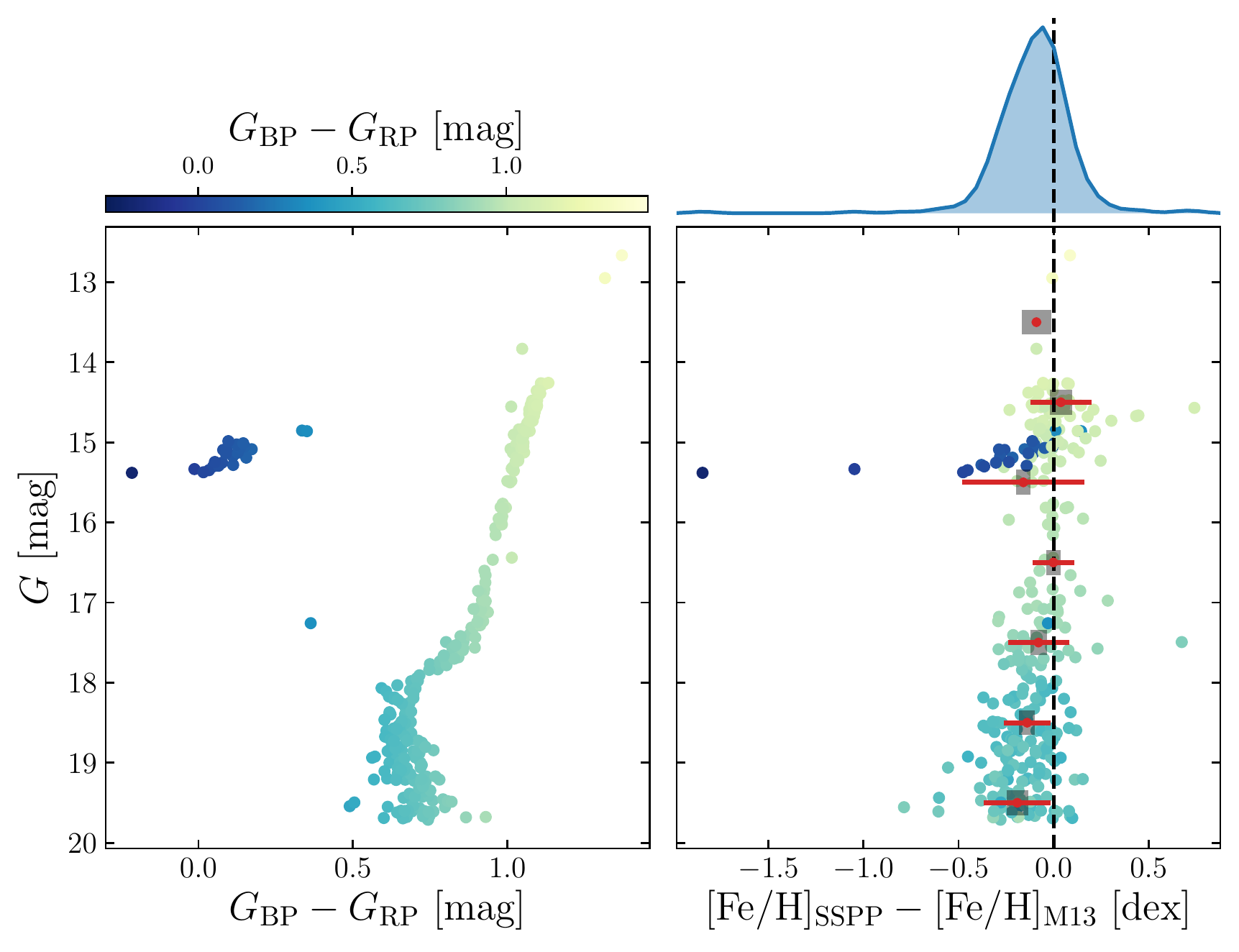}}
      \caption{Precision and accuracy validation of [Fe/H$]_\mathrm{SSPP}$ for M\,13 member stars. \textit{Lower left panel}: Proper-motion-cleaned CMD in the \gaia $G$, $G_\mathrm{BP}$, and $G_\mathrm{RP}$ passbands. For reference to the \textit{lower right panel}, $G_\mathrm{BP} - G_\mathrm{RP}$ is additionally indicated by colored symbols. \textit{Right panels}: Marginalized distribution of deviations (\textit{top}) and behavior with evolutionary state (\textit{bottom}) of SSPP metallicities with respect to the literature value: [Fe/H$]_\mathrm{M 13}=-1.53$~dex. Red points and error bars denote the means and standard deviations in bins of 1~mag  width in $G$, whereas underlying gray bars represent the median of the SSPP uncertainties on [Fe/H].
              }
      \label{Fig:SDSS_systematics}
\end{figure}
While performing an in-depth evaluation of different error sources, we compiled a sample of stars with a high probability of being associated with the GC \object{M\,13}. This cluster was chosen because it is well-studied and, due to its proximity, has been targeted with many SDSS fibers. Only SDSS targets that fall within a projected separation of one tidal radius from the cluster center were considered. Furthermore, we employed \gaia parallaxes and $v_\mathrm{r}$ from SDSS as a means to reject fore- and background stars by selecting only those with insignificant ($<5\sigma$) deviations from the mean values of the cluster. We note that at this point neither information about the chemical composition from SDSS nor \gaia proper motions entered the selection procedure. Nevertheless, as can be seen in Fig.~\ref{Fig:SDSS_systematics}, the resulting color--magnitude diagram (CMD) based on \gaia DR2 $G$ and $G_\mathrm{BP}-G_\mathrm{RP}$ photometry appears clean with a low degree of contamination (likely less than five stars out of 283). On this account, the vast majority of the sample can be assumed to be cluster members. When comparing SSPP results for [Fe/H] with the literature value \citep[$-1.53$~dex,][]{Harris96}, systematic discrepancies as a function of the evolutionary state of the individual stars (Fig.~\ref{Fig:SDSS_systematics}) become apparent. Specifically, stars on the upper main sequence (MS) deviate by $-0.18$~dex with a decreasing trend toward the subgiant branch ($\Delta\mathrm{[Fe/H]}\approx-0.10$~dex), while the RGB is consistent with a zero difference. Stars on the horizontal branch (HB) on the other hand differ by about $-0.20$~dex. Given the fact that M\,13 shows no signs of intrinsic metallicity spread \citep[e.g.,][]{Johnson12}, any such difference can be attributed to SSPP inaccuracies.

Again taking advantage of the fact that there are effectively no detectable metallicity spreads in M\,13, we can estimate the internal SSPP precisions for [Fe/H] at different evolutionary states from the observed scatter. Typically, this scatter is larger by a factor of three compared to the provided SSPP uncertainties. The latter comparison is illustrated in the right panel of Fig. \ref{Fig:SDSS_systematics}.

Given the caveats discussed above, we quadratically added a global systematic error of 0.15~dex to individual errors on [Fe/H] measurements\footnote{We emphasize that in doing so we are, strictly speaking, mixing a statistical source of error with an unrelated systematic component. Nonetheless, we treat this combined error as if it were of an entirely stochastic nature.} throughout this work. This drastically increases the majority of the quoted (internal) SSPP uncertainties. We further caution that the total error budget on $v_r$ cannot be less than 5~km~s$^{-1}$ \citep[see the comparison of SSPP radial velocities to high-resolution results by][]{Smolinski11}, which we accordingly adopted as another systematic error that is added in quadrature.

\subsubsection{\gaia astrometric solution}
For stars with an available five-parameter astrometric solution
\begin{equation}
 \vec{x}= \left( \alpha, \delta, \varpi, \mu_{\alpha}\cos{\delta}, \mu_\delta \right)^T,
\end{equation}
the \gaia DR2 data tables enable the computation of the full covariance matrix, $\Cov{(\vec{x})}$, for the solution of each individual star. Incorporating its off-diagonal entries is crucial for this work, in that the measurements of $\varpi$, $\mu_{\alpha}\cos{\delta}$, and $\mu_{\delta}$ may be correlated to varying degrees and thus must not be considered independent. Furthermore, we follow the technical note GAIA-C3-TN-LU-LL-124-01\footnote{\url{http://www.rssd.esa.int/doc_fetch.php?id=3757412}} and scale the covariance matrices by the squared re-normalized unit weight error (RUWE),
\begin{equation}
 \mathrm{RUWE}^2 = \frac{\chi^2}{(N-5)\cdot u_0^2(G, G_\mathrm{BP} - G_\mathrm{RP})},
\end{equation}
where $\chi^2$ is the chi-square value of the astrometric fit to all $N$ \textit{Gaia} measurements in the direction  along the scan that are considered ``good''. The factor $u_0$ is an empirically calibrated quantity that can be extracted from dedicated lookup tables as a function of $G$ and $G_\mathrm{BP} - G_\mathrm{RP}$\footnote{\url{https://www.cosmos.esa.int/web/gaia/dr2-known-issues}}. We emphasize that all employed covariance matrices involving \gaia data were scaled by $\mathrm{RUWE}^2$. In this work we moreover corrected for the quasar-based $\varpi$ zero point of $-0.029$~mas \citep{Lindegren18} and assumed an additional global systematic error for proper motions of 0.035~mas~yr$^{-1}$ \citep{Helmi18}.

\begin{figure}
    \centering
    \resizebox{\hsize}{!}{\includegraphics{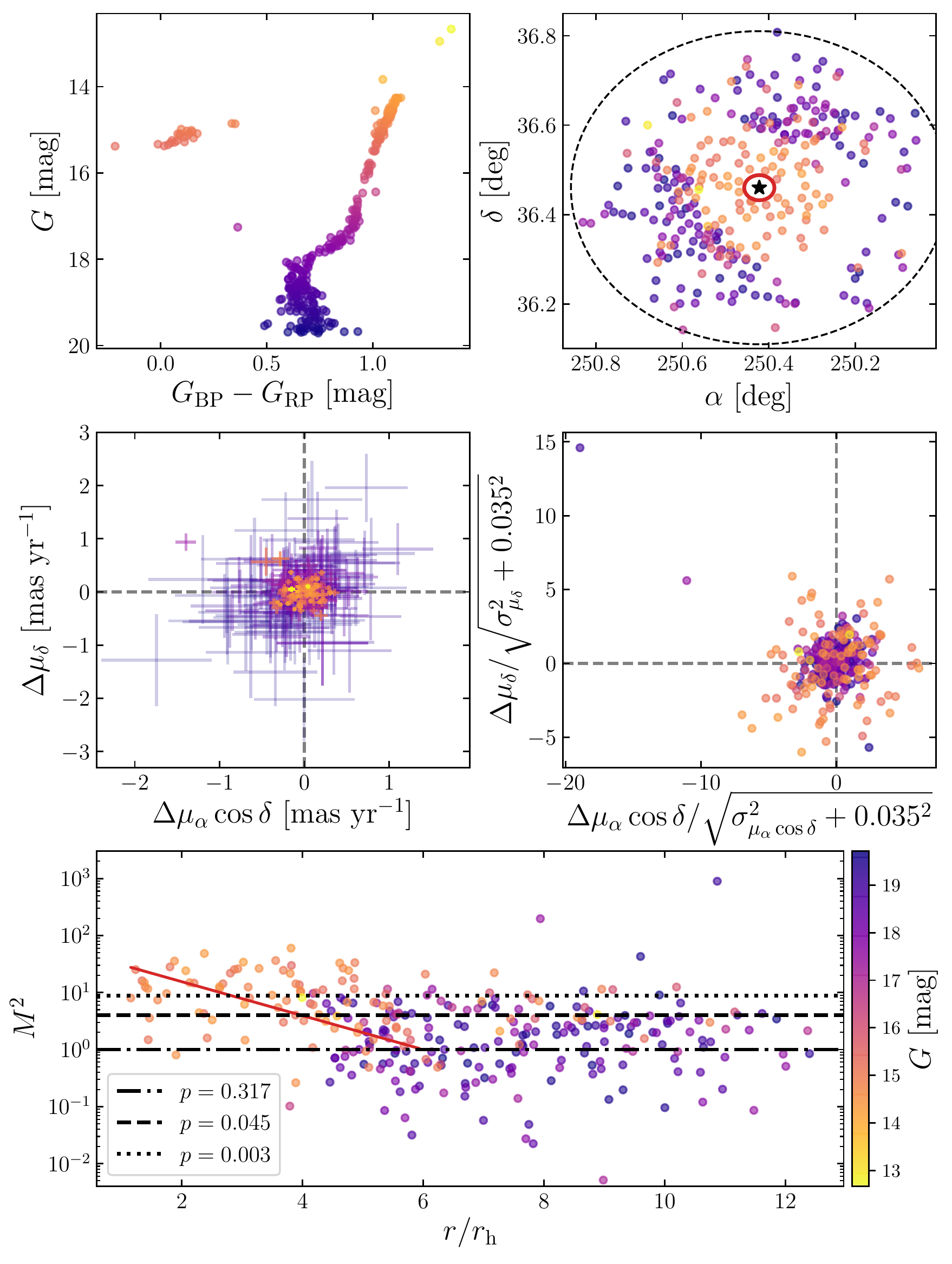}}
      \caption{\textit{Upper left}: Same CMD as in Fig.~\ref{Fig:SDSS_systematics}, but with color coding indicating the brightness (see \textit{lower right} color bar); the same color code is used in all other panels. \textit{Upper right}: Spatial distribution of the M\,13 validation sample. The cluster core, half-light radius (1.69~arcmin), and tidal radius (21~arcmin) are represented by a black star, red circle, and black-dashed line, respectively. We emphasize that the tidal radius together with $\varpi$ and $v_r$ were the only selection criteria applied to investigate proper motion systematic errors (main text). \textit{Middle panels}: Absolute (\textit{left}) and relative (\textit{right}) deviation in proper motion of the sample from the cluster mean value. \textit{Lower panel}: Squared Mahalanobis distance, $M^2$, from proper motions only versus projected distance from the cluster center. In analogy to the $1\sigma$,  $2\sigma$, and  $3\sigma$ significances of a normal distribution, dash-dotted, dashed, and dotted lines represent the corresponding $p$-values (see legend). The red line denotes the scaling relation introduced in Eq. \ref{Eq:pm scaling relation}.
              }
      \label{Fig:Gaia_pm_systematics}
\end{figure}
In elaborating on the formal DR2 proper-motion errors we again employ the sample of M\,13 stars introduced earlier in this chapter. Their spatial distribution is presented in Fig.~\ref{Fig:Gaia_pm_systematics} next to the corresponding CMD. We further show the distribution of the selection in absolute and relative (i.e., scaled by standard errors) proper-motion differences with respect to the mean values for M\,13. While bright stars are highly clustered in an absolute sense, from Fig.~\ref{Fig:Gaia_pm_systematics} it becomes apparent that there is an inversion of the relative distribution widths of bright and faint targets when going from absolute to relative proper motion differences. It is noteworthy that the RUWE remains below 1.5 for all but nine targets in the selection, four of which are fainter than $G=17$~mag. Hence, for the vast majority of our bright stars, inseparable blends at the edge of the detection limit can be excluded with high confidence. We therefore attribute our observation to an additional, hitherto unexplained systematic error component of the proper motions in excess of the already applied 0.035~mas~yr$^{-1}$.  

Although we neglected covariances for the middle panels of Fig. \ref{Fig:Gaia_pm_systematics} for illustrative purposes, owing to the discussed correlations among the components of $\vec{y}=(\mu_{\alpha}\cos{\delta},~\mu_\delta)^T$, it is not straightforward to estimate the statistical significance of the deviation from the cluster mean, $\vec{y}_\mathrm{GC}$. Hence, for each star, we computed the squared Mahalanobis distance
\begin{equation}\label{Eq: Mahalanobis distance}
M^2 = (\vec{y}-\vec{y}_\mathrm{GC})^T \Cov{(\vec{y}-\vec{y}_\mathrm{GC})}^{-1} (\vec{y}-\vec{y}_\mathrm{GC})
,\end{equation}
which respects the combined covariance. Under the assumption of normally distributed errors, the latter quantity is chi-square distributed. Therefore, the $p$-value -- that is, the probability of finding a value of $M^2$ or more extreme under the null hypothesis that the star is not kinematically distinct from the cluster -- can be computed from a chi-square distribution with one degree of freedom. In Fig.~\ref{Fig:Gaia_pm_systematics}, $M^2$ is plotted against the projected angular distance to the cluster center, $r$, normalized by the half-light radius, $r_\mathrm{h}$. We identify two distinct groups of highly significant outliers that would be rejected in Sect. \ref{Subsec: Extratidal escapee candidates around clusters} based on their proper motions alone. First, starting from $r/r_\mathrm{h}\approx6$, there is a trend of increasingly significant deviation with decreasing $r$. Secondly, irrespective of the separation from the cluster core, bright stars ($G\lesssim16$~mag) tend to deviate more significantly than fainter ones. Based on the clean CMD and the $v_\mathrm{r}$-based association, we exclude the possibility that the majority of the strong proper motion outliers do not in fact belong to the cluster and conclude that the origin is genuinely to be found in underestimated \gaia errors. We compensate for the latter by introducing a distance-dependent factor
\begin{equation}\label{Eq:pm scaling relation}
    \alpha(r/r_\mathrm{h})= 
\begin{cases}
    10^{-0.3(r/r_\mathrm{h}-6)},              & \text{if } r/r_\mathrm{h}\leq 6\\
    1,              & \text{otherwise,}
\end{cases}
\end{equation}
to be applied to the covariances. Though optimized for the M\,13 stars, we note that we found the relation
to hold true for several other GCs from our sample  within an acceptable amount of variance.

\section{Methods}\label{Sec: Methods}
In the following sections, we introduce the three approaches that were used to identify potential former GC members. The first method is tailored to test for stellar associations in the immediate cluster surroundings and does not require all of the three space motions and positions to be measured with the same precision. The second approach ties field CN-strong stars to clusters and bears the advantage of not being limited to a comparatively small patch of the sky, though at the expense of obtaining results that are strongly impacted by the lowest-precision entry in the phase-space vector. The third and last method is an adaptation of the first method, where we search for chemically normal stellar populations that share the same metallicity and kinematic properties as CN-strong stars in the halo field, possibly indicating a common origin.

In all three approaches we employed variations of the Bayesian approach described in \citet{Anderson13}. In brief, the association probability (posterior) of a star being attributable to a reference object (be it a CN-strong star or a GC) is given by
\begin{equation}\label{Eq:Bayes theorem}
 P(A|B) = \frac{P(B|A)\times{P(A)}}{P(B)}\propto P(B|A)\times P(A).
\end{equation}
Here, $P(B|A)$ is the conditional probability (likelihood) of encountering the star at its observed position in information space assuming it is indeed associated with (or used to be an intratidal member of) the reference. Further, $P(B)$ is the probability of observing the data and $P(A)$ is the initial degree of belief in association (prior). 

\subsection{Method I: Stars in the neighborhood of clusters}\label{Subsec: Stars in the neighborhood of clusters}
The potential GC origin of individual stars in the cluster vicinity (out to a few degrees separation) was evaluated on a cluster-by-cluster basis. Stars that escaped the GC potential and show a significant spatial separation from their previous host -- while conserving their metallicity -- no longer necessarily share the same space-motion vector as the cluster. We instead presume that their currently observed motion should show a stream-like behavior and thus be similar (though not exactly identical; see Sects. \ref{Subsec: Associating CN-strong stars with clusters} and \ref{Subsubsec: Associations to the Sagittarius stream}) to the closest point along the hypothesized stream.

To implement this behavior, we extrapolated the orbits of all GCs in the list by \citet{Vasiliev19} to 2~Gyr in the past and future by performing point-particle integrations employing the python library \texttt{galpy} \citep{Bovy15} and its standard Galactic potential \texttt{MWPotential2014}. As a means to track the behavior of stars lost from the clusters, we further used \texttt{galpy} for the modeling of dynamically cold ($\sigma_v=1$~km~s$^{-1}$) leading and trailing tidal streams. We introduced tracer particles along the streams with parameter covariances that account for the simulated distribution functions of the streams. From these, the likelihood terms in Eq. \ref{Eq:Bayes theorem} were calculated individually for each star based on the difference $\vec{\Delta z}$ between the five-component vector $\vec{z}=(\mathrm{[Fe/H]}, v_\mathrm{r}, \varpi, \mu_{\alpha}\cos{\delta},\mu_\delta)^T$ and the corresponding closest stream anchor point, $\vec{z}_\mathrm{s}$. Again following \citet{Anderson13}, in analogy to Eq. \ref{Eq: Mahalanobis distance}, we computed the squared Mahalanobis distance $M^2$. Here we used the \gaia covariance entries for $\varpi$, $\mu_{\alpha}\cos{\delta}$, and $\mu_\delta$ and assumed no correlation between SDSS and \gaia quantities. Furthermore, covariances between cluster mean proper motions were taken from \citet{Vasiliev19}. Complete independence of the cluster quantities from the stellar quantities was presumed, such that the cross-covariances between $\vec{z}$ and $\vec{z_\mathrm{s}}$ are zero. With respect to the second assumption, while our potential extratidal candidates were not used by \citet{Vasiliev19} to constrain mean cluster parallaxes and proper motions, this does not necessarily hold for our supposedly bound cluster members. \citet{Vasiliev19} on the other hand commonly employed several orders of magnitude more stars from the \gaia tables than there are counterparts in the SDSS catalog. As a consequence, the overlap and therefore cross-covariances are minor. Interdependencies between $\varpi$, [Fe/H], $v_r$, and their respective mean cluster values can be excluded, as the cluster parallaxes (by means of inverse distances) and [Fe/H] are based on the Harris catalog of GC parameters \citep{Harris96}, and $v_r$ stems from the collection of ground-based measurements compiled by \citet{Baumgardt19}. None of the former are in any way connected to SDSS or \textit{Gaia}.

Finally, the likelihood can be expressed as 
\begin{equation}\label{Eq:likelihood}
 P(B|A) = 1-p(M^2),
\end{equation}
where $p(M^2)$ is the $p$-value of a $\chi^2$ distribution with five degrees of freedom. \citet{Anderson13} emphasize that large errors -- translating into low-significance values -- cannot rigidly exclude a large portion of their corresponding phase-space dimensions and thus do not limit the high-likelihood regime to a confined range. A prime example of this behavior is $\varpi$, which -- in light of typically large cluster distances -- hardly exceeds the $2\sigma$ significance level for most of the stars that are deemed cluster members below. Nonetheless, $\varpi$ is a powerful means to reject the large number of foreground stars, exposing significant parallaxes that are inconsistent with cluster association.

In contrast to \citet{Anderson13}, who demanded their targets be gravitationally bound to the clusters, we cannot use this criterion for extratidal stars and hence assume a loose prior of the form
\begin{equation}\label{Eq: spatial prior}
    P(A)= \mathrm{exp}\left(-\frac{1}{2}\left(\frac{d_\mathrm{s}}{7r_\mathrm{t}}\right)^2\right),
\end{equation}
with $d_\mathrm{s}$ being the projected distance of a star to the closest tracer stream particle and $r_\mathrm{t}$ being the tidal radius of the cluster. This prior formulates the initial belief that former members at arbitrarily large angular separation will not share the same $\vec{z}$.

\begin{figure*}
    \centering
    \resizebox{0.8\hsize}{!}{\includegraphics{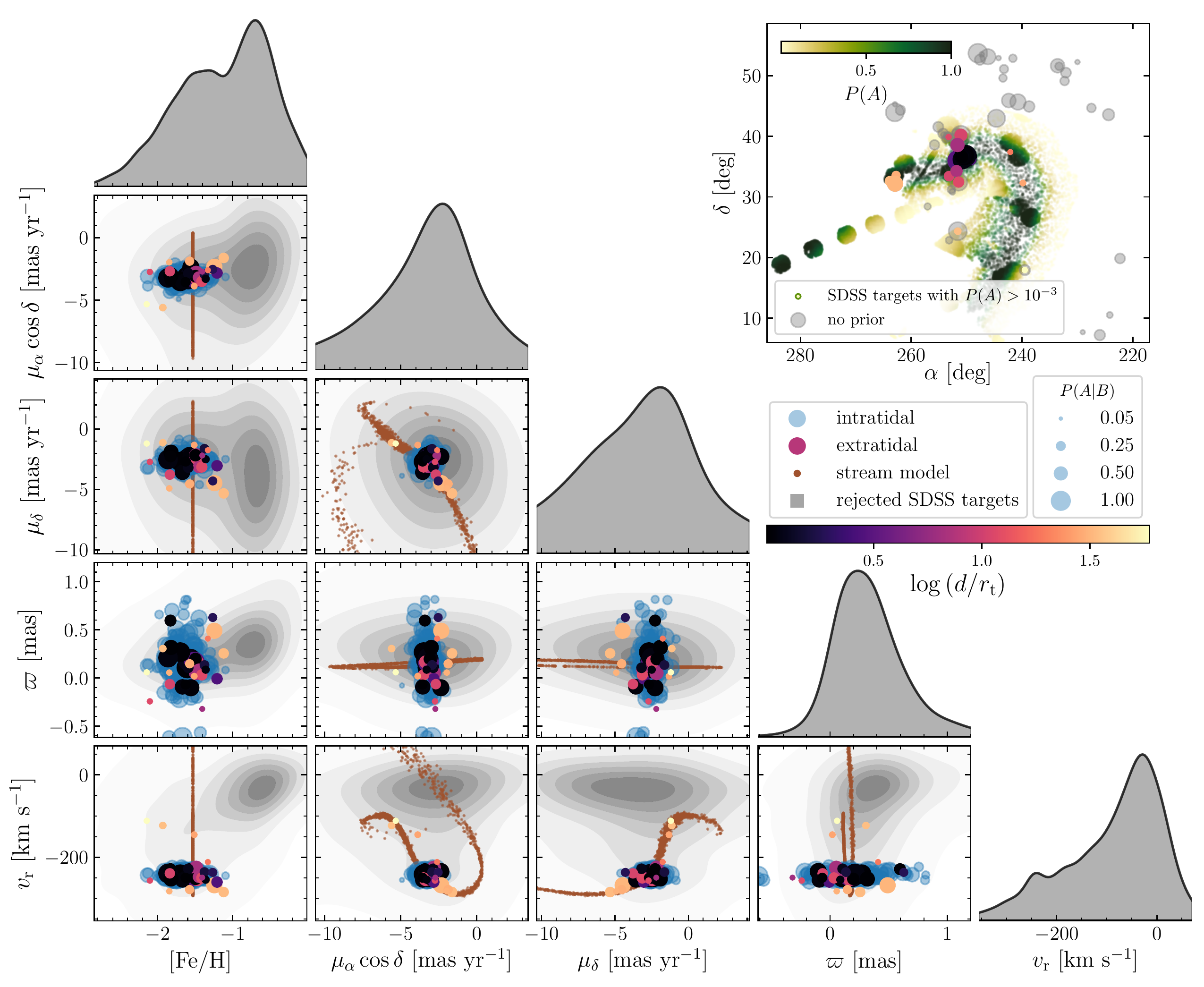}}
      \caption{Five-parameter chemodynamical association criteria used to constrain the association with the GC M\,13. The rejected back- and foreground population of stars with SSPP parameters and \gaia DR2 motions is indicated with gray density contours. Brown dots indicate the simulated stream (see main text). Intratidal and extratidal stars are depicted with blue and colored circles (see also Fig.~\ref{Fig:M13_extradtidal}), respectively. Circle sizes indicate the association probability and colors resemble the projected distance to the cluster center (see legend and color bar). The marginalized distributions of each coordinate on the abscissa are shown on top of the two-dimensional representations. The adopted spatial prior as well as the stars that are rejected due to its usage (gray circles with no colored counterpart) are illustrated in the {upper right-hand} panel.
              }
      \label{Fig:M13_chemodyn_selection}
\end{figure*}
Figure~\ref{Fig:M13_chemodyn_selection} illustrates a graphical representation of the multivariate association procedure for the exemplary case of the cluster M\,13. Candidates are reported if their associated $P(A|B)$ exceeds the threshold of 0.05. Both intra- and extratidal stars are treated in the same way.

\subsection{Method II: Integrals of motion}\label{Subsec: Integrals of motion}
It is possible that GC escapees that did not 
recently become unbound from their host cluster may no longer be found in the immediate cluster vicinity; such GCs would not be recovered by the approach in the previous section. Thus, for the manageably small sample of CN-strong stars used here, we resort to the fact that for most orbits in axisymmetric potentials there exists a set of three conservative integrals of motion \citep[e.g.,][]{Henon64}; in other words, despite being spatially separated, escapees almost completely retain the integrals of motion\footnote{We emphasize that this is only approximatively true, because for a star to become unbound its phase-space position already has to be distinct from the main body of the cluster.} of  their host \citep[e.g.,][]{Savino19}. We used the \texttt{galpy} implementation of the St\"ackel approximation by \citet{Binney12} in order to integrate the axisymmetric \texttt{MWPotential2014} for the radial and vertical actions, $J_r$ and $J_z$, as well as the azimuthal component of the angular momentum, $L_z$. With respect to the previous approach, such integrations bear the main disadvantage of relying on the full six-component phase-space vector to initialize an orbit, such that uncertainties are strongly driven by the least constrained observational quantity. While the latter restriction is not a major concern for our sample of GCs, the lack of highly significant measurements for the heliocentric distance, $D$, to the CN-strong stars ultimately drives the confidence for rejecting or accepting a potential cluster association. \gaia
parallax significances $\varpi/\sigma_\varpi<1$   have been attributed to 61\% (68 stars) of our sample whereas only three measurements exceed the $4\sigma$ level. It is evident that any distance inferred from $\varpi$ alone \citep[e.g.,][]{Bailer-Jones18} will result in critical errors in the actions.

\begin{figure*}
    \centering
    \resizebox{\hsize}{!}{\includegraphics{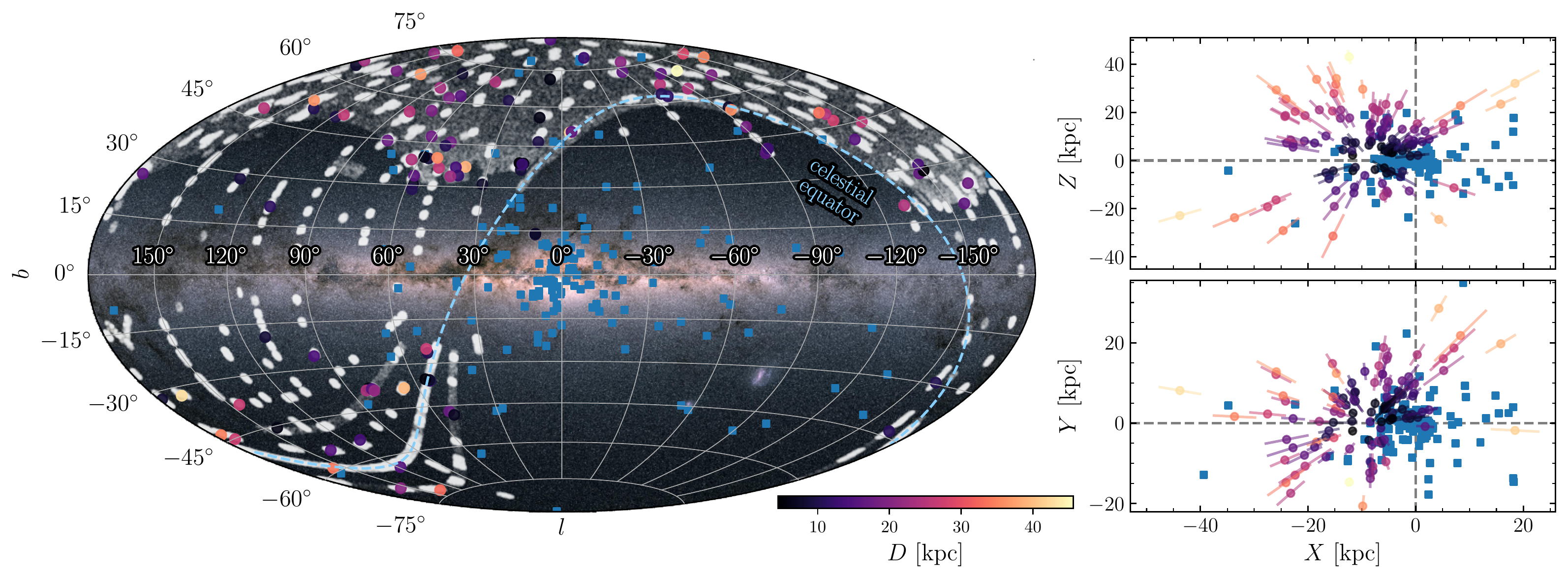}}
      \caption{Spatial distribution of GCs (blue squares) and CN-strong halo giants (colored according to their distance) in Galactic coordinates (\textit{left}) and the Cartesian Galactocentric frame (\textit{right}). The dashed line in the \textit{left panel} denotes the celestial equator. The SDSS footprint area is depicted in white while the \gaia all-sky map is illustrated in the background for orientation. \textbf{Image credit.} \textit{Gaia Data Processing and Analysis Consortium (DPAC); A. Moitinho / A. F. Silva / M. Barros / C. Barata, University of Lisbon, Portugal; H. Savietto, Fork Research, Portugal.} 
              }
      \label{Fig:CN_strong_spatialdist}
\end{figure*}
\begin{figure}
    \centering
    \resizebox{\hsize}{!}{\includegraphics{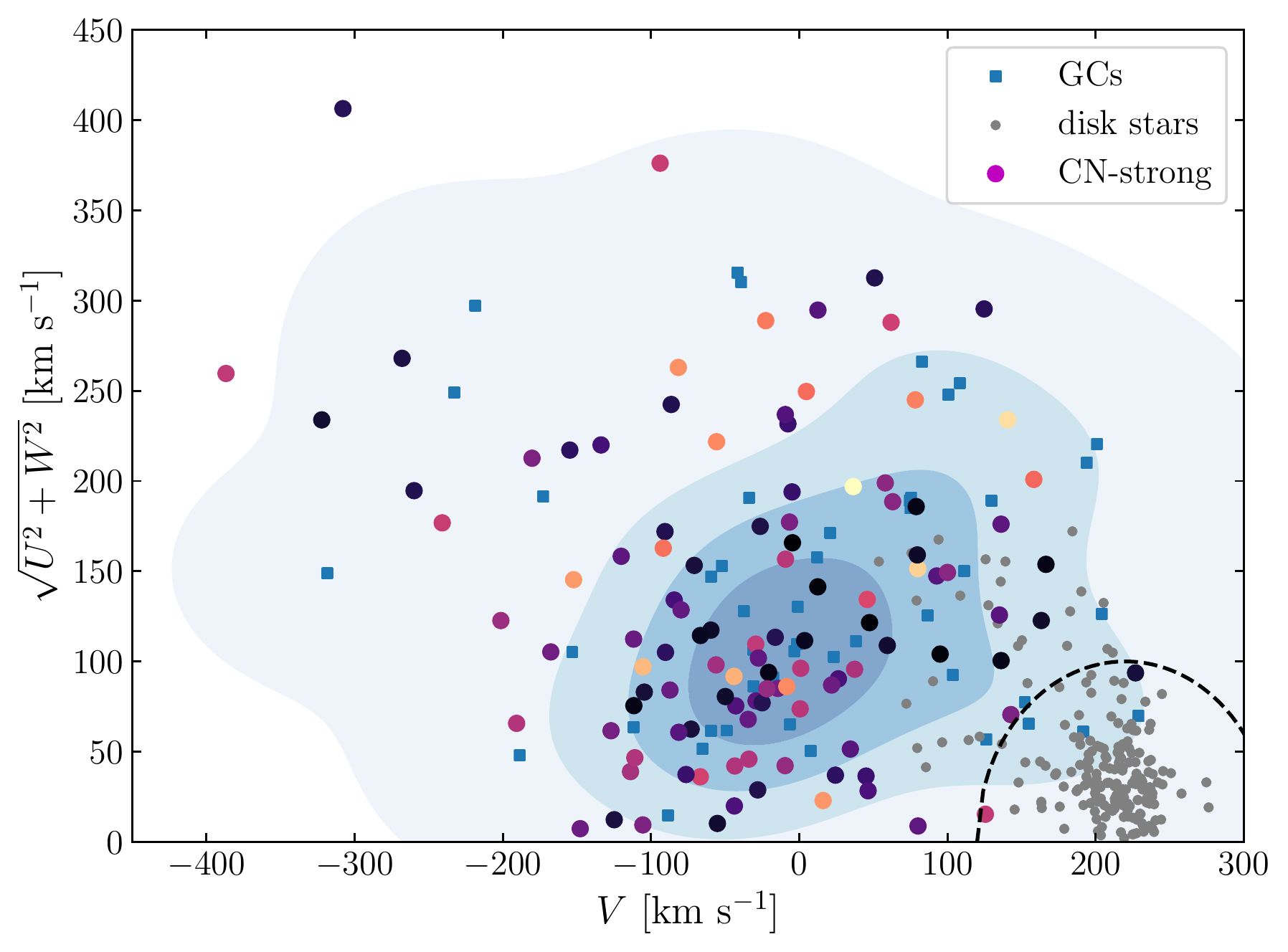}}
      \caption{Toomre diagram for our CN-strong giants (colored  circles) and the Galactic GC population (blue squares). The latter is only shown in the regime $-1.8<\mathrm{[Fe/H]}<-1.3$~dex, representing the coverage of the analyzed stars. The color coding of the CN-strong stars is the same as in Fig.~\ref{Fig:CN_strong_spatialdist}. For better comparability, an additional Gaussian kernel density estimate of the GC sample is provided where increasing color darkness resembles increasing density. To show the clear contrast to the Galactic disk population, we also indicate all targets from our \gaia - SDSS crossmatch that obey $-5^\circ<b<5^\circ$ and have a spectrophotometric distance as determined by SDSS of less than 1~kpc (gray dots). Moreover, the dashed line encircles the region of absolute velocities of less than 100~km~s$^{-1}$ with respect to the local standard of rest.
              }
      \label{Fig:Toomre_diagram}
\end{figure}
In order to obtain more precise estimates for $D$, we refined the spectrophotometric formalism from Paper~I by adopting the Bayesian inference method for stellar distances by \citet{Burnett10} using photometry and stellar parameters from SDSS and $\varpi$ as an additional constraint \citep[cf., e.g.,][]{Savino19}. Therefore, a likelihood was computed from the residuals between the observed quantities (color, apparent magnitude, $T_\mathrm{eff}$, [Fe/H], $\log{g}$, and $\varpi$) and their theoretical counterparts from a grid of PARSEC \citep{Marigo17} isochrones. As prior, we used the three-component Galactic model adopted by \citet{Burnett10} for the thin and thick disks and the stellar halo, the latter being constructed from the parameters stellar age, [Fe/H], and mass. The magnitude and variance of $D$ were deduced by means of the first and second moments of the star's posterior probability density function (pdf) marginalized in all other dimensions \citep[cf., e.g., ][]{Savino19}. Even though the nominal errors on the inferred distances are small and consequently the statistical significances are high (nearly 50\% reside above $D/\sigma_D=10$), we caution that there is ample room for various systematic errors that are not captured by the mentioned treatment. One example is the SSPP surface gravity -- a quantity that is notoriously hard to obtain with any accuracy from low-resolution spectra -- which may strongly favor much larger distances in cases where a star was erroneously classified as a low-gravity giant whilst in fact being a giant of intrinsically higher gravity or even a dwarf star.

Adopting the former distances, we present the resulting spatial distribution of our CN-strong giants in the Galactocentric frame on the right-hand side of Fig.~\ref{Fig:CN_strong_spatialdist}, while a Hammer projection in the Galactic frame is shown on the left-hand side. Furthermore, in Fig.~\ref{Fig:Toomre_diagram}, we show a Toomre diagram of our CN-strong stars and GCs of the same [Fe/H], from which, qualitatively, a remarkable distribution overlap can be seen. Nonetheless, we envision a more quantitative approach, exploiting the kinematic memory of our targets. 

\begin{figure}
    \centering
    \resizebox{\hsize}{!}{\includegraphics{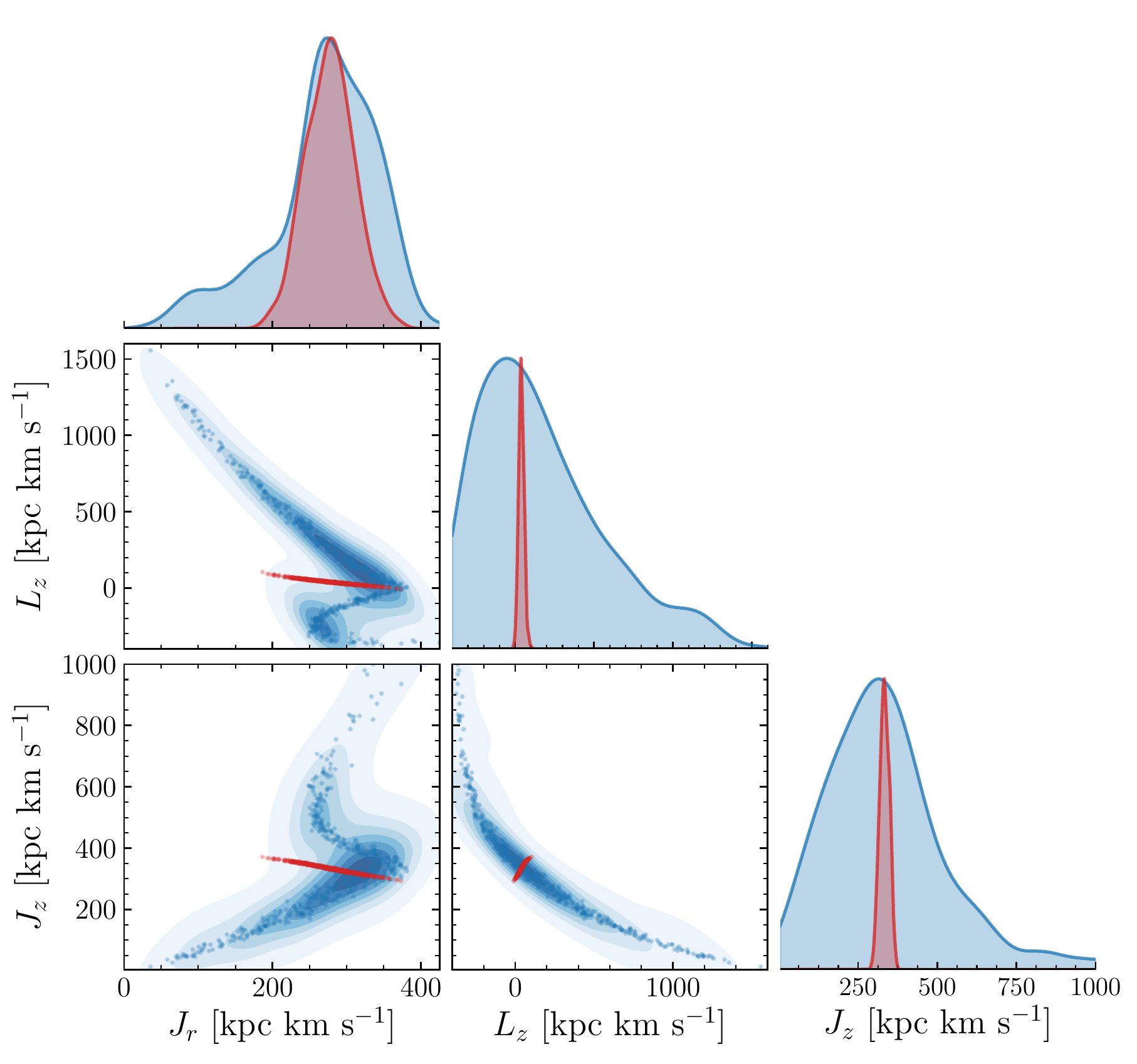}}
      \caption{Distribution of Monte Carlo samples for the integrals of motion for M\,70 (red) and the star \gaia DR2 3833963854548409344 (blue). The marginalized two-dimensional and one-dimensional kernel density estimates are indicated as contours in the frames and histograms at the top, respectively. For illustrative purposes, the histograms were normalized by their maxima. 
              }
      \label{Fig:actions_corner_plot}
\end{figure}
To this end, all a priori known statistical error sources on the phase space vectors for both GCs and CN-strong stars were propagated in the action integration by means of a Monte Carlo analysis. Each orbit was initialized 500 times with values randomly drawn from a multivariate normal distribution respecting the covariance matrix, where we assumed cross-covariances between $\alpha$, $\delta$, $D$, and $v_\mathrm{r}$ to be negligible while maintaining the covariances for \gaia proper motions. As shown by the exemplary comparison in Fig.~\ref{Fig:actions_corner_plot}, the obtained distributions in action space are highly coupled and non-normal, that is, the relations between action coordinates show a strong nonlinear behavior. It is apparent that, whilst cluster (\object{M\,70}) and star (\gaia DR2 3833963854548409344) indicate a perfect match in the marginalized, one-dimensional distributions,  in the multidimensional representation there is only a much smaller distribution overlap. If not properly accounted for, this effect would overestimate the association probability in most cases. 

To test the possibility of former membership of a CN-strong halo star to a cluster, we required both the actions and  [Fe/H]  to be consistent. Hence, we compared the cluster vector $(J_r,J_z,L_z,\mathrm{[Fe/H]})_\mathrm{GC}^T$ to the corresponding vector for the star. Due to the non-normality of the MC samples mentioned above, we cannot simply compute the covariance matrices of the samples and proceed with the Mahalanobis distance employed in Sect. \ref{Subsec: Extratidal escapee candidates around clusters}. Therefore, we approximated the four-dimensional probability density function, $\psi(\vec{v})$, of the difference vector, $\vec{v}$, using a nonparametric Gaussian kernel density estimate. We note that this is the main difference between our approach and that of \citet{Savino19}, who did not account for nonlinear correlations between actions. As a means to accept or reject association, we obtained
\begin{equation}
 p = \int\displaylimits_{\psi(\vec{v}') < \psi(\vec{0})} \psi(\vec{v}') d\vec{v}'
\end{equation}
through Monte Carlo integration.

\subsection{Method III: Chemodynamical matches in the fields surrounding CN-strong stars}\label{Subsec: Chemodynamical matches in the fields surrounding CN-strong stars}
Following the predictions made by evolutionary GC models \citep[e.g.,][]{Dercole08, Bastian13}, a substantial number of first-generation stars are lost early on in the cluster formation phase. This unenriched population is chemically no different from the standard collection of halo stars and thus remains unidentifiable for classical tagging methods (e.g., Paper I). Yet, by exploring chemodynamical similarities between the enriched, bona fide second-generation cluster stars in the Galactic halo and their surrounding field, we can search for links that could indicate a common origin. Therefore, we once again employed the formalism of Sect. \ref{Subsec: Stars in the neighborhood of clusters} with CN-strong stars taking over the role of the GCs. The main difference between this latter method and the one described here is that we did not attempt to integrate the orbit of the CN-strong stars because the much larger errors on the astrometry of these individual stars do not allow for the computation of the tightly constrained orbits required to confidently reject field interlopers. As a spatial prior, in analogy to Eq. \ref{Eq: spatial prior}, we adopted a Gaussian centered on the CN-strong star with a standard deviation of seven degrees in projected distance.

\section{Results and Discussion}\label{Sec: Results}
\subsection{Extratidal escapee candidates around clusters}\label{Subsec: Extratidal escapee candidates around clusters}
\begin{table*}
\caption{Clusters for which extratidal associations were singled out by method I.}
\label{Table: cluster information}
\centering
\resizebox{\textwidth}{!}{%
\begin{tabular}{lcccccrrcrcrr}
\hline\hline
\\[-5pt]
Name & $\alpha^{(a)}$ & $\delta^{(a)}$ & $r_\mathrm{h}^{(b)}$ & $r_\mathrm{t}^{(b)}$ & $D^{(b)}$ & \multicolumn{1}{c}{$v_\mathrm{r}^{(a)}$} & \multicolumn{1}{c}{$\mu_{\alpha}\cos{\delta}^{(a)}$} & $\mu_\delta^{(a)}$ & \multicolumn{1}{c}{$r_\mu^{(a)}$} & [Fe/H]$^{(b)}$ & $N_{<r_t}^{(c)}$ & $N_{>r_t}^{(c)}$\\
 & [deg] & [deg] & [arcmin] & [arcmin] & [kpc] & \multicolumn{1}{c}{[km s$^{-1}$]} & \multicolumn{1}{c}{[mas yr$^{-1}$]} & [mas yr$^{-1}$] & & [dex] & &\\[3pt]
\hline\\[-8pt]
NGC 4147                       & 182.526 & $18.543$ &  0.5 &  6.1 &  19.3 & $  179.1\pm 0.3$ & $ -1.702\pm0.016$ & $ -2.108\pm0.010$ & $-0.147$ & $-1.80$ &   3 &  19\\
NGC 5024 (M 53)                & 198.230 & $18.168$ &  1.3 & 18.4 &  17.9 & $  -63.1\pm 0.2$ & $ -0.151\pm0.006$ & $ -1.350\pm0.004$ & $-0.261$ & $-2.10$ &  36 &   7\\
NGC 5053                       & 199.113 & $17.700$ &  2.6 & 11.4 &  17.4 & $   42.5\pm 0.2$ & $ -0.353\pm0.009$ & $ -1.257\pm0.007$ & $-0.334$ & $-2.27$ &  26 &  24\\
NGC 5272 (M 3)                 & 205.548 & $28.377$ &  2.3 & 28.7 &  10.2 & $ -147.2\pm 0.2$ & $ -0.094\pm0.004$ & $ -2.626\pm0.003$ & $-0.064$ & $-1.50$ & 181 &  18\\
NGC 6205 (M 13)                & 250.422 & $36.460$ &  1.7 & 21.0 &   7.1 & $ -244.4\pm 0.3$ & $ -3.172\pm0.004$ & $ -2.586\pm0.005$ & $ 0.152$ & $-1.53$ & 260 &  32\\
NGC 6341 (M 92)                & 259.281 & $43.136$ &  1.0 & 12.4 &   8.3 & $ -120.7\pm 0.3$ & $ -4.923\pm0.005$ & $ -0.556\pm0.006$ & $ 0.085$ & $-2.31$ &  75 &  35\\
NGC 7078 (M 15)                & 322.493 & $12.167$ &  1.0 & 27.3 &  10.4 & $ -106.5\pm 0.2$ & $ -0.622\pm0.006$ & $ -3.782\pm0.006$ & $-0.037$ & $-2.37$ &  91 &   5\\
NGC 7089 (M 2)                 & 323.363 & $-0.823$ &  1.1 & 12.4 &  11.5 & $   -3.6\pm 0.3$ & $  3.531\pm0.007$ & $ -2.139\pm0.007$ & $ 0.027$ & $-1.65$ &  66 &  11\\

\hline
\end{tabular}}
\tablefoot{
\tablefoottext{a}{\citet{Vasiliev19}.}
\tablefoottext{b}{\citet{Harris96}, 2010 version.}
\tablefoottext{c}{This study.}
}
\end{table*}
In this section, we present the discovery of new candidates for extratidal stars around GCs in the halo field based on their position in multi-dimensional information space. We started by cleaning the SSPP catalog for stars lacking a [Fe/H] or $v_r$ detection. The remaining list was subsequently cross-matched with the \gaia DR2 source catalog. This vetting process -- due to the SDSS sky coverage being limited to the northern sky (cf., Fig. \ref{Fig:CN_strong_spatialdist}) -- naturally rejected the southern GCs and left us with about $3.7\cdot 10^5$ stars for further analysis. The relevant cluster properties entering the analysis can be found in Table \ref{Table: cluster information}.

We consider those stars with $P(A|B)>0.05$ as bona fide cluster associates. A further sub-classification into extra- and intratidal candidates is performed based on whether the projected distances of the stars do or do not exceed the tidal radius of the cluster, $r_\mathrm{t}$. We note the major caveat of this distinction to be the potential exclusion of truly extratidal stars that, owing to their line-of-sight position, happen to be projected into the tidal radius of the cluster. Unfortunately, the already mentioned low significance of $\varpi$ for these distant objects cannot yield additional constraints and prohibits their identification.

In the following sections, we comment on our  151 extratidal candidates with likely associations with eight GCs. The discussion is separated by individual host clusters. A summary of the properties of all halo stars with potential GC origin is provided in Table \ref{Table: Extratidal target information}. 

\subsubsection{M 13 (NGC 6205)}\label{Subsubsec: M 13 (NGC 6205)}
\begin{figure}
    \centering
    \resizebox{\hsize}{!}{\includegraphics{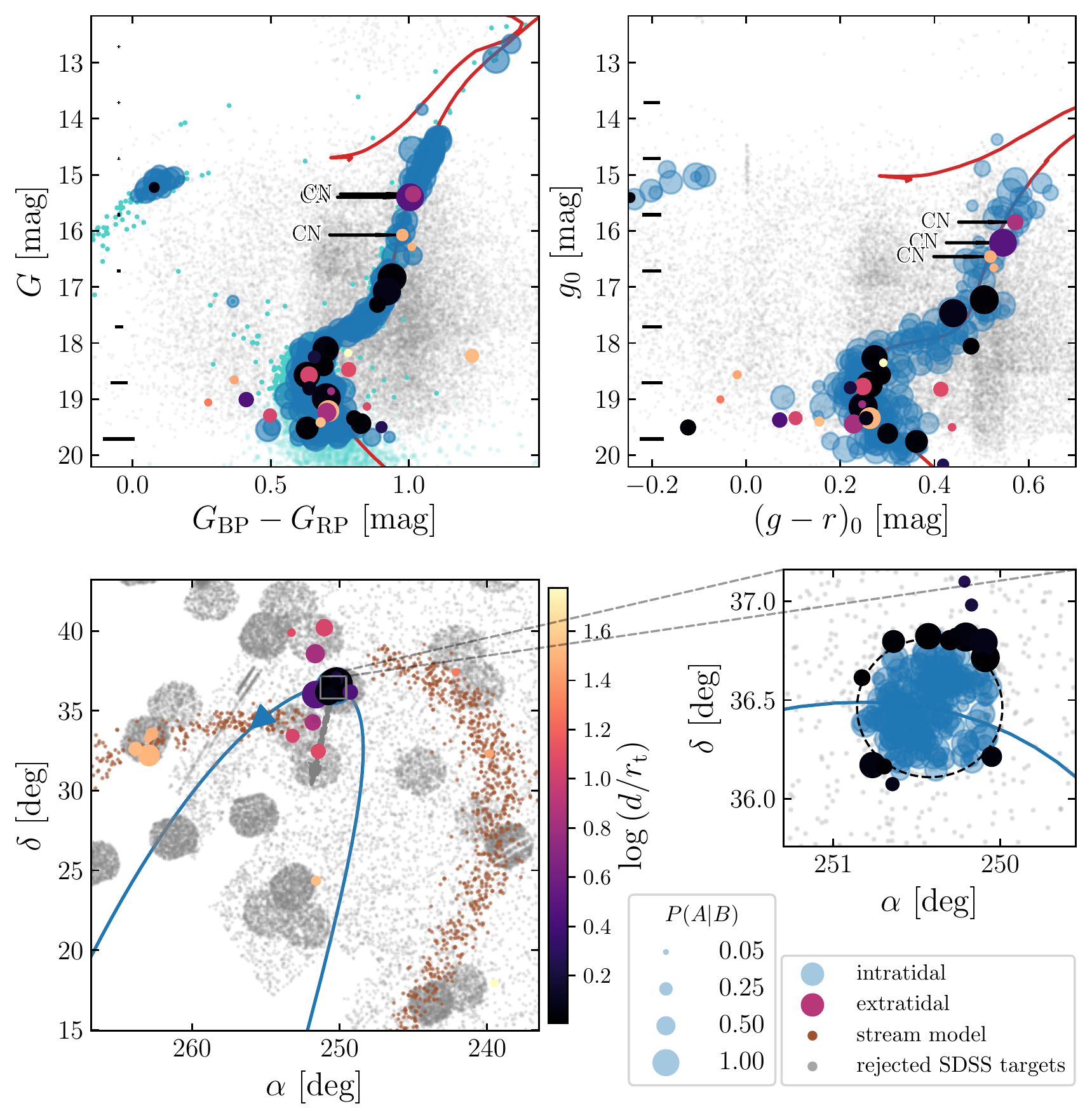}}
      \caption{Parameter distributions of intra- (blue) and extratidal (black to light yellow) stars associated with M\,13 on top of the field star population covered by the SDSS footprint (gray dots). The sizes of the colored circles are directly proportional to the membership probability $P(A|B)$ as indicated in the legend, while the color of the extratidal candidates denotes the projected distance from the cluster core in units of $r_\mathrm{t}$ (see color bar). \textit{Upper panels}: CMDs from \gaia (\textit{left}) and dereddened SDSS (\textit{right}) bands. Cyan dots indicate proper motion and $\varpi$-selected \gaia sources without spectroscopic SDSS counterparts (see main text for details). Typical photometric errors are specified in each case by error bars on the left-hand side. For a subset of bright intratidal stars, reliable color information from SDSS is not available, which is why those targets were omitted in the \textit{right} CMD. Red lines are 12~Gyr PARSEC isochrones \citep{Marigo17} matching the parameters from Table \ref{Table: cluster information} with an applied manual shift in color for the \gaia CMD. The CN-strong giants classified in Paper~I are labeled ``CN'' in both panels. \textit{Lower panels}: Distribution on the sky (\textit{left}) and a zoomed-in view of the immediate cluster vicinity (\textit{right}). The black-dashed ellipse indicates the tidal radius, $r_\mathrm{t}$, and the direction to the Galactic center is marked by a gray arrow. Properties for the integrated cluster orbit and the simulated stream are indicated by the blue curve and brown dots, respectively.
              }
      \label{Fig:M13_extradtidal}
\end{figure}
\begin{figure*}
    \centering
    \resizebox{\hsize}{!}{\includegraphics{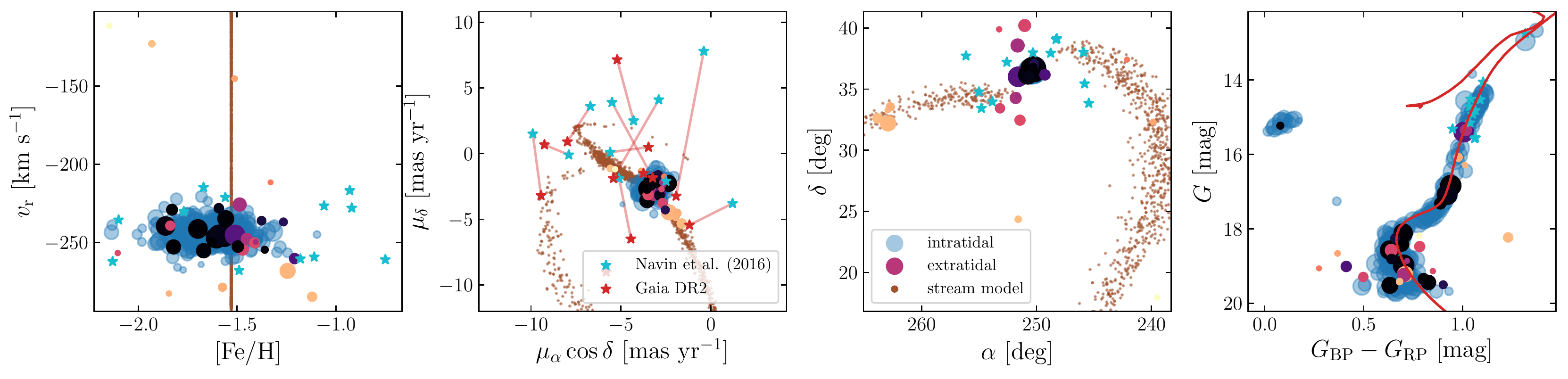}}
      \caption{Comparison of the chemodynamical (\textit{left two panels}) and spatial (\textit{third panel}) distributions and the CMD (\textit{right}) for M\,13 between \citeauthor{Navin16} (\citeyear{Navin16}; cyan stars) and the present study (same color and size scheme as in Fig.~\ref{Fig:M13_extradtidal}). In the proper motion diagram, the values from the original study are connected by red lines with their \gaia DR2 counterparts (red stars).
              }
      \label{Fig:M13_extradtidal_Navincomp}
\end{figure*}
M\,13 is an old \citep[12.3~Gyr,][]{VandenBerg13}, intermediately metal-poor \citep[$-1.5$~dex,][]{Johnson12} GC known to host extreme light-element abundance variations \citep[e.g.,][]{Sneden04, Johnson12}. It has been shown by \citet{Cordero17} that the most strongly enriched population of stars possesses a stronger degree of rotation with respect to the other samples, while \citet{Savino18} were not able to find significant display of spatial segregations. \citet{Jordi10} reported on an extratidal stellar halo in the immediate vicinity of this GC with a slight elongation aligned with its motion that was later extended to even larger distances out to $13.8~r_\mathrm{t}$ by \citeauthor{Navin16} (\citeyear{Navin16}; however, we note the reevaluation of these associations later in this section).

For M\,13, we identified a total of 292 candidates that -- according to our classification scheme -- are candidate cluster associates. Out of those, 260 are still bound to the cluster and 32 are most likely extratidal. A multi-parametric representation of the chemodynamical associations against the rejected background population is presented in Fig.~\ref{Fig:M13_chemodyn_selection}. The clustering of the associates around the mean values for M\,13  and the simulated stream is evident. Interestingly, M\,92 (Sect. \ref{Subsubsec: M 92 (NGC 6341)}) can be identified as a slight background overdensity. Nonetheless, considering the full five-dimensional information space,  M\,13 can be discerned from M\,92 stars with high confidence.

An independent and powerful constraint for the validity of our method can be obtained from the color--magnitude information for our candidates, which is presented in Fig.~\ref{Fig:M13_extradtidal} along with the spatial distribution. Though they are not part of the initial analysis, the distributions in the \gaia and SDSS passbands constitute well-behaved structures, which -- with few exceptions and considering observational uncertainties -- are consistent with a single $\sim12$~Gyr PARSEC isochrone that matches the age, [Fe/H], and distance modulus of M\,13 (see Table \ref{Table: cluster information}). We note that the reddening value for the \gaia isochrone displayed in Fig.~\ref{Fig:M13_extradtidal} had to be manually adjusted to achieve a better fit. When using reddening coefficients provided in \citet{Yuan13} for SDSS colors and from \citet{Casagrande18} for \gaia colors, this reddening is not consistently described by the same visual extinction $A(V)$ as the SDSS colors. A detailed investigation of this behavior is beyond the scope of the present study. Due to this and the effects of varying extinction throughout the substantial sky coverage of more than 10~deg away from the cluster center, we do not attempt to incorporate the photometry in our mathematical treatment, but use it to qualitatively assess membership likelihoods. To guide the eye in the \gaia CMD, we furthermore show all \gaia targets -- not just the ones with a counterpart in SDSS -- that lie within the tidal radius and that do not deviate by more than $3\sigma$ in $\varpi$, $\mu_\alpha \cos{\delta}$, and $\mu_\delta$. Additionally, in order to overcome contamination by stars that might be affected by crowding and therefore might be blueshifted from the MS, the MS turnoff (MSTO), and the RGB, we required the \gaia color excess factor\footnote{The color excess factor encodes the comparison of the summed BP and RP fluxes in a fixed $3.5\arcsec\times2.1\arcsec$ area to the $G$ flux, which itself is estimated from spatially much more stringently confined PSF photometry. Hence, a high color excess indicates that blending sources affect the colors (see section 5.5.2 of the \gaia DR2 documentation).} to remain below 1.4. 

The projected angular distances of 13 out of the 32 extratidal candidates are only slightly larger than the tidal radius ($d<1.2~r_\mathrm{t}$) and thus may still be considered loosely bound to M\,13. Their distribution qualitatively matches the photometrically determined contours by \citet{Jordi10}. At $d = 1.6~r_\mathrm{t}$ and above, the remaining 19 candidates can safely be denoted extratidal. Looking at their spatial distribution (Fig.~\ref{Fig:M13_extradtidal}), it is tempting to claim evidence for an overdensity towards the leading portion of the cluster orbit and/or stream. However, owing to the nonisotropic coverage of the SDSS footprint around the cluster, this observation should be treated with caution. 

Apart from one star on the blue HB, the  three brightest extratidal associations are affiliated with the RGB and were classified as  CN-strong in Paper~I. Furthermore, the stars are also attributable to M\,13 by the method outlined in Sect. \ref{Subsec: Associating CN-strong stars with clusters}. Therefore, it appears very likely that the three stars in question are escapees that can be attributed to the second -- or enhanced -- population of M\,13 stars. We emphasize that among all 32 candidates presented here, only four giants were studied in Paper~I due to strict parameter limitations of the method therein. In other words, 75\% of the stars in the region of parameter overlap are CN-strong. This being the case, it is intriguing to argue that a substantial fraction of all M\,13 escapees might be second-generation stars, but due to low-number statistics, this remains rather speculative (see Sect. \ref{Subsec: The fraction of chemically altered stars among bona-fide escapees} for further discussion). 

\citet{Navin16} performed a pre-\textit{Gaia}-era search for cluster associations both within and outside of the tidal radius of M\,13. The study was based solely on space motions and photometry. The authors obtained initial membership estimates relying on radial velocities from the first data release of the LAMOST spectroscopic survey \citep{Luo15} and subsequently refined the selection criteria using cuts in color--magnitude and stellar-parameter space in order to end up with probable cluster giants and to exclude foreground dwarf stars. A further constraint was established by employing UCAC4 \citep{Zacharias13} proper motions and demanding deviations from the cluster means of less than 10~mas~yr$^{-1}$. We note that the proper motions used in this latter study are barely significant in any of the cases. Due to large uncertainties in LAMOST DR1, [Fe/H] was not used by the authors to further eliminate candidates.

Here, we cross-matched the 12 candidate extratidal halo stars for M\,13 from \citet{Navin16} with the \gaia DR2 catalog (see Fig. \ref{Fig:M13_extradtidal_Navincomp}). Given the bright nature of the stars in question, corresponding \gaia proper motions are highly significant\footnote{Here, the effects of crowding that were discussed in Sect. \ref{Subsec: Precision and accuracy assessment} should not play a role, as the stars in question are far from the cluster center.} and from a visual inspection of the proper motion comparison in Fig. \ref{Fig:M13_extradtidal_Navincomp} it is already evident that most claimed candidates can be excluded with high confidence. Moreover, our membership formalism (Eqs. \ref{Eq:Bayes theorem} and \ref{Eq:likelihood}) provides a zero probability and thus rejects all 12 candidates. Here, we assumed the sample mean and median errors quoted by \citet{Navin16} on the LAMOST quantities of 17.1~km~s$^{-1}$ and 0.86~dex for $v_r$ and [Fe/H], respectively. The latter values are relatively conservative, such that our strong exclusion confidence is almost entirely driven by the high-precision \gaia proper motions, which could not even remotely be equalled by available catalogs at the time of publication of the original study. At statistically significant parallax-based radial distances of 4.1~kpc and 4.7~kpc -- as opposed to a cluster distance of 7.1~kpc -- two stars can be ascribed to the foreground population and are excluded by this parameter alone.

\subsubsection{M 92 (NGC 6341)}\label{Subsubsec: M 92 (NGC 6341)}
\begin{figure}
    \centering
    \resizebox{\hsize}{!}{\includegraphics{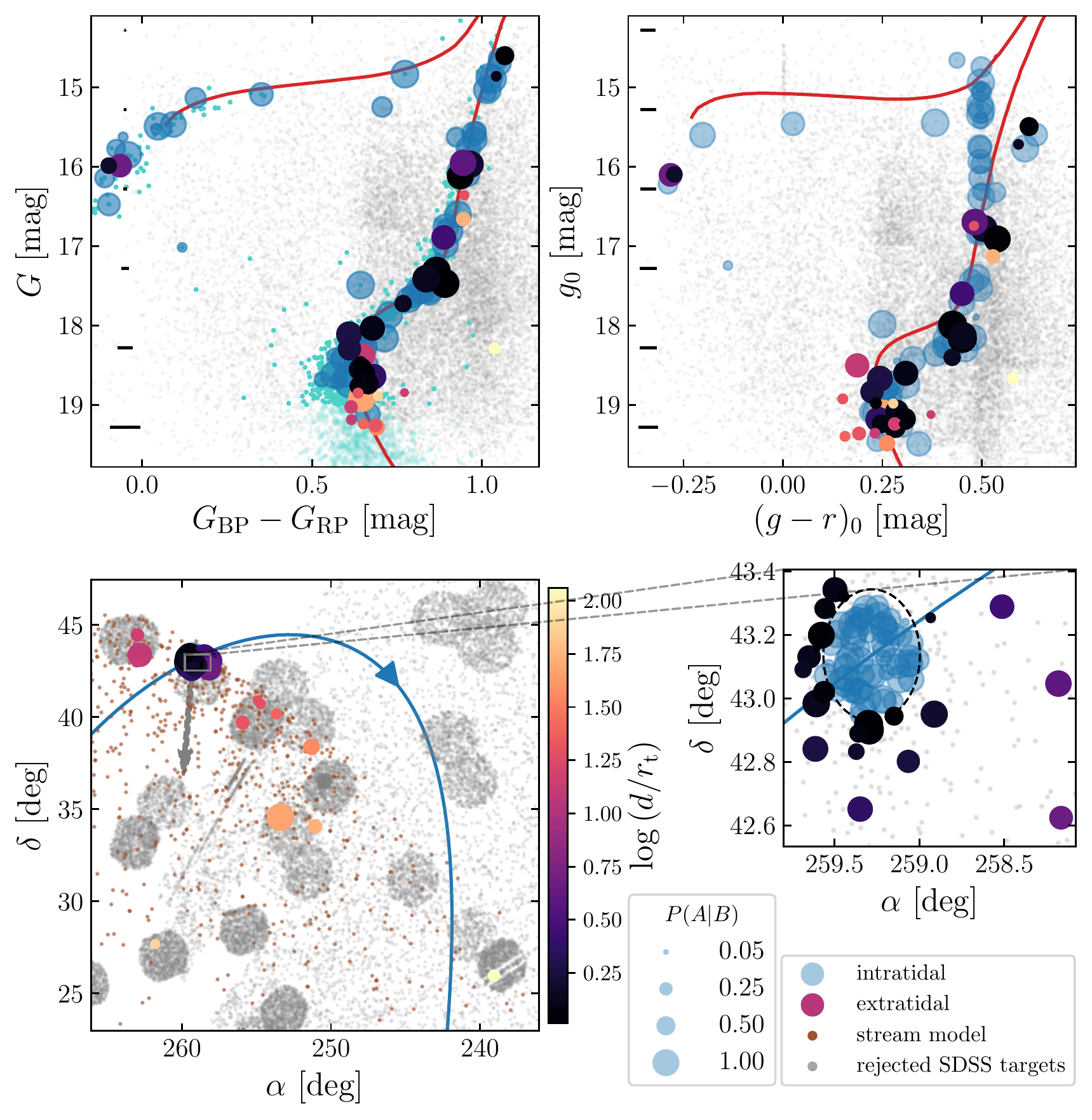}}
      \caption{Same as Fig. \ref{Fig:M13_extradtidal}, but for M\,92.
              }
      \label{Fig:M92_extradtidal}
\end{figure}
At [Fe/H$]=-2.35$~dex \citep{Carretta09}, the GC \object{M\,92} is amongst the most metal-poor Galactic clusters known. Using photometric data, \citet{Testa00} and \citet{Jordi10} consistently reported on an extratidal halo for M\,92. It is noteworthy that our debris model for this cluster is exceptionally scattered because of its rather close pericentric passages ($R_\mathrm{peri}\approx0.17$~kpc) that result in a strong tidal field. 

Figure \ref{Fig:M92_extradtidal} indicates the distribution in parameter space of our 110 potential M\,92 associates, of which 35 are likely to be extratidal. Due to the low metallicity of the cluster, and its considerable proper motion in $\alpha\cos{\delta}$ direction ($-4.9$~mas~yr$^{-1}$), it is straightforward to chemodynamically discern bona fide cluster associates from the fore- and background field stars and hence effectively diminish the false-positive detection probability. From the CMDs we find that -- with the exception of one low-probability case at large angular separation -- our candidates are consistent with the cluster isochrone and that they are distributed between the upper MS ($G\sim19$~mag) and the upper RGB ($G\sim14.5$~mag). Two stars fall on top of the blue HB of  M\,92. While nine candidates are borderline cases in terms of their distance to the cluster core ($d < 1.30\,r_\mathrm{t}$), 26 stars, mostly positioned south of the cluster, are clearly unbound at core separations of more than 1.3 tidal radii. Unfortunately, M\,92 was placed on the northern edge of the SDSS plate such that any potential extratidal candidates in the immediate surroundings north of the cluster are missed by our analysis. The eastward elongation of the distribution of chemodynamically associated stars in the near-field surrounding the cluster shows a striking similarity to the photometric overdensity reported by \citet{Jordi10}. 

Since our approach is density insensitive and merely limited by the SDSS selection function, we can expand the search for cluster members out to larger separations. Moreover, omitting the clear contaminant from the CMD inspection, we associate 12 targets at $12.5<d/r_\mathrm{t}<75.2$ with M\,92, thus expanding the previously known extension of the tidal debris from this GC.

\subsubsection{M 3 (NGC 5272)}\label{Subsubsec: M 3 (NGC 5272)}
\begin{figure}
    \centering
    \resizebox{\hsize}{!}{\includegraphics{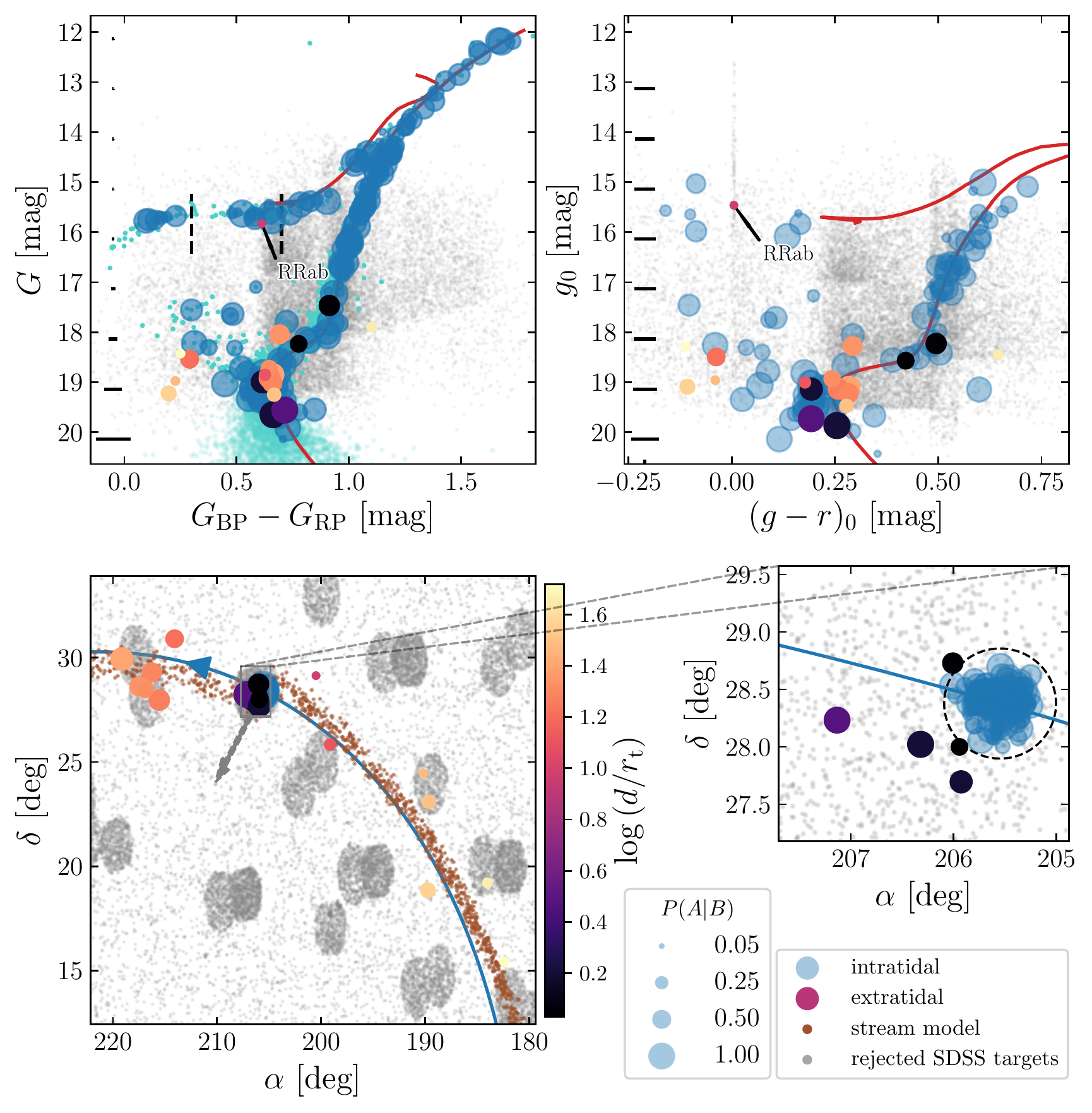}}
      \caption{Same as Fig. \ref{Fig:M13_extradtidal}, but for M\,3. The approximate color range of the instability strip from \citet{Clementini19} is indicated in the \textit{upper left panel} by vertical dashed lines and the pulsating variable (see main text) is labeled ``RRab''. We note that the illustrated color and brightness for the RR Lyrae star are merely mean quantities, which may vary substantially over the pulsation cycle.
              }
      \label{Fig:M3_extradtidal}
\end{figure}
\begin{figure*}
    \centering
    \resizebox{\hsize}{!}{\includegraphics{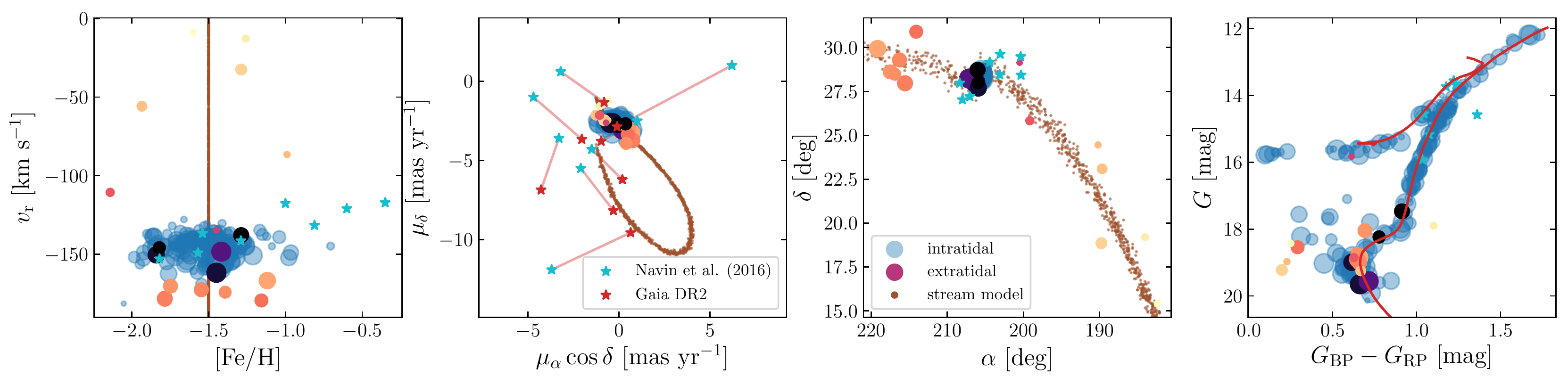}}
      \caption{Same as Fig. \ref{Fig:M13_extradtidal_Navincomp}, but for M\,3.
              }
      \label{Fig:M3_extradtidal_Navincomp}
\end{figure*}
Of the 199 associates for \object{M\,3}, we report on 17 probable candidates in the surrounding field star population that coincide with evolutionary stages between the cluster's upper MS, base of the RGB, and HB. The brightest extratidal star ($G=15.8$~mag) shares the same region as the locus of intratidal HB stars. In fact, both \citet{Abbas14} and \citet{Clementini19} list this star as being a pulsating RR Lyrae star of type ab. The former authors provide a distance estimate of 10.99~kpc, which -- lacking a proper attached error margin on that value -- is hard to compare with the cluster distance of 10.2~kpc, although an association seems feasible. We performed our own analysis of the photometric time series available in the literature and furthermore corrected for the fact that, due to their pulsating nature, RR Lyrae do not straightforwardly reveal their systemic velocity from single-epoch spectra alone. A detailed description can be found in Appendix \ref{Sec: RR Lyrae analysis}. Our inferred value of $10.9\pm2.6$~kpc is in excellent agreement with the cluster distance. \citet{Kundu19} present another two RR Lyrae stars that they associate with M\,3. However, the search radius of these latter authors was restricted to $2/3<d/r_\mathrm{t}<3$, a circumstance that prohibited them from finding the variable presented here at $d=9.4\,r_\mathrm{t}$.   

Among the presumably bound and unbound candidate stars, several reside blueward of the MSTO and subgiant branch (SGB) as can be seen in Fig. \ref{Fig:M3_extradtidal}. These could either be connected to false-positive associations or be genuine blue straggler stars (BSS). We favor the latter interpretation, as we do not see any particular reason for all random associations to be preferentially found on the blue- as opposed to the red side of the isochrone, even though the field star population is much more numerous in the red part. Our explanation is bolstered by the fact that the number of true BSS in M\,3 is much larger than in M\,13 \citep{Ferrraro03} for example, where our treatment did not associate any potential blue straggler candidate. 

Intriguingly, we found a strong degree of spatial alignment between the high-probability extratidal stars and the leading arm of the simulated tidal stream both in the near- and far-field around the cluster. Unfortunately, there are no SDSS plates covering the in-between regions and so the existence of a stream remains uncertain.

\citet{Navin16} also found potential extratidal associates for M\,3. Following the same approach and reasoning as in Sect. \ref{Subsubsec: M 13 (NGC 6205)}, we exclude previous membership for all eight with high confidence. Two stars can even be discarded based on their highly significant \gaia parallaxes which render them much closer than the cluster (7~kpc and 5.7~kpc compared to 10.2~kpc distance to M\,3). The same graphical representation shown for M\,13 in Fig. \ref{Fig:M13_extradtidal_Navincomp} is presented in Fig. \ref{Fig:M3_extradtidal_Navincomp} for M\,3. 

\subsubsection{M 2 (NGC 7089)}\label{Subsubsec: M 2 (NGC 7089)}
\begin{figure}
    \centering
    \resizebox{\hsize}{!}{\includegraphics{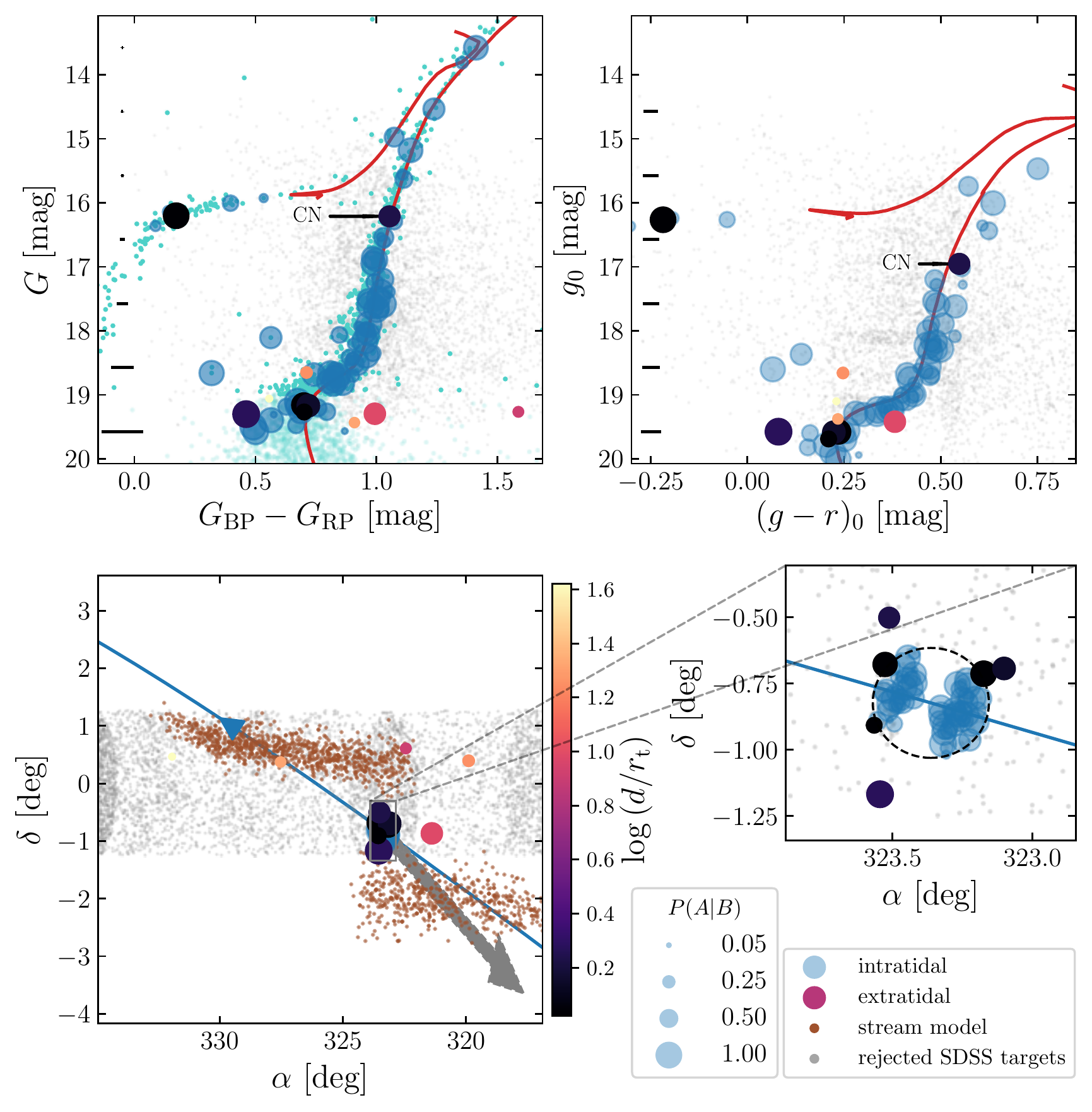}}
      \caption{Same as Fig. \ref{Fig:M13_extradtidal}, but for M\,2. The CN-strong star from Paper~I is labeled ``CN''. The red outlier in the \textit{upper left panel} is outside of the plotting range in the \textit{upper right panel} as showing it ($g_0=20.71$~mag, $(g-r)_0=1.05$~mag) would strongly distort the diagram.
              }
      \label{Fig:M2_extradtidal}
\end{figure}
\object{M\,2} is a halo ($R_\mathrm{GC}\sim10.1$~kpc) GC with a complex chemical-enrichment history. \citet{Yong14} and \citet{Lardo16}, for example, suggested an accretion origin of this cluster, which may be the stripped core of a dwarf spheroidal galaxy based on observed anomalous spreads in iron and neutron-capture elements. \citet{Kuzma16} detected a photometric stellar envelope surrounding M\,2 beyond its tidal radius. 

Our formalism yielded 77 associates including 11 extratidal candidates for M\,2; three of which are uncertain to be truly unbound because of their small projected separations from the GC ($<1.2\,r_\mathrm{t}$). While ten of the stars might be associated from a photometric point of view (see Fig. \ref{Fig:M2_extradtidal}), one candidate is consistently reported by both \gaia and SDSS to be redder than the cluster population at comparable brightness on the MSTO. A further two to three stars reside in the BSS region; at this point we can neither confirm nor reject their BSS nature. Owing to the sparse coverage of SDSS plates around M\,2, we cannot draw firm conclusions with respect to the spatial distribution of extratidal halo stars for this GC. However, in the immediate surroundings, we can chemodynamically confirm the finding of \citet{Kuzma16}, who reported on a diffuse stellar envelope that extents out to at least $5\,r_\mathrm{t}$.

The only identified extratidal giant coincides with the parameter range of -- and thus was analyzed in -- Paper~I; it was found to be CN-enhanced, again suggesting a large fraction of 2P stars in the group of escapees (see Sect. \ref{Subsec: The fraction of chemically altered stars among bona-fide escapees} for further discussion).

\subsubsection{M 15 (NGC 7078)}\label{Subsubsec: M 15 (NGC 7078)}
\begin{figure}
    \centering
    \resizebox{\hsize}{!}{\includegraphics{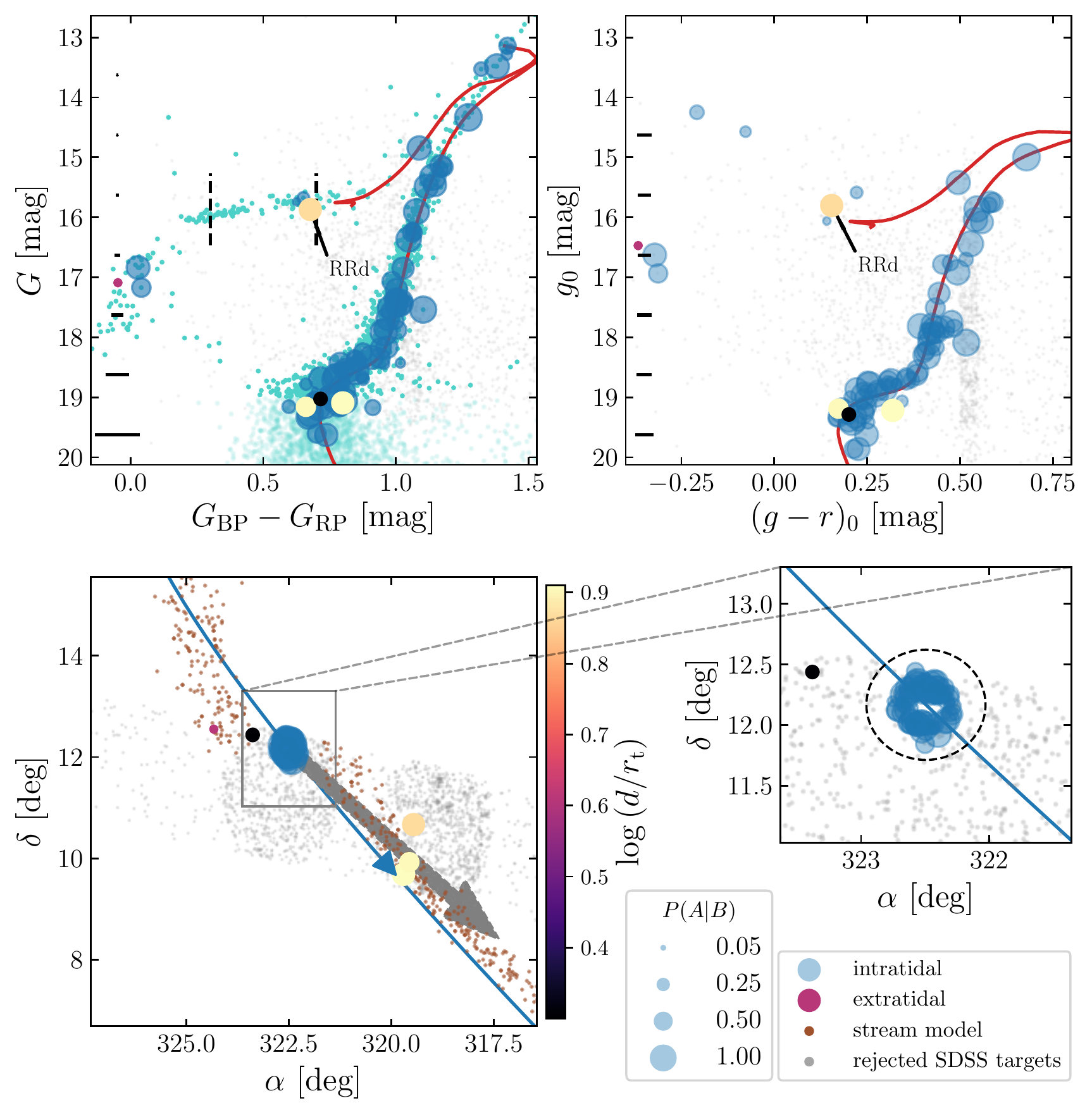}}
      \caption{Same as Fig. \ref{Fig:M13_extradtidal}, but for M\,15. 
              }
      \label{Fig:M15_extradtidal}
\end{figure}
At a slightly lower [Fe/H] than M\,92, the cluster \object{M\,15} is the most metal-poor cluster for which we could identify associations. The region around M\,15 is barely covered by SDSS plates (see Fig. \ref{Fig:M15_extradtidal}). Nevertheless, we find five halo field stars that are a chemodynamical match to this GC. All of them inherit an excellent photometric consistency with the cluster CMD on the MSTO and the blue and red HB. Thus, a cluster origin appears highly feasible.

The star falling in the instability strip is listed as variable in \gaia DR2. Nonetheless, it is not tabulated in either of the works by \citet{Drake13, Drake13b, Drake14}, nor in \citet{Abbas14}. We therefore performed our own light curve analysis (Appendix \ref{Sec: RR Lyrae analysis}) and found this star to be a double-mode RR Lyrae of type d. A distance of $9.5\pm2.3$~kpc was deduced, which places the star at a distance in good agreement with that of M\,15. 

\subsubsection{M 53 (NGC 5024) and NGC 5053}\label{Subsubsec: M 53 (NGC 5024) and NGC 5053}
\begin{figure}
    \centering
    \resizebox{\hsize}{!}{\includegraphics{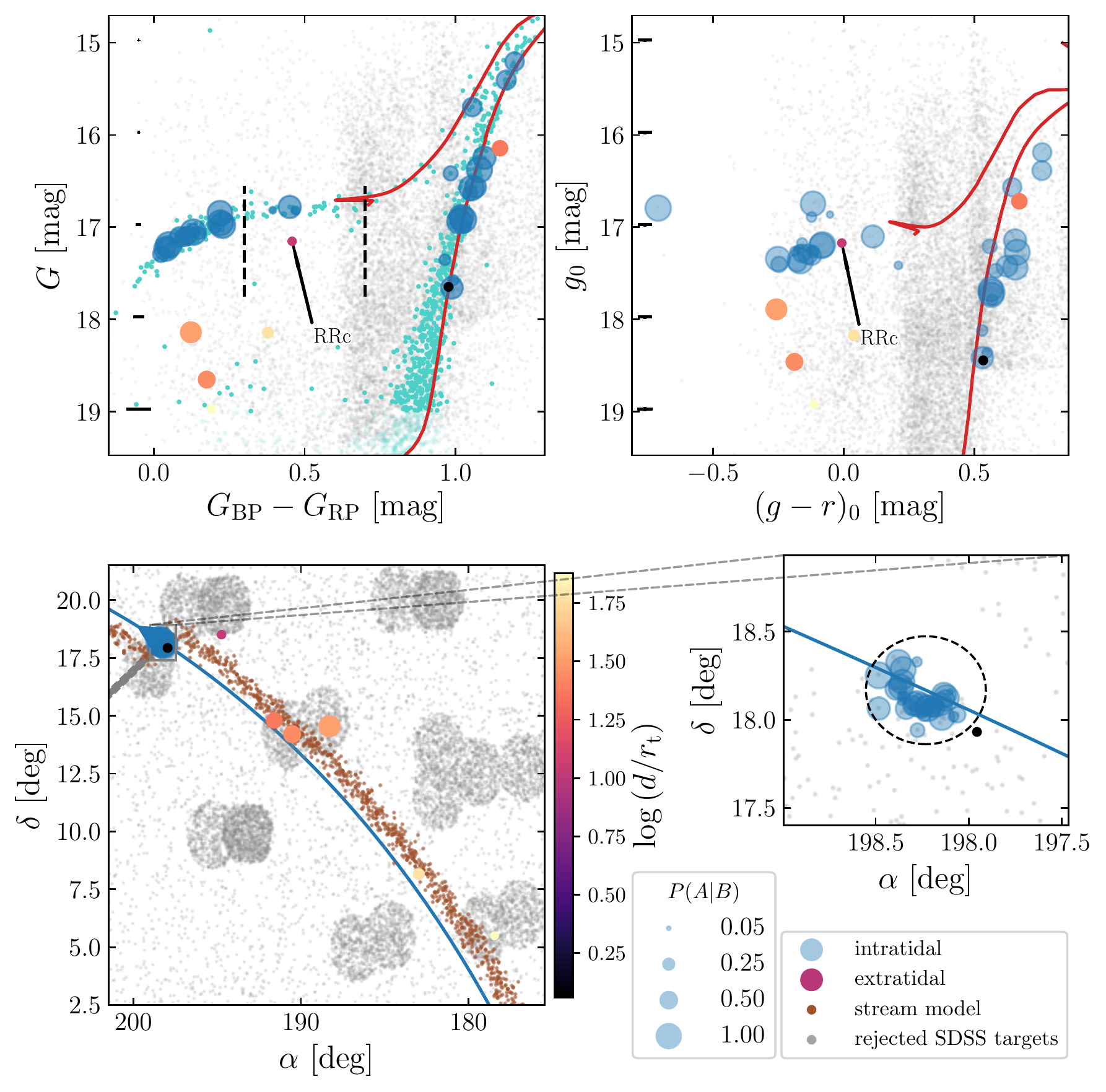}}
      \caption{Same as Fig. \ref{Fig:M3_extradtidal}, but for M\,53. 
              }
      \label{Fig:M53_extradtidal}
\end{figure}
\begin{figure}
    \centering
    \resizebox{\hsize}{!}{\includegraphics{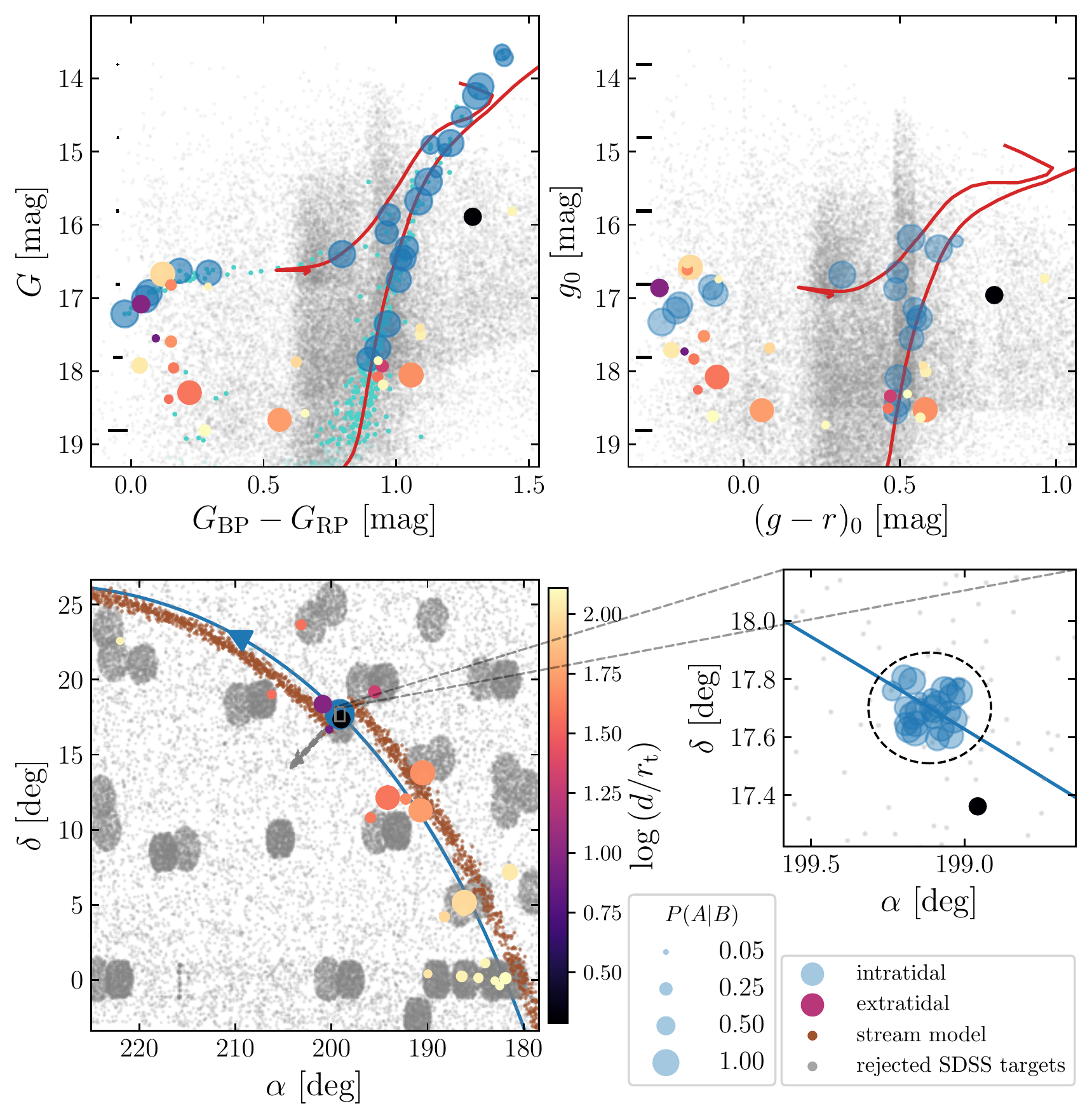}}
      \caption{Same as Fig. \ref{Fig:M13_extradtidal}, but for NGC\,5053.
              }
      \label{Fig:NGC5053_extradtidal}
\end{figure}
\object{M\,53} and \object{NGC\,5053} are two GCs at a low angular separation of 0.96~deg, even though their real spatial separation is about 1~kpc \citep{Jordi10}. \citet{Law10b} suggested that both clusters could be possible associates of the Sagittarius stream, while \citet{Forbes10} speculate over the possibility of one or both of them being the nucleus of a disrupted dwarf galaxy. The former hypothesis has been refuted with high confidence by \citet{Sohn18} using \textit{Hubble} Space Telescope proper motions. \citet{Chun10} report on extended overdensities for both clusters and claim the detection of a tidal bridge between them. However, \citet{Jordi10} could not reproduce that finding.    

Among the extratidal field star population around M\,53, we found seven stars that -- based on their kinematics and metallicity -- may have originated in M\,53. As can be seen in Fig. \ref{Fig:M53_extradtidal}, photometrically, two of the stars can be readily associated with the RGB. One star is fainter than the HB in $G$ while being a perfect match in $g_0$, thus indicating photometric variability. Indeed, \gaia DR2 lists this star as variable and \citet{Abbas14} classify it as an RR Lyrae star of type c and provide a distance of 24.07~kpc. This appears marginally consistent with the cluster's distance of 17.9~kpc. Nevertheless, as for the RR Lyrae in Sect. \ref{Subsubsec: M 3 (NGC 5272)}, a comparison is prohibited in light of missing error margins. Our own analysis showed a distance of $20.1\pm4.8$~kpc, which is in agreement with that of M\,53 within one error margin. \citet{Kundu19} also reported on five extratidal RR Lyrae for M\,53. Again, due to their restricting themselves to the immediate cluster vicinity, our target is not part of their list.

The remaining four extratidal associations with M\,53 cannot be photometrically matched to either the RGB or the HB. Nevertheless, they cover the same colors and magnitudes as the comparably large number of intratidal stars of the BSS population that we attributed from \gaia kinematic properties alone (cyan dots in Fig. \ref{Fig:M53_extradtidal}).

For NGC\,5053, we report on 24 plausible extratidal detections, 12 of which have their chemodynamical membership bolstered by a reasonable photometric match to the populations of intratidal members on the blue HB (five stars) and RGB (seven stars, see Fig. \ref{Fig:NGC5053_extradtidal}). Two stars are clearly too red to be attributed to the cluster, while ten others lie in the BSS region of the CMDs with their affiliation nature remaining spurious.

We emphasize that none of the associations are shared between the two clusters considered in this section. This does not necessarily exclude the possibility of a tidal bridge for two reasons: Firstly, we do not model such tidal interactions when simulating the streams. Hence, in the near-field of the clusters, the streams are spatially almost parallel and do not cross. The differences in $v_r$ between the clusters exclude mutual associations. The second, more important reason is that the intercluster field was only covered by a few ($\sim10$) SDSS fibers, thereby tremendously reducing the chance of finding a potential bridge associate.     

\subsubsection{NGC 4147}\label{Subsubsec: NGC 4147}
\begin{figure}
    \centering
    \resizebox{\hsize}{!}{\includegraphics{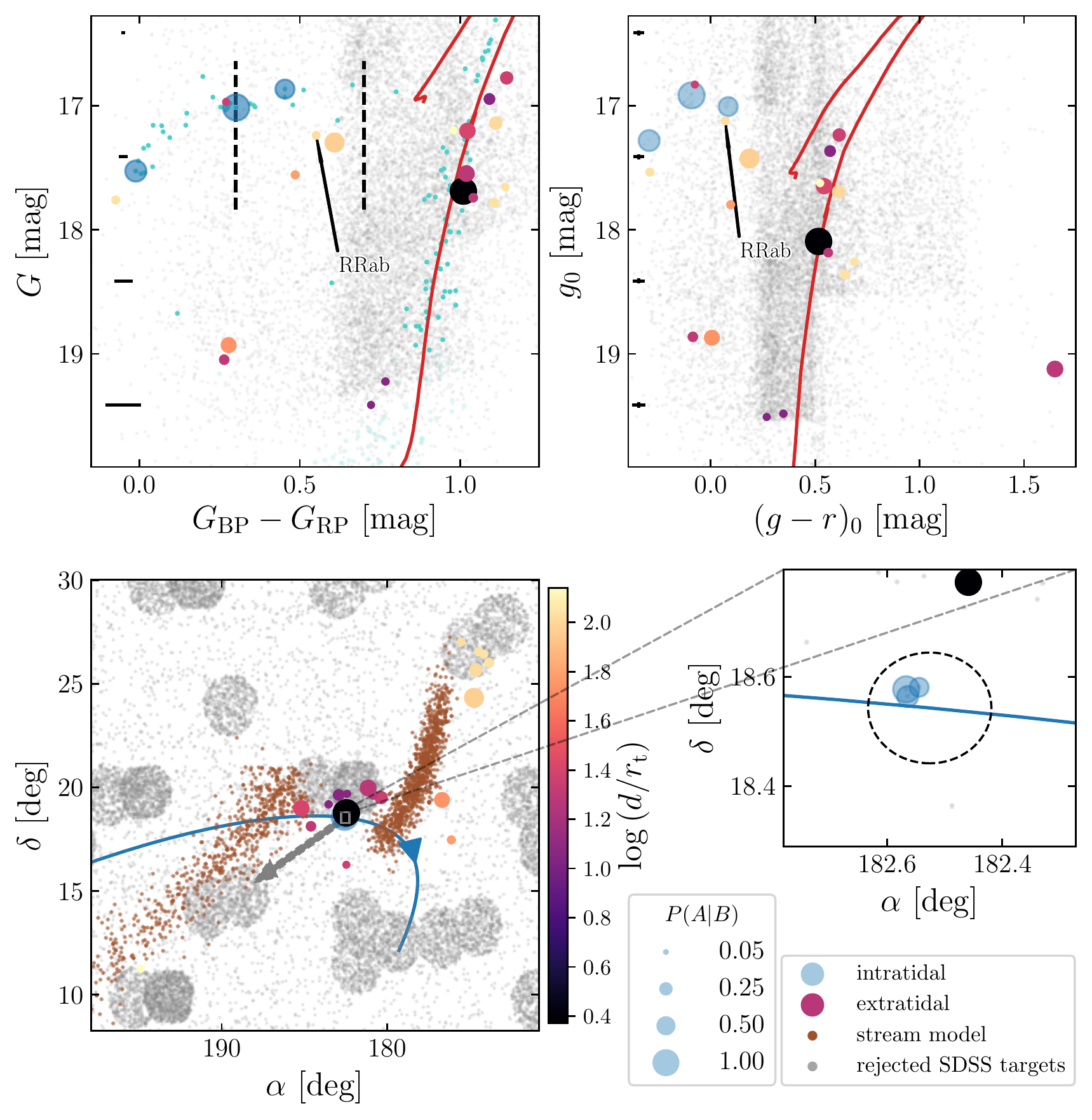}}
      \caption{Same as Fig. \ref{Fig:M13_extradtidal}, but for NGC\,4147. 
              }
      \label{Fig:NGC4147_extradtidal}
\end{figure}
\object{NGC\,4147} is an outer-halo ($R_\mathrm{GC}=21.4$~kpc) GC that has been suggested to be associated with the Sagittarius stream by \citet{Mackey05}. However, \citet{Villanova16} cast some doubt on this by showing that the cluster is a closer match to the Galactic halo in terms of their respective chemical abundances. More recently, proper motion studies firmly rejected an association since the cluster is on a counter-rotating orbit with respect to Sagittarius \citep{Sohn18, Riley20}.

Owing to the  faint nature of NGC\,4147 \citep[horizontal branch magnitude, $V_\mathrm{HB}=17.02$~mag,][]{Harris96}, our approach can merely test targets that are as bright as the RGB of this GC. Of 19 extratidal candidates, we identified three associations that fall on top of the  HB of the cluster. The reddest of these three is an RRab star at a distance of $22.0\pm5.2$~kpc\footnote{\citep{Drake13} reported a distance of 21.36~kpc for this star.}. The latter distance renders it consistent with NGC\,4147 at 19.3~kpc. Most of the remaining associations (8 stars) were found to be on the upper RGB, whereas two stars below the HB, two stars redward of the RGB, and the five faintest candidates are likely to be false-positive detections judging from their photometric discrepancy from the intratidal \gaia sources.

\subsection{Associating CN-strong stars with clusters and major merger events}\label{Subsec: Associating CN-strong stars with clusters}
We explored chemodynamical links of the sample of 112 CN-strong field stars from Paper~I with the Galactic GC population through method II. Similar efforts have recently been made by \citet{Savino19} who used the similar but smaller sample of \citet{Martell11}. We emphasize that only 27 of the 63 cluster-star pairs reported by \citet{Savino19} can be directly compared to our study because the remaining ones lie outside of our stellar metallicity restriction of $-1.8 < \mathrm{[Fe/H]}<-1.3$~dex\footnote{We note that this rejection criterion excludes all potentially CN-strong stars that \citet{Savino19} reported to be on highly circular prograde orbits.} (Paper I). Of these 27, only one is marginally attributable to their sample of best association candidates (<65\% rejection confidence). The reason for the low overall overlap is that we find generally lower probabilities because we use the full distribution of actions instead of marginalized distributions for each of the three components.

\begin{figure}
    \centering
    \resizebox{\hsize}{!}{\includegraphics{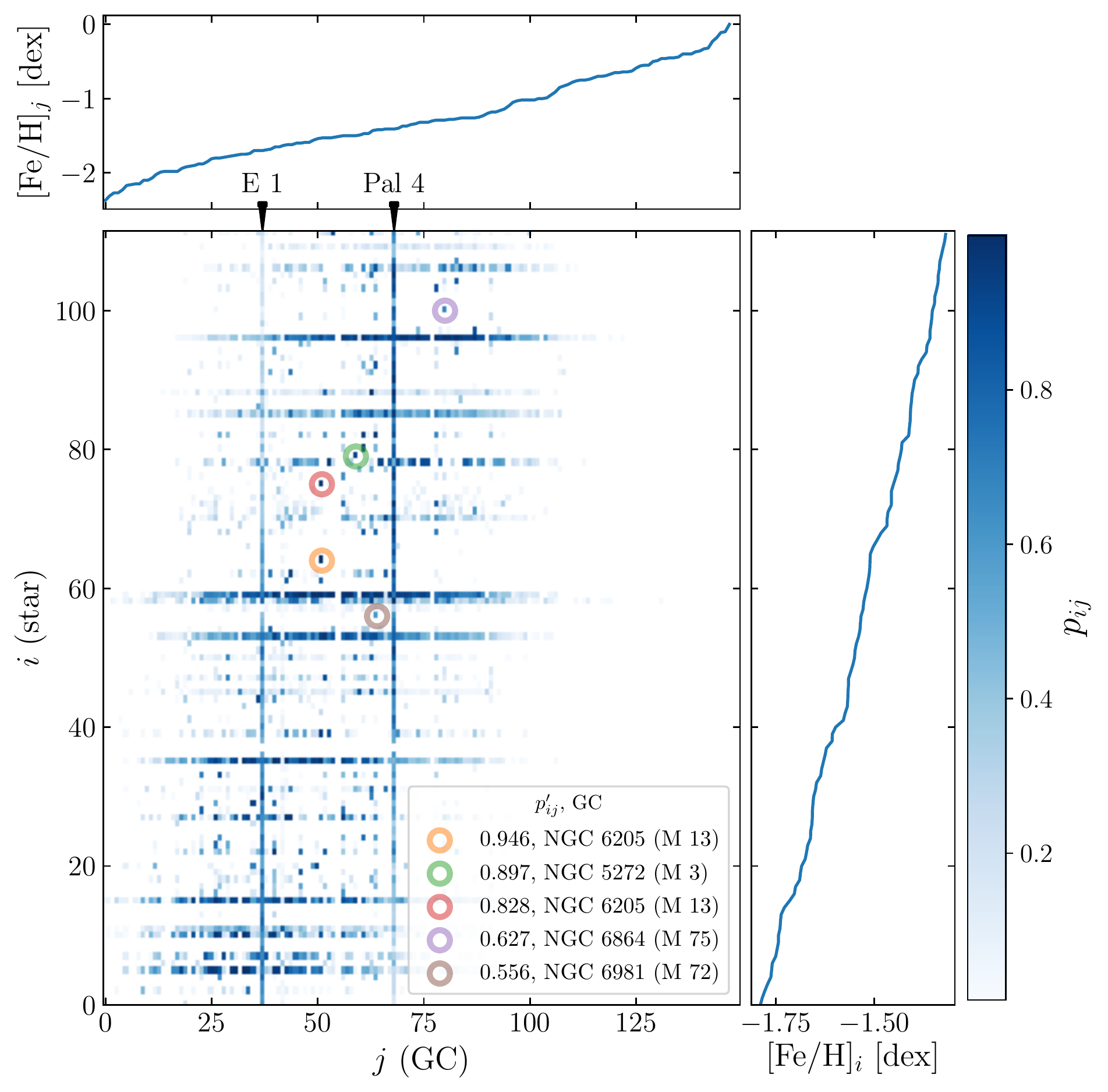}}
      \caption{Graphical representation of the matrix $p_{ij}$ with clusters on the abscissa and CN-strong stars on the ordinate. Both coordinates are sorted by increasing [Fe/H] as indicated by the \textit{top} and \textit{right panels}. The five pairs with highest modified confidence, $p_{ij}'$ (Eq. \ref{Eq: modified confidence}), are highlighted by colored circles and presented in the legend.
              }
      \label{Fig:CNstrong_association_matrix}
\end{figure}
Comparing each CN-strong halo star $i$ with each cluster $j$ yields a matrix $p_{ij}$ with 16\,800 entries. Figure \ref{Fig:CNstrong_association_matrix} depicts this matrix in a representation where both rows and columns are sorted by increasing metallicity. Once again, we highlight the fact that a value for $p$ close to unity does not necessarily imply high confidence in association, but rather that association cannot be rigidly excluded. In case of considerable ambiguities for the orbital parameters of a particular star -- for example due to a highly uncertain distance -- almost no cluster can be kinematically excluded from being a former host. In such cases, $p$ is entirely driven by [Fe/H] alone. In Fig. \ref{Fig:CNstrong_association_matrix} those manifest in rows essentially showing broad normal distributions with peak positions at the intersect of the stellar and cluster [Fe/H]. The latter are only occasionally interrupted by low- and high-$p$ columns representing GCs in the outermost regions of the Galaxy and the GCs E~1 and Pal~4, respectively. The latter have loosely constrained orbits themselves, and therefore we do not consider these two clusters in the following.

Inverting the above reasoning implies that pairs $(i,j)$ are good association candidates if they have a high attributed $p,$ while at the same time $p_{ij}/\sum_j{p_{ij}}$ is strongly peaked. Hence, each cluster can in principle have an arbitrary number of stellar associations whilst to be considered as part of a good pair, each star should have a strongly limited number of attributable clusters. This leads us to the modified quantity:
\begin{equation}\label{Eq: modified confidence}
 p_{ij}'=\frac{p_{ij}^2}{\sum_j{p_{ij}}}.
\end{equation}
In total, 145 pairs were found to satisfy $p_{ij}'> 0.05$, whereas only 15 posses values above 0.32. The five pairs with highest $p'$ are highlighted in Fig. \ref{Fig:CNstrong_association_matrix} and the whole list can be found in Table \ref{Table: CN-strong cluster pairs}. We recover all four associations of CN-strong stars to the GCs M\,13 and M\,2 that were presented in Sect. \ref{Subsec: Extratidal escapee candidates around clusters}. In case of the association with M\,2, there are two additional clusters that come into question given their action integrals (M\,22 and ESO~280-06). Given the spatial coincidence with M\,2 -- which did not enter the present analysis -- we favor association with this cluster.

Of the 145 reported star-cluster association pairs, 26 involve GCs that were proposed to have possibly been accreted as part of either Gaia-Enceladus \citep{Myeong18} or the Sequoia merger event \citep{Myeong19}; an observation that represents tentative evidence that the involved CN-strong stars were donated by those galaxies. The pairs are indicated in Table \ref{Table: CN-strong cluster pairs}. Moreover, among the 15 strongest ($p^{'}_{ij}\geq0.32$), three pairs involve the bona-fide Enceladus clusters \object{M\,75} and \object{NGC\,1261}. Overall, we found that most, namely 23, pairs are accounted for by Enceladus clusters such as M\,2, while only three pairs with low association probability ($\leq0.08$) involve Sequoia GCs. We caution that several stars have associations not only with clusters from the two merger events but with other clusters, too. One star, \gaia DR2 603202356856230272, at the same time shows associations with two Enceladus GCs (M\,79 and NGC\,1851) and one Sequoia cluster (NGC\,3201), although with generally low probabilities.

We confirm the finding by \citet{Savino19} that a substantial fraction (here 38\%) of the CN-strong stars from the sample seemingly cannot be chemodynamically associated with any of the known GCs. We recall here the three reasons these latter authors proposed for this observation under the restrictive assumption that GCs are the only birth site of these chemically peculiar stars: The first, rather unlikely option is that the hitherto unassociated stars could originate from yet-to-be-discovered clusters that are concealed by the high-extinction regions of the Galactic disk or bulge. Secondly, the stars in question could stem from already entirely dissolved clusters. Finally, the stars could have originated from one of the known GCs without retaining the orbital characteristics of the cluster. For this latter case, \citet{Savino19} offer two highly plausible explanations involving high ejection velocities due to three-body interactions or an early escape from the cluster followed by a drastic change of the Galactic potential due to, for example, a major merger. In order to address the former scenarios, it is of utmost importance to establish a ground truth by rigidly confirming the CN-strong nature of the targets using spectroscopic follow-up analyses that could reveal other light-element anomalies.

\subsection{Associations around CN-strong field stars}\label{Subsec: Associations around CN-strong field stars}
\begin{figure*}[!ht]
    \centering
    \resizebox{0.85\hsize}{!}{\includegraphics{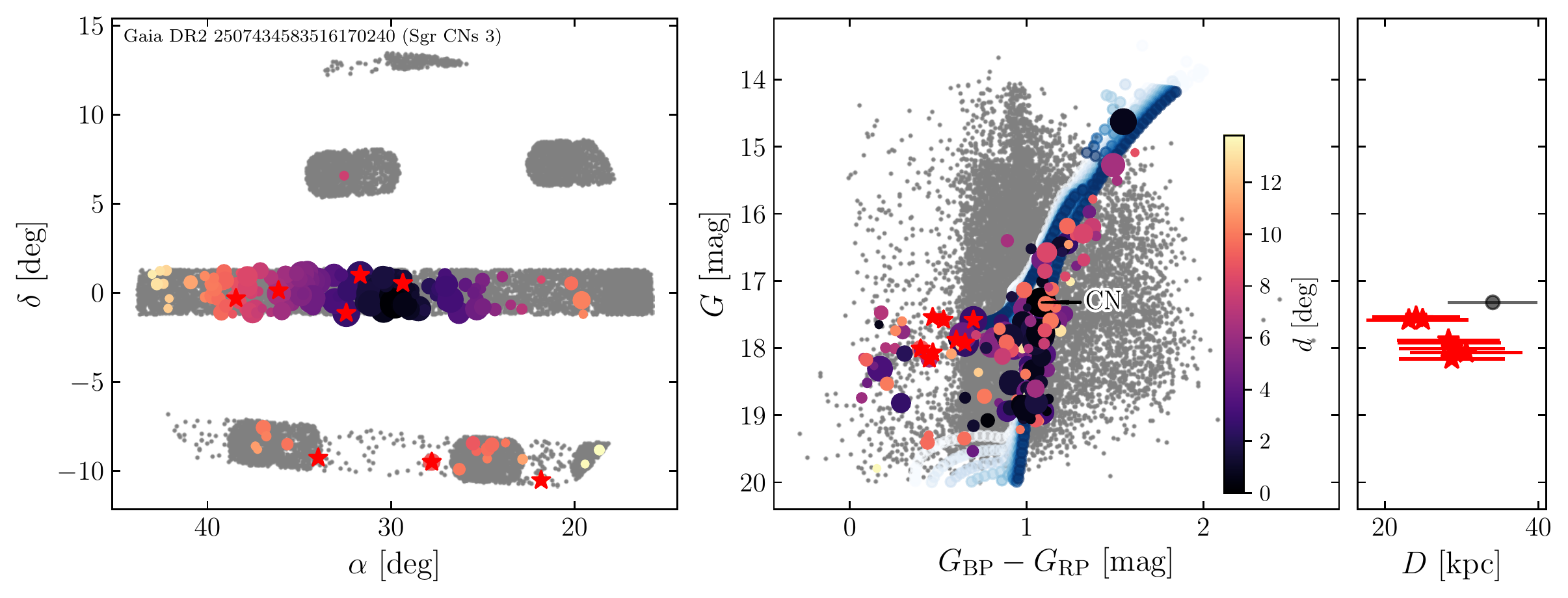}}
      \caption{Graphical representation of the population that was chemodynamically associated with the CN-strong star \gaia DR2 2507434583516170240 (Sgr CNs 3). The color coding and circle sizes have the same meaning as in Fig. \ref{Fig:M13_extradtidal}. Additional red stars indicate the identified RR Lyrae variables (see main text). \textit{Left panel}: Position on the sky on top of rejected stars from the SDSS footprint (gray dots). \textit{Middle panel}: Photometric associations in the \gaia CMD. To guide the eye, isochrones that represent ages from 4 to 13~Gyr (white to dark blue) at the mean [Fe/H] of the population are shown for reference. The CN-strong star is labeled ``CN``. \textit{Right panel}: RR Lyrae $G$ magnitudes and their inferred distances (red stars) in comparison to the spectrophotometric distance of the CN-strong star (black circle, see Sect. \ref{Subsec: Integrals of motion}).
              }
      \label{Fig:SgrCNs3_method3}
\end{figure*}
Of the 112 investigated stars, method III revealed 81 groups with at least two -- possibly, but not necessarily CN-normal -- additionally associated stars. Of these groups, 69 comprise five or more stars, whereas 51 still consist of at least ten stars in excess of their respective CN-strong candidate. Naturally, we recover a large number of intratidal targets for the four extratidal CN-strong stars in the vicinity of M\,13 and M\,2 that were reported in Sect. \ref{Subsec: Extratidal escapee candidates around clusters}.

All remaining CN-strong stars have no obvious direct connection to any cluster. Nevertheless, many of the chemodynamically linked groups of stars can be photometrically attributed to the same stellar population even in the absence of a cluster (see Fig. \ref{Fig:SgrCNs3_method3} for an example). Since our selection function is not only spatially inhomogeneous but also strongly biased towards the brighter evolutionary stages, many associated stars coincide with the HB in the CMDs of the latter populations. For candidate variable stars that fall in the instability strip, we performed the analysis outlined in Appendix \ref{Sec: RR Lyrae analysis} and recomputed the Mahalanobis distance using updated values for $v_r$. This way, we were able to associate 31 already known RR Lyrae stars \citep{Drake13, Drake13b, Drake14, Abbas14} and add one more RRc-type pulsator. 

Unfortunately, a number of factors prevented a straightforward automated classification for the degree of belief in true photometric association. Among these are the occasionally high rate of obviously spurious associations and the ambiguity of the distance modulus in the absence of significant parallaxes and/or attributed RR Lyrae stars. We present those 17 groups that passed a visual inspection. Their CMDs and spatial distributions can be found in Figs. \ref{Fig:SgrCNs3_method3} and \ref{Fig: C1} through \ref{Fig: C16}, while the relevant information about individual stars is presented in Table \ref{Table: Associations to CN-strong stars}. Among all CN-strong stars, only \gaia DR2 615481011223972736 and \gaia DR2 634777096694507776 (cf., Figs. Appendix \ref{Fig: pair1} and \ref{Fig: pair2}) were pairwise attributed to each other using method III, which provides evidence that they may share the same birth place. 

For five of the groups of associated stars, the CN-strong star was attributed to at least one of the Enceladus clusters through method II. However, four of those only have very low association probabilities, which is why we do not propose an Enceladus membership. The remaining star, \gaia DR2 3696548235634359936,  generates the sixth strongest among all star-cluster associations ($p^{'}_{ij}=0.54$), the involved cluster being NGC\,1261. Nonetheless, it is noteworthy that \citet{Myeong18} classified this cluster as  merely a possible member of Enceladus whilst eight additional GCs were labeled probable members.

\subsubsection{Associations to the Sagittarius stream and M 54}\label{Subsubsec: Associations to the Sagittarius stream}
\begin{figure}[!h]
    \centering
    \resizebox{0.95\hsize}{!}{\includegraphics{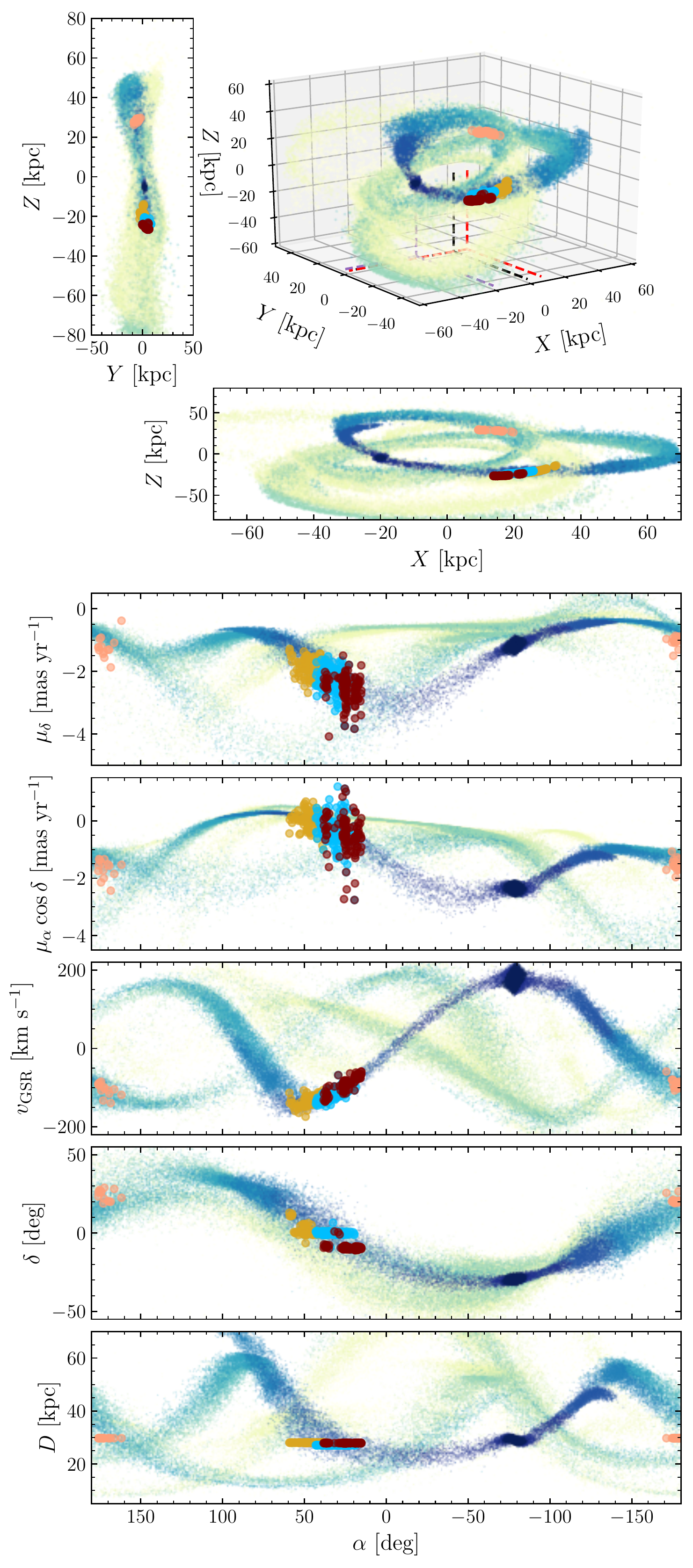}}
      \caption{Multi-parametric representation of four populations of chemodynamical associations with CN-strong giants (salmon, yellow, dark red, and light blue circles) on top of the debris model of the Sgr stream in a triaxial halo by \citet{Law10}. The latter is depicted by dots where increasing color lightness denotes a longer elapsed time since becoming unbound from the Sgr main body. \textit{Top}: 3D position in Cartesian, Galactocentric coordinates. Red-, black-, and violet-dashed lines indicate the positions of the Sun, the Galactic center, and the Sgr main body, respectively. \textit{Bottom panels}: Projections in all six dimensions of positional and kinematic observables. For the \textit{bottom panel}, we assumed all stars to reside at exactly the same $D$ as inferred from the mean RR Lyrae distance for each population. 
              }
      \label{Fig:Sgr_association}
\end{figure}
We found four of our high-confidence groups of associations to coincide with the Sagittarius (Sgr) stream. For further reference, we label the groups Sgr CNs 1 to 4. Fortunately, all four have at least one attributed RR Lyrae star, thus enabling a meaningful 6D phase-space characterization. Figure \ref{Fig:SgrCNs3_method3} exemplarily illustrates the tight photometric consistency for Sgr CNs 3, while in Fig. \ref{Fig:Sgr_association} we present various population representations on top of the N-body simulation of the tidal disruption of the Sgr dwarf spheroidal galaxy in a triaxial MW halo by \citet{Law10}. The phase-space comparison places our agglomerations right on top of the dynamically young portion of the leading and trailing arms of the stream. On the basis of the simulated model parameters -- under the assumption of true association -- the stars became unbound from the Sgr main body no longer than 3~Gyr ago. 

From a chemical point of view, the metallicities of the four involved CN-strong stars cover a narrow range from $-1.51$ to $-1.40$~dex and thus -- within their errors -- are fully consistent with the mean [Fe/H] of \object{M\,54} \citep[$-1.45$~dex, e.g.,][]{Bellazzini08}. This observation suggests that four chemically altered giants and potentially a considerable fraction of their associates identified here originated from the massive GC M\,54. Indeed, one of the four involved CN-strong stars -- namely \gaia DR2 2507434583516170240, giving rise to the association Sgr CNs 3 -- has already been linked to this cluster based on its position in action space (method II), though with very low confidence (0.08). We suspect that the reason for the latter is to be found in the main flaw of the assumptions in Sect. \ref{Subsec: Associating CN-strong stars with clusters}, that is, the fact that actions of potential cluster escapees do not necessarily have to be exactly identical to the cluster itself; otherwise, they would never have been able to become unbound in the first place.

\subsection{The fraction of chemically altered stars amongst bona-fide escapees}\label{Subsec: The fraction of chemically altered stars among bona-fide escapees}
\begin{figure*}[h]
    \centering
    \resizebox{0.65\hsize}{!}{\includegraphics{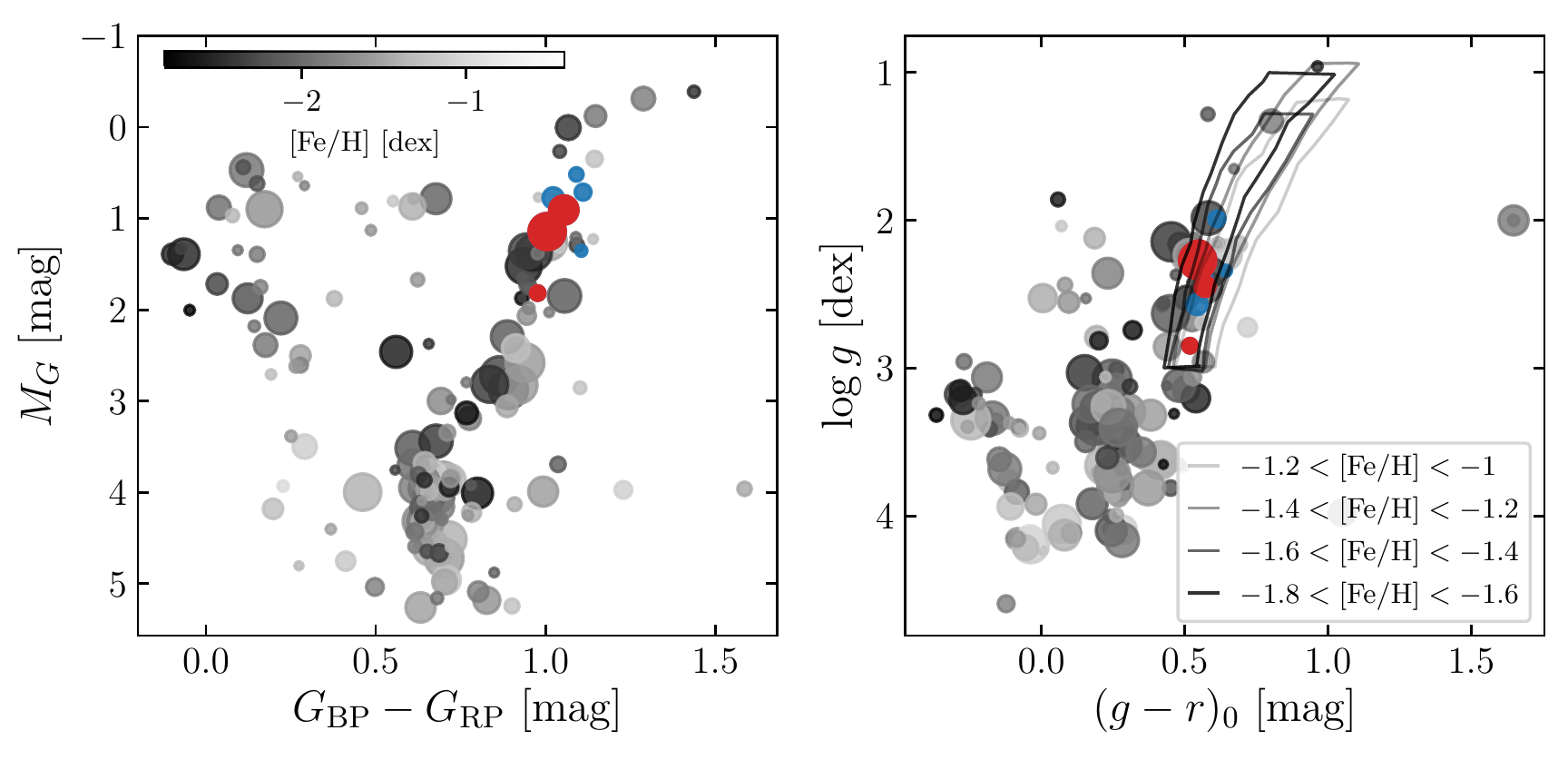}}
      \caption{\gaia CMD (\textit{left}) and Kiel diagram (\textit{right}) of all association candidates from method I. For the absolute magnitude, $M_G$, we computed a distance modulus under the assumption that all stars reside at exactly the same distance as their potential former hosts. Dark gray to light gray colors indicate the stars' metallicity according to the color bar, while the point size mirrors the calculated association probability. The metallicity-dependent polygons used for pre-selection in Paper I are depicted in the \textit{right panel}. Associated stars which at the same time obey these selection cuts are shown in blue. Red circles depict the four CN-strong stars that are likely associated with M\,13 and M\,2, respectively.  
              }
      \label{Fig:methodI_2pvs1p}
\end{figure*}
\begin{figure*}[h]
    \centering
    \resizebox{0.65\hsize}{!}{\includegraphics{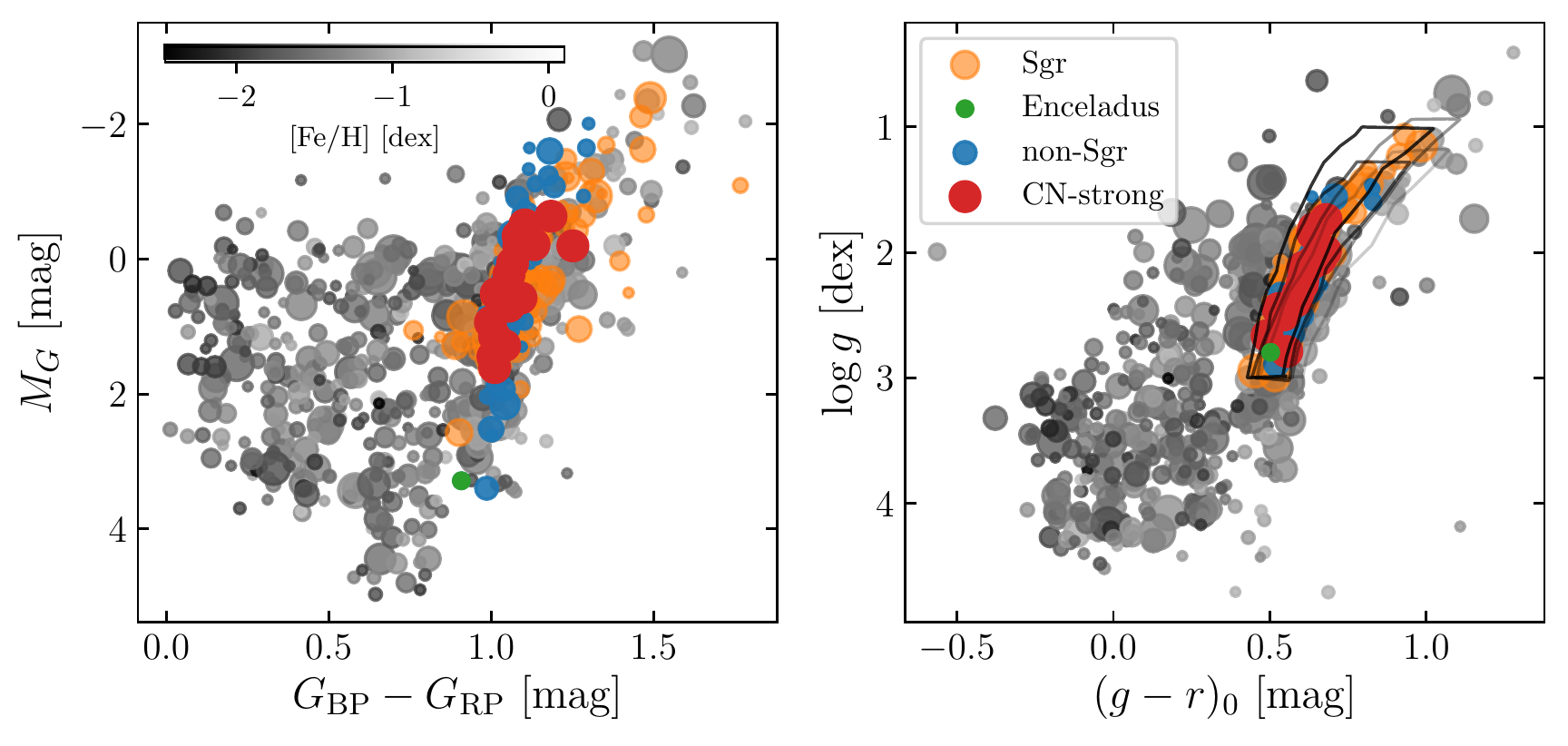}}
      \caption{Same as Fig. \ref{Fig:methodI_2pvs1p} but for associations around CN-strong stars (method III). Here, $M_G$ was computed using the spectrophotometric distances of the CN-strong stars as outlined in Sect. \ref{Subsec: Integrals of motion}. Orange and green circles show the associations with the Sgr stream and Gaia-Enceladus that obey the selection criteria of Paper I but were not considered when estimating the fraction $N_\mathrm{1P}/N_\mathrm{tot}$.
              }
      \label{Fig:methodIII_2pvs1p}
\end{figure*}
\begin{figure}
    \centering
    \resizebox{0.9\hsize}{!}{\includegraphics{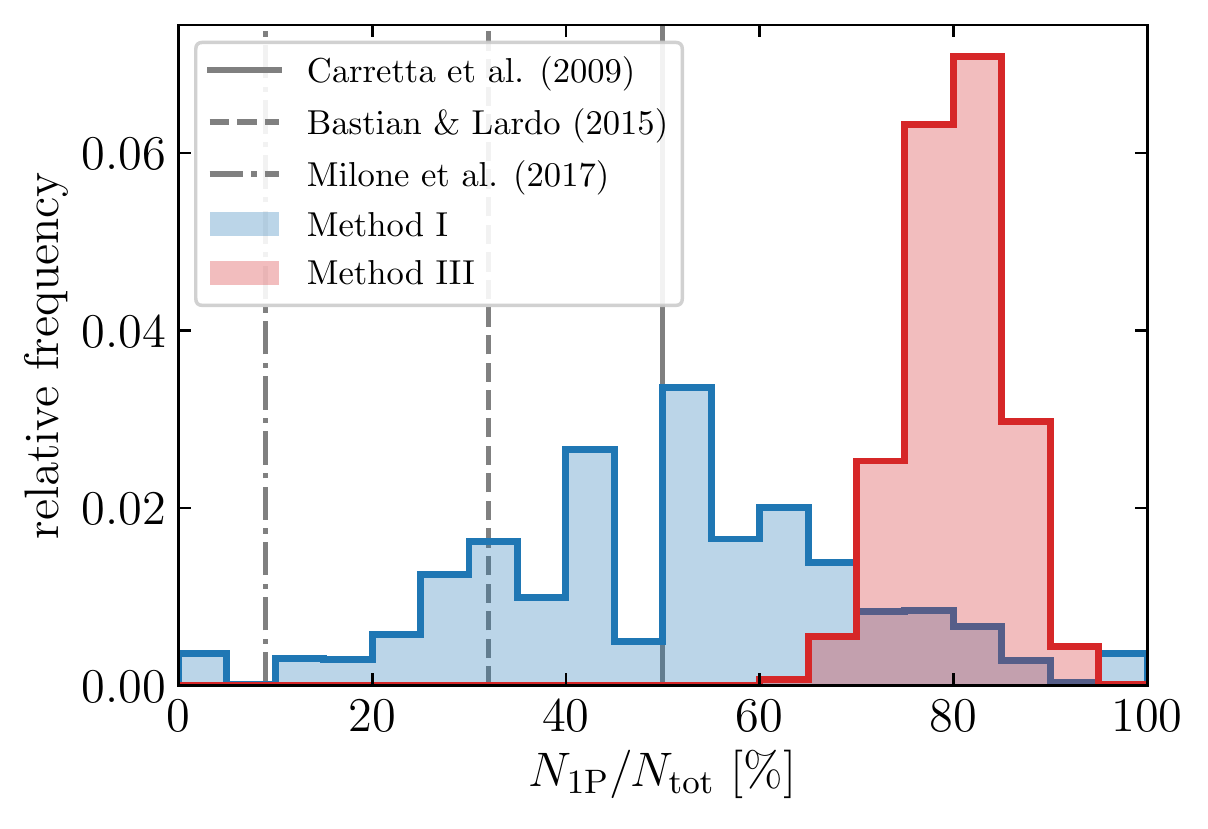}}
      \caption{Result of an MC simulation with $10^7$ realizations for the deduced fraction of first-generation stars among all potential GC escapees obeying the selection function in Paper I (Figs. \ref{Fig:methodI_2pvs1p} and \ref{Fig:methodIII_2pvs1p}). Here, the blue and red histograms represent the findings from methods I and III, respectively. For comparison, we indicate the fractions for bound cluster stars as reported by \citet[][solid vertical line]{Carretta09}, \citet[][dashed vertical line]{Bastian15}, and the lowest value from \citet[][dash-dotted vertical line]{Milone17}.   
              }
      \label{Fig:2pvs1p_hist}
\end{figure}
In Paper I we investigated the fraction of CN-strong stars in our sample of analyzed halo stars to estimate the fractional GC contribution to the stellar inventory of the Galactic halo, $f^\mathrm{GC}_\mathrm{h}$. The outcome of this assessment, among other factors, depends on the assumed fractions of chemically normal 1P stars with respect to the total number of escapees ($N_\mathrm{1P}/N_\mathrm{tot}$), resulting in a corridor for the overall GC contribution to the halo between 8.5\% and 12.5\%. As discussed in detail in Paper I, several values for $N_\mathrm{1P}/N_\mathrm{tot}$ have been proposed for the still bound stellar populations of GCs. For instance, \citet{Carretta09} and \citet{Bastian15} reported 50\% and 32\%, respectively, irrespective of GC properties. \citet{Milone17} found a dependency on the present-day mass of the clusters. Here, for the first time, we elaborate on the average $N_\mathrm{1P}/N_\mathrm{tot}$ in the collection of escaped stars in the Galactic halo by employing our established chemodynamical links.  

Methods I and III revealed new potential GC escapees in the halo field star population that are either directly chemodynamically linked to a cluster or -- adopting the assumption that GCs are the only production sites of CN-strong stars -- can be associated with a CN-strong star. Here, we account for the fact that the candidates cover a broad range of evolutionary stages, while Paper I focused exclusively on the RGB. Hence, the newly discovered sample was cleaned to match the selection criteria outlined in section 2.1. of Paper~I. Briefly, the targets were split into metallicity bins and the exact same fiducial regions for the RGB in a $\log{g}$-$(g-r)_0$ diagram were employed. Most RGB stars with an established chemodynamic link in  this study are lost for the present analysis as they fall outside of the stringent metallicity corridor applied in Paper I ($-1.8<\mathrm{[Fe/H]}<-1.3$~dex\footnote{In hindsight, this should have accounted for the tremendously underestimated errors in SSPP since even a mono-metallic population (e.g., a GC) at exactly [Fe/H$]=-1.55$~dex would lose a substantial amount of stars just based on this cut.}). As a result, our calculations exclude, for example, the metal-poor cluster M\,92 with its many associations on the RGB. Despite reducing the statistical significance, restricting the sample to the RGB confers the coincidental advantage that spurious associations can be considered less likely as the involved stars are brighter than, for example, the upper MS. This results in higher precision in the astrometric quantities [Fe/H] and $v_r$ and therefore stronger discriminatory power. 

Figures \ref{Fig:methodI_2pvs1p} and \ref{Fig:methodIII_2pvs1p} illustrate the subsample obtained in this way for methods I and III. We note that we see no reason to suspect that 1P and 2P stars would have been subject to different selection functions in the SDSS fiber placement, which was solely based on pre-selection from SDSS photometric broad-band filters which in turn are not sensitive to light-element anomalies. Given that method III cannot firmly distinguish between former 1P GC stars and stars stripped from the Sgr dwarf spheroidal galaxy, we excluded the four involved CN-strong stars and their associations. The same argument holds for the population that was tentatively associated with Gaia-Enceladus (Sect. \ref{Subsec: Associations around CN-strong field stars}), which is why we exclude the CN-strong star and its one associated giant obeying the selection criteria, too. Having identified four bona-fide 1P and four 2P stars through method I and 48 1P and 12 2P stars from method III, we calculated $N_\mathrm{1P}/N_\mathrm{tot}=50.0\pm16.7\%$ and $80.2_{-5.2}^{+4.9}\%$, respectively. The values and error margins are taken from the median, 15.9$^\mathrm{th}$, and 84.1$^\mathrm{th}$ percentiles of distributions that were generated in a Monte Carlo (MC) analysis using $10^7$ random draws from Poisson distributions for the involved counting statistics. The two resulting distributions can be found in Fig. \ref{Fig:2pvs1p_hist} where we also show the findings reported for bound cluster stars that were mentioned earlier in this section. 

Interestingly, in the near-field around clusters (i.e., using method I), the first-generation fraction appears to balance the second-generation fraction, and is therefore broadly consistent with the present-day cluster findings by \citet{Carretta09} and \citet{Bastian15}, and with the majority of the analyzed clusters of \citet{Milone17}. Moving away from the GCs, that is, to regions in the halo without a direct spatial link to any GC, $N_\mathrm{1P}/N_\mathrm{tot}$ seems to increase considerably. While it is tempting to claim a solid finding, in particular in light of the low attributed statistical errors from method III, we discuss here two effects that may lead to biases in the estimates. 

The first possibility is that some of the CN-strong stars could have originated in a cluster that was part of an accreted, now entirely disrupted dwarf galaxy (other than Sgr, Gaia-Enceladus, and Sequoia) with its own system of GCs. Therefore, much like already seen in the five CN-strong stars of the Sgr stream and (possibly) Gaia-Enceladus, the stars associated with those CN-strong stars would not only share the chemodynamical information of the  cluster escapees but that of the  stars of the galaxy as well\footnote{Since the now disrupted galaxy was not necessarily chemically homogeneous, this statement is restricted to those stars from the galaxy's potentially broad metallicity distribution function that share the CN-strong stars' [Fe/H].}. Vice versa, while being characterizable as ex-situ, associates of the CN-strong stars would not necessarily be of GC origin. As a consequence, the number of chemically normal stars that are erroneously classified as 1P stars would be artificially increased, which in turn would lead to a higher $N_\mathrm{1P}/N_\mathrm{tot}$. Evidence for CN-bimodalities amongst the dSph field star populations is sparse \cite[e.g.,][]{Smith83,Lardo16b,Norris17,Salgado19}. However, several GCs in the MW have been associated with dwarf-accretion events. It is equally likely that GCs that were part of dSphs have dissolved into their field star populations (prior to their merger with the MW; see \citealt{Malhan19}) meaning that the presence of CN-strong stars within dSphs could be expected, provided that the GCs of  dSphs follow the same trends as the in-situ population of the MW \citep[e.g.,][]{Hendricks16}. 

Another source of increased false-positive rates for 1P stars are spurious associations from the in-situ population of halo stars. However, this latter possibility seems rather unlikely given the independent finding of tight photometric consistency even outside of the selection bounds, whereas halo interlopers could in principle occupy any position in the CMD.  

Bearing in mind all the caveats mentioned above, taken at face value, our high $N_\mathrm{1P}/N_\mathrm{tot}$ for the halo far away from GCs could imply that 1P stars were preferentially lost during early cluster dissolution while 2P stars were more easily retained. This would provide support for early mass-loss scenarios that commonly assume that only 1P stars are lost (e.g., \citeauthor{Dercole08}~\citeyear{Dercole08}, see also the review by \citeauthor{Bastian18}~\citeyear{Bastian18} and the discussions in Paper I). Alternatively, assuming that GC mass loss affects both 1P and 2P stars to the same extent \citep{Kruijssen15} and considering the inverse mass dependency of the now-observed occurrence rate of 1P stars following \citet{Milone17}, low-mass clusters could have contributed the majority of the former GC stars now found in the halo. For the fraction of halo stars donated by GCs, the findings presented here, in concert with the formalism outlined in section 5.1. of Paper I, yield $f^\mathrm{GC}_\mathrm{h}=10.5\pm0.7\%$ and $11.8\pm0.2\%$, assuming either the 1P fraction for the cluster vicinities or for the general halo, where we adopted an early mass-loss rate of 56\% and an average present-day cluster mass of $2.2\cdot10^5\,M_\sun$ (see Paper I, for detailed discussions).

\section{Summary and Conclusions}\label{Sec: Summary and Conclusions}
Here, we investigate the connection between GCs and the stellar component of the Galactic halo in order to contribute to our understanding of how this old component of the Galaxy was formed. To this end, we explore chemodynamical links between halo field stars and the present-day cluster population using combined data from the SDSS/SEGUE and BOSS surveys as well as from the \gaia mission. We created a dataset that allows for the characterization of about $3\cdot10^5$ stars through their full phase-space vectors and chemistry (by means of metallicity). As realistic error budgets are key to our probabilistic approaches, uncertainties on SDSS radial velocities and metallicities were reassessed using well-constrained cluster populations and as a consequence considerably increased. The most important ingredient of our study is the characterization of 112 giant stars in the halo that in our previous study \citep{Koch19} were found to be chemically peculiar in the sense that they show atypically high nitrogen abundances. In the present investigation, we build upon the assumption that the environmental conditions required for the production of such peculiar chemical characteristics are unique to GCs, the only sites where an abundant occurrence of these 2P stars has been found so far.    

Our analysis focuses on three techniques that can be distinguished by the stringency they impose on the precision of the input data. The first approach is restricted to the immediate volumes around current cluster positions, where a solid distance estimate -- made using significant parallaxes -- is not strictly required. In this case, field stars were associated using only six parameters (two spatial and three velocity coordinates as well as [Fe/H]). The confidence in true association was independently solidified by the fact that the associated stars show a remarkably close match to the still bound populations of stars. The second method uses action integrals to tie CN-strong stars to clusters under the presumption that cluster escapees retain their kinematic memory. A spatial coincidence with the clusters is not required at the expense of a much stronger demand in terms of well-constrained distance estimates. This prohibits a wider applicability for distant halo stars in excess of the 112 CN-strong stars, for which we inferred spectrophotometric distances. Nevertheless, we discuss several persistent systematic uncertainties that may have considerable impact on such measurements. The third and final tagging approach adopted throughout this work was used to single out stellar populations in the near-field of CN-strong stars that share similar chemodynamical properties. It may therefore be feasible that those populations stem from the same birth place, that is, GCs.

In total, we established direct chemodynamical links in the near field of eight clusters (NGC\,4147, M\,53, M\,3, M\,13, M\,92, M\,15, M\,2) for 789 stars. While the projected distances of 638 stars are found to reside within the clusters' tidal radii -- thereby rendering them likely bound cluster members -- 151 can be denoted extratidal. Among the escapees, we qualitatively recover structures that resemble the results of previous photometric studies identifying extratidal envelopes \citep[e.g.,][]{Jordi10}. The associations are spread over a wealth of evolutionary stages and metallicities (e.g., four of the eight GCs are more metal-poor than $-2$~dex). Nonetheless, eight giant stars share the same selection function as required in \citet{Koch19}, and four of these are recognized as being CN-strong. From this, based on the low number statistics at hand, we estimate the fraction of 1P stars among the close-by escapees to be $N_\mathrm{1P}/N_\mathrm{tot}=50.0\pm16.7\%$, which is broadly in line with the results of studies that were dedicated to estimating this fraction in nowadays observable cluster populations \citep{Carretta09, Bastian15, Milone17}.

We further report on 145 possible ($p>0.05$) pairs of CN-strong stars in the halo field star population and individual GCs. Of these, 15 are presented as probable star-to-cluster connections ($p\geq0.32$). About 18\% of the association pairs involve clusters that were proposed to be attributable to the major merger events Gaia-Enceladus and Sequoia. Three of the high-probability pairs involve Enceladus GCs (M\,75 and NGC\,1261). We show that 38\% of the CN-strong stars cannot be linked to any surviving cluster, which is in marginal agreement with the finding of $\sim50\%$ by \citet{Savino19}. Among the possible explanations is a scenario whereby these unassociated stars originate from already destroyed clusters.

The third tagging approach revealed 17 populations of stars that share the same portion of the chemodynamical information space as CN-strong stars. For nine such agglomerations of stars, we are able to attribute already known and newly discovered RR Lyrae stars, for which our refined approach for the distance inference provides more reliable distance estimates than the spectrophotometric distances with their attached caveats. Of particular interest, using this method, four CN-strong targets together with their associated stars are found to be probable members of the Sgr stream. Based on the metallicity overlap with M\,54, a GC in the central part of the Sgr dwarf galaxy, we suggest that the four stars may have been stripped from this cluster. 

From the identified RGB stars among the stellar populations around CN-strong stars without spatial coincidence with GCs, we again determined $N_\mathrm{1P}/N_\mathrm{tot}$ and find a value of $80.2_{-5.2}^{+4.9}\%$, which should be regarded as tentative because the statistical uncertainties do not account for all the potential caveats discussed. Nevertheless, it is tempting to argue that a substantial increase of $N_\mathrm{1P}/N_\mathrm{tot}$ from the vicinity of clusters to the overall halo field points towards either a preferential loss of 1P stars during early cluster dissolution or  the birth of a considerable fraction of former GC stars of the halo in since-dissolved clusters. In any case, our findings provide a powerful observational benchmark for theoretical studies of cluster disruption.

As already emphasized in our earlier work, unambiguous confirmation that the identified CN-strong stars are indeed agents of the GC populations showing light-element anomalies remains to be attained. Evidence for imprints from high-temperature proton burning could be gathered through spectroscopic analyses of other key tracer elements such as O, Na, Mg, and Al. Another layer of complexity is added by the open question of whether GCs are genuinely the only sites providing the necessary conditions for the development of abundance anomalies. \citet{Bekki19}, for example, presented a scenario that could explain the substantial fraction of N-rich stars found amongst the Galactic bulge field stars; namely, that these N-rich stars may have  formed in the field from an ISM that has been pre-enriched by AGB ejecta. This latter author further envisions a similar mechanism for the halo. However, our -- admittedly tentative -- observation of an overabundance of CN-strong stars in the cluster surroundings as compared to the halo field challenges this explanation and suggests a mixture of both GC contribution and in-situ formation in the case of the halo.

Our method that is capable of identifying peculiar stars (Paper I) was restricted to the metallicity range $-1.8<\mathrm{[Fe/H]}<-1.3$~dex. Nevertheless, most of the associates reported in the present study, in particular the ones directly surrounding clusters, are more metal-poor than the lower [Fe/H] limit used in Paper I. For instance, the very metal-poor GC M\,92 with its many RGB associates both in the neighboring extratidal envelope and at larger separations remained untouched by our previous approach. Hence, a high-resolution follow up of these moderately faint ($14.5<G<17.5$~mag) targets is strongly called for in order to determine the extratidal $N_\mathrm{1P}/N_\mathrm{tot}$ in the metal-poor regime. Such investigations are paramount to observationally test simulations predicting that the fraction of 1P stars in the halo should increase with decreasing metallicity \citep[][]{ReinaCampos19}. In addition, not being restricted by the accessibility of CN band indices, high-resolution follow-up studies could target our associated MS, MSTO, SGB, and HB targets as well. This would considerably increase the statistical basis of the fractional estimates. The availability of improved number statistics would allow trends with other parameters such as the cluster age to be investigated \citep[e.g.,][]{Martocchia19}. Finally, in the very near future, ongoing and upcoming large-scale spectroscopic surveys such as APOGEE \citep{Majewski17}, WEAVE \citep{Dalton12}, and 4MOST \citep{deJong12, Helmi19, Christlieb19} will substantially enlarge the sample of chemodynamically characterized stars in the Galactic halo. The exploration of those vast datasets that are not restricted to a handful of chemical elements will ultimately lead to a much refined picture of the connection of GC stellar populations and the Milky Way halo. 

\begin{acknowledgements}
M.H., A.K., and E.K.G. gratefully acknowledge support by the Deutsche Forschungsgemeinschaft (DFG, German Research Foundation) -- Project-ID 138713538 -- SFB 881 (``The Milky Way System'', subprojects A03, A05, A08, and A11). Z.P. acknowledges the support of the Hector Fellow Academy. The authors are grateful to G. Parmentier for fruitful discussions. We highly appreciate the constructive report by the anonymous referee. Funding for the Sloan Digital Sky Survey IV has been provided by the Alfred P. Sloan Foundation, the U.S. Department of Energy Office of Science, and the Participating Institutions. SDSS-IV acknowledges
support and resources from the Center for High-Performance Computing at
the University of Utah. The SDSS web site is \url{www.sdss.org}. SDSS-IV is managed by the Astrophysical Research Consortium for the Participating Institutions of the SDSS Collaboration including the Brazilian Participation Group, the Carnegie Institution for Science, Carnegie Mellon University, the Chilean Participation Group, the French Participation Group, Harvard-Smithsonian Center for Astrophysics, Instituto de Astrof\'isica de Canarias, The Johns Hopkins University, Kavli Institute for the Physics and Mathematics of the Universe (IPMU) / University of Tokyo, the Korean Participation Group, Lawrence Berkeley National Laboratory, Leibniz Institut f\"ur Astrophysik Potsdam (AIP), Max-Planck-Institut f\"ur Astronomie (MPIA Heidelberg), 
Max-Planck-Institut f\"ur Astrophysik (MPA Garching), Max-Planck-Institut f\"ur Extraterrestrische Physik (MPE), National Astronomical Observatories of China, New Mexico State University, New York University, University of Notre Dame, Observat\'ario Nacional / MCTI, The Ohio State University, Pennsylvania State University, Shanghai Astronomical Observatory, 
United Kingdom Participation Group, Universidad Nacional Aut\'onoma de M\'exico, University of Arizona, University of Colorado Boulder, University of Oxford, University of Portsmouth, University of Utah, University of Virginia, University of Washington, University of Wisconsin, Vanderbilt University, and Yale University. This work presents results from the European Space Agency (ESA) space mission \textit{Gaia}. \gaia data are being processed by the \gaia Data Processing and Analysis Consortium (DPAC). Funding for the DPAC is provided by national institutions, in particular the institutions participating in the \gaia MultiLateral Agreement (MLA). The \gaia mission website is \url{https://www.cosmos.esa.int/gaia}. The CSS survey is funded by the National Aeronautics and Space Administration under Grant No. NNG05GF22G issued through the Science Mission Directorate Near-Earth Objects Observations Program.  The CRTS survey is supported by the U.S.~National Science Foundation under grants AST-0909182 and AST-1313422.
\end{acknowledgements}

\bibliographystyle{aa}
\bibliography{sources}

\begin{appendix}

\section{RR Lyrae analysis}\label{Sec: RR Lyrae analysis}
RR Lyrae stars are pulsating variables residing on the horizontal giant branch within the instability strip. Their pulsation periods are tightly connected to their luminosity through period--metallicity--luminosity relations \citep[see, e.g.,][]{Catelan04,Muraveva18}, which makes them invaluable distance indicators within the Local Group. Furthermore, they are part of the old stellar population with ages $>$\,10\,Gyr due to their exclusive occurrence in old stellar systems such as globular clusters \citep{VandenBerg13}. RR~Lyrae variables are divided into three classes based on the pulsation mode; fundamental mode stars (RRab), first overtone mode pulsators (RRc), and double-mode variables (RRd, pulsating simultaneously in the fundamental and first overtone). 

A few hundred stars in our analysis fall inside the fundamental mode RR~Lyraes' instability strip boundaries provided by \citet{Clementini19}, intrinsically making them candidates for possible variability. In order to investigate their periodic alternation, we retrieved their photometry from the time domain Catalina sky survey \citep[CSS,][]{Drake09}. In addition, we used  datasets assembled using the data from the CSS \citep{Drake13,Drake13b,Drake14,Abbas14} as a bona
fide catalog of RR~Lyrae stars in the Galactic halo. The crossmatch of the RR~Lyrae candidates with the halo RR~Lyrae sample resulted in matches for 65 cases with determined pulsation periods. To verify the nonvariability of the remaining stars, we elaborated on their possible periodicity using the \texttt{upsilon} package\footnote{Available at \url{https://github.com/dwkim78/upsilon}.} \citep{Dae-Won16}, which searches for periodic behavior among photometric data and provides a classification and light curve parameters of identified variables. To ensure reliable classification, we required at least 50\,\% class probability for stars identified as RR~Lyrae pulsators which were fulfilled by nine stars that we denote as newly discovered RR~Lyrae variables (three RRab, five RRc and one RRd). 

\begin{figure}
    \centering
    \resizebox{0.9\hsize}{!}{\includegraphics{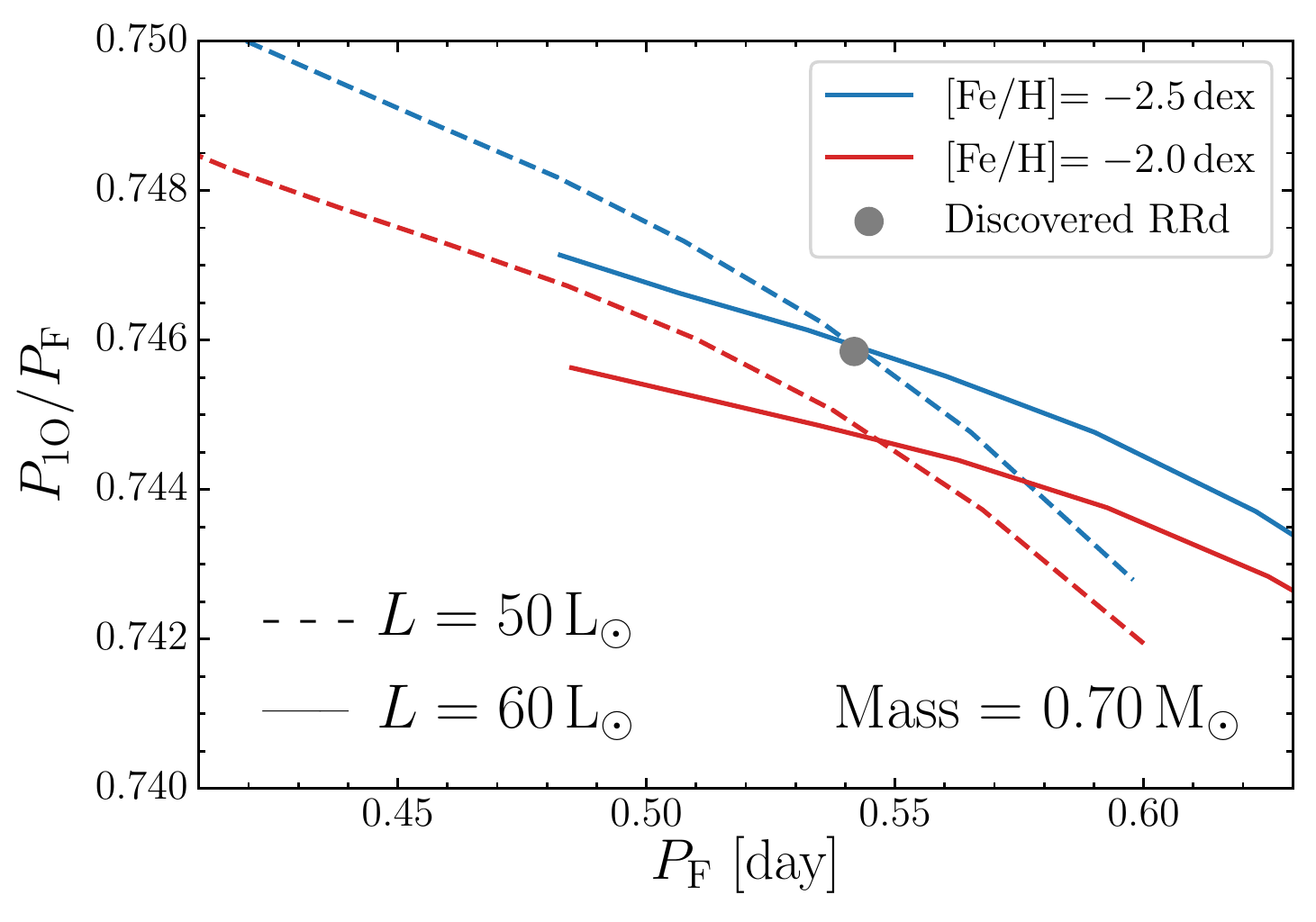}}
      \caption{Period ratio vs. fundamental pulsation period (errors on depicted quantities are smaller than the point size) for a discovered RR~Lyrae star (gray point) compared to the pulsation models \citep[blue and red colors,][]{Smolec08} calculated for a fixed mass ($0.7$\,M$_{\odot}$), metallicity ($-2.0$ and $-2.5$\,dex), and a grid of stellar parameters ($L$ and $T_{\rm eff}$) typical for RR~Lyrae stars.   
              }
      \label{fig:AppPetersen}
\end{figure}
The one RRd variable in our sample is a newly discovered double-mode RR~Lyrae star with a dominant first overtone mode (pulsation period $P_{\rm 1O}=0.4040992$\,d) and a secondary fundamental mode at $P_{\rm F}=0.5417994$\,d. The period ratio between both modes is $P_{\rm 1O} / P_{\rm F} = 0.7458465$, which together with the long fundamental mode period suggests a low metallicity between $-2.0$ and $-2.5$\,dex (see Fig.~\ref{fig:AppPetersen} for details) in line with M\,15's [Fe/H].

The assembled RR~Lyrae light curves (both known and newly discovered) were phased using the stars' ephemerids and decomposed using the Fourier series to determine the time of maximum brightness, the mean magnitude, and the pulsation amplitude:
\begin{equation} \label{eq:FourierSeries}
m\left ( t \right ) = A_{0} + \sum_{k=1}^{n} A_{k} \cdot \text{cos} \left (2 \pi k \vartheta + \varphi_{k} \right ),
\end{equation}
where $\vartheta$ represents the phase function defined as $\left(MJD-M_{0}\right)/P$, $MJD$ denotes the observation in the Modified Julian Date, and $P$ and $M0$ are ephemerids of a given star (pulsation period and time of brightness maximum). $A_{k}$ and $\varphi_{k}$ stand for amplitudes and phases, respectively, with $n$ denoting the degree of the Fourier series, and $A_{0}$ representing the mean magnitude. Each phased light curve was visually inspected to verify the ephemerids and pulsation amplitudes derived from the Fourier fits.

The absolute magnitudes of RR~Lyrae stars in the visual part of the spectrum strongly correlate with their metallicities \citep[see, e.g.,][]{Catelan04,Muraveva18}, thus acquiring homogenous absolute magnitudes for all three RR~Lyrae classes in our sample can be troubling due to the lack of reliable metallicities \citep[cf., e.g.,][]{Smolinski11,Hanke18,Fabrizio19}. Hence, for the distance estimates we decided to assume a single absolute magnitude for the entire RR~Lyrae sample and allowed a large dispersion $M_{\rm V} = 0.6 \pm 0.5$\,mag.  

The single value for the absolute magnitude of all RR~Lyrae variables in our sample was derived using a sample of RR~Lyrae stars from \citep{Muraveva18}, {\it Gaia} parallaxes, 3D dust maps from \citet{Green19}, and mean $V$-band magnitudes from the ASAS survey \citep{Pojmanski02}. For the aforementioned quantities, we ran an MC error simulation of the distance modulus assuming two different offsets in {\it Gaia} parallaxes \citep[$-0.057$\,mas, and $-0.029$\,mas, respectively,][]{Muraveva18,Lindegren18}, and estimated absolute magnitudes for individual RR~Lyrae variables. The $\langle M_{\rm V}\rangle$ magnitudes clustered around $0.68 \pm 0.35$\,mag and $0.56 \pm 0.37$\,mag for {\it Gaia} parallax offsets, $-0.057$\,mas and $-0.029$\,mas, respectively. Pursuing a conservative approach, we used $M_{\rm V} = 0.6$\,mag with a large dispersion of 0.5\,mag to account for the offset in both absolute magnitude values derived from parallaxes. 

The distances for our sample of RR~Lyrae variables were calculated using the MC simulation with 1000 realizations assuming Gaussian error distributions in the reddening \citep[using dust maps from][]{Schlafly11}, absolute magnitudes, and apparent magnitudes\footnote{We used the average photometric error of a given RR~Lyrae variable as error estimate for the mean magnitude.}. The resulting distances come with errors of around 20-25\,\%, larger than generally reported based on optical data for RR~Lyrae stars \citep[around 8\,\%,][]{Neeley17}. This is mainly caused by the conservative assumption for absolute RR~Lyrae magnitudes.

Periodic luminosity changes are driven by the variation of the stellar radius and effective temperature. This inevitably affects the radial velocity measurements that comprise a combination of systemic and pulsation velocity. The pulsation effect can be removed from a single radial velocity measurement using radial velocity templates \citep{Sesar12}\footnote{ \citet{Sesar12} provides radial velocity templates for measurements based on H$\alpha$, H$\beta$, H$\gamma$, and metallic lines. We note that in our analyses of SDSS spectra we used only the H$\alpha$ profile due to its high prominence even in the low-quality spectra of faint RR~Lyrae stars in our dataset.} that scale with their photometric counterparts. 

We note that for RRc and RRd type RR~Lyrae stars, \citet{Sesar12} does not provide a scaling relation between the amplitudes of $v_r$ and photometric light curves. Thus, we used a sample collected by \citep{Sneden17} for radial velocity amplitudes (based on the H$\alpha$ line) of the first overtone pulsators in the field and compared them with their photometric amplitudes using data from the ASAS survey \citep{Pojmanski02}. Applying the latter comparison we obtained the following scaling relation between the radial and photometric amplitudes for the first overtone pulsators:
\begin{equation} \label{eq:RadialPhotomet}
A_{\rm rv}^{\text{H}\alpha} = 95.8 A_{V} - 5.2.
\end{equation}
The aforementioned relation was also used for the determination of the systemic velocity of the one RRd star in our sample.

In order to determine the systemic velocity of RR~Lyrae stars in our sample we extracted single-epoch spectra from the SDSS and measured the radial velocity for the H$\alpha$ line by cross-correlation with a synthetic template\footnote{Synthesized spectrum of a star with stellar parameters typical for an RR~Lyrae star; $T_{\rm eff}=6700$\,K, log$g=2.15$, [Fe/H]$=-1.5$\,dex} using the \texttt{iSpec} package routines \citep{Blanco-Cuaresma14,Blanco-Cuaresma19} and MOOG \citep[][February 2017 version]{Sneden73}. Individual radial velocity measurements were phased using the stars' ephemerids with MJDs centered in the middle of the exposure. We scaled the H$\alpha$ radial velocity template based on the stars amplitude and subtracted the pulsation velocity from the measured radial velocity. The final systemic velocity for a given RR~Lyrae star was estimated through the median of individual, pulsation-corrected radial velocity measurements. Errors on the systemic velocities were estimated following the prescriptions by \citep{Sesar12}. 

\section{Additional figures}
\begin{figure*}
    \centering
    \resizebox{0.73\hsize}{!}{\includegraphics{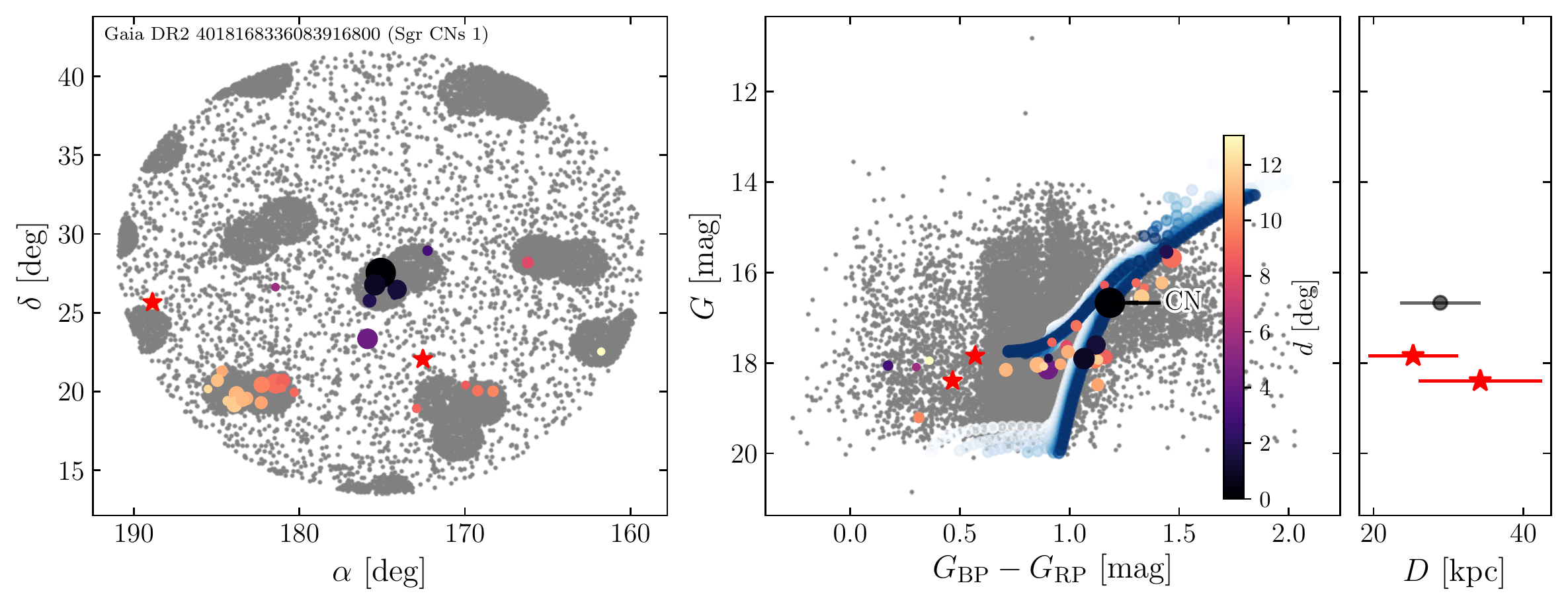}}
      \caption{Same as Fig. \ref{Fig:SgrCNs3_method3} but for associations to the CN-strong star \gaia DR2 4018168336083916800 (Sgr CNs 1).
              }
     \label{Fig: C1}
\end{figure*}
\begin{figure*}
    \centering
    \resizebox{0.73\hsize}{!}{\includegraphics{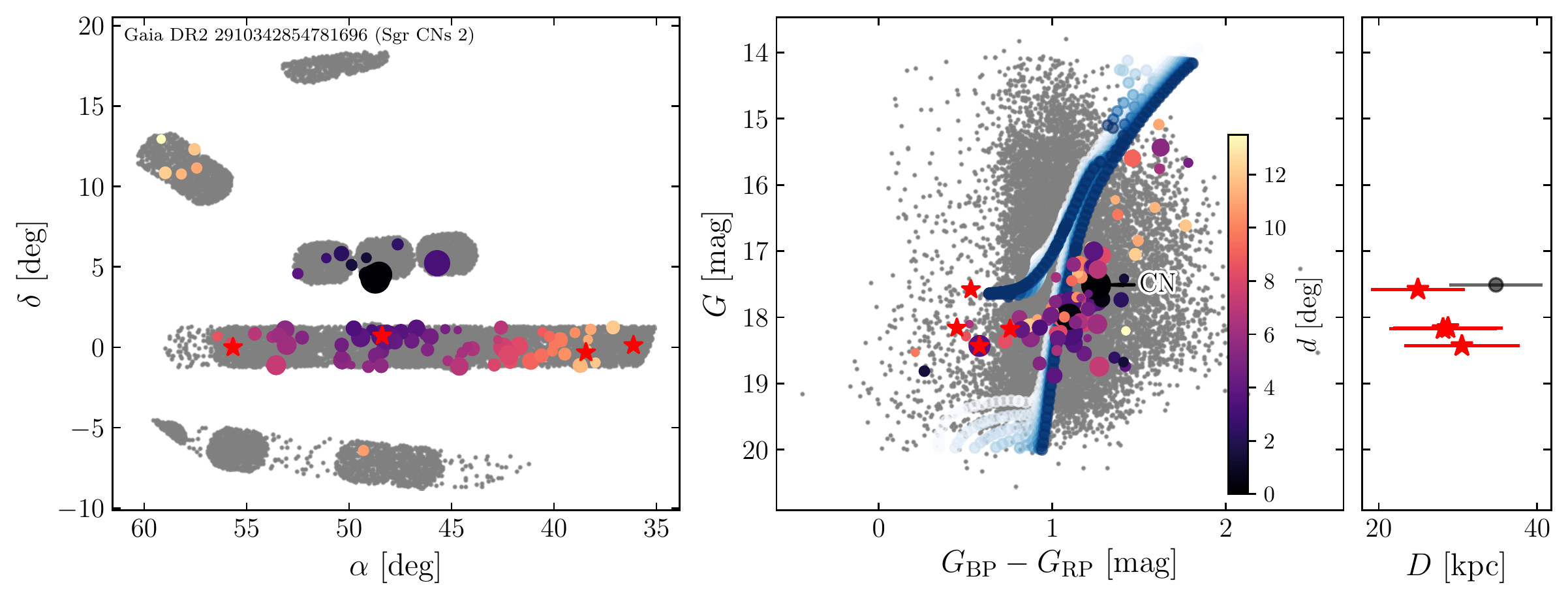}}
      \caption{Same as Fig. \ref{Fig:SgrCNs3_method3} but for associations to the CN-strong star \gaia DR2 2910342854781696 (Sgr CNs 2).
              }
\end{figure*}
\begin{figure*}
    \centering
    \resizebox{0.73\hsize}{!}{\includegraphics{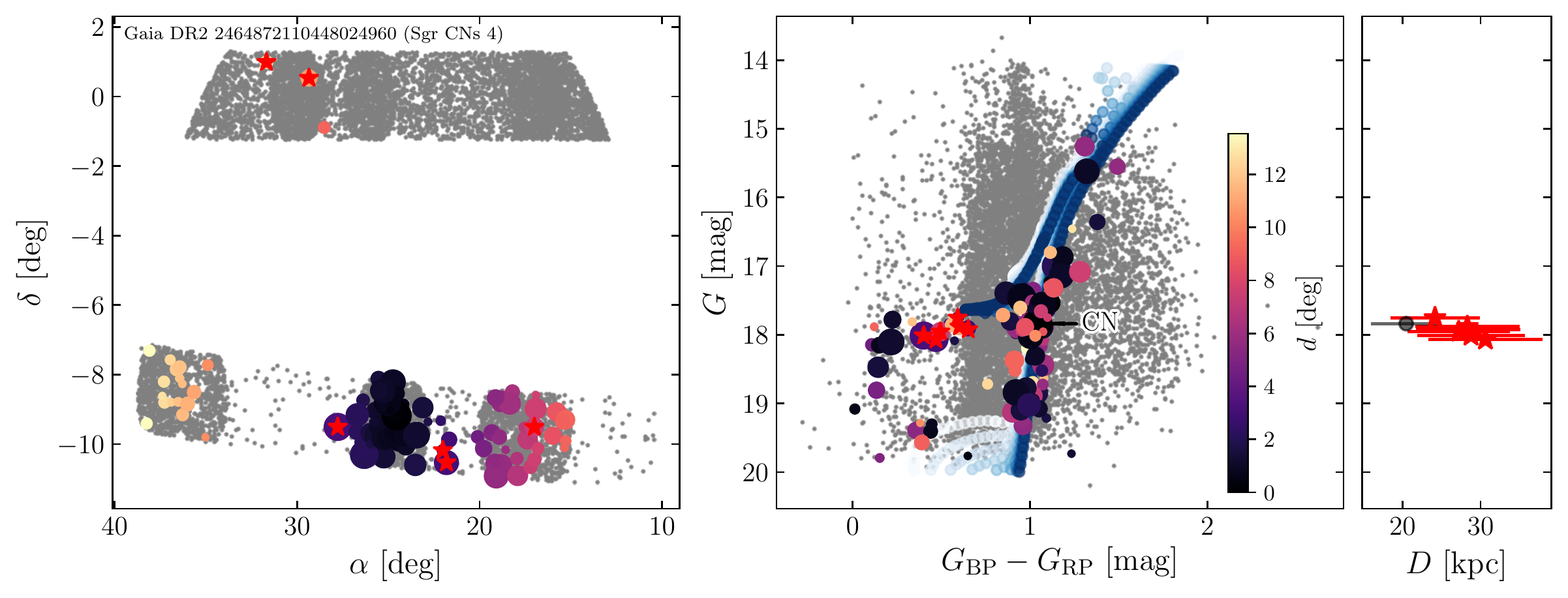}}
      \caption{Same as Fig. \ref{Fig:SgrCNs3_method3} but for associations to the CN-strong star \gaia DR2 2464872110448024960 (Sgr CNs 4).
              }
\end{figure*}
\begin{figure*}
    \centering
    \resizebox{0.73\hsize}{!}{\includegraphics{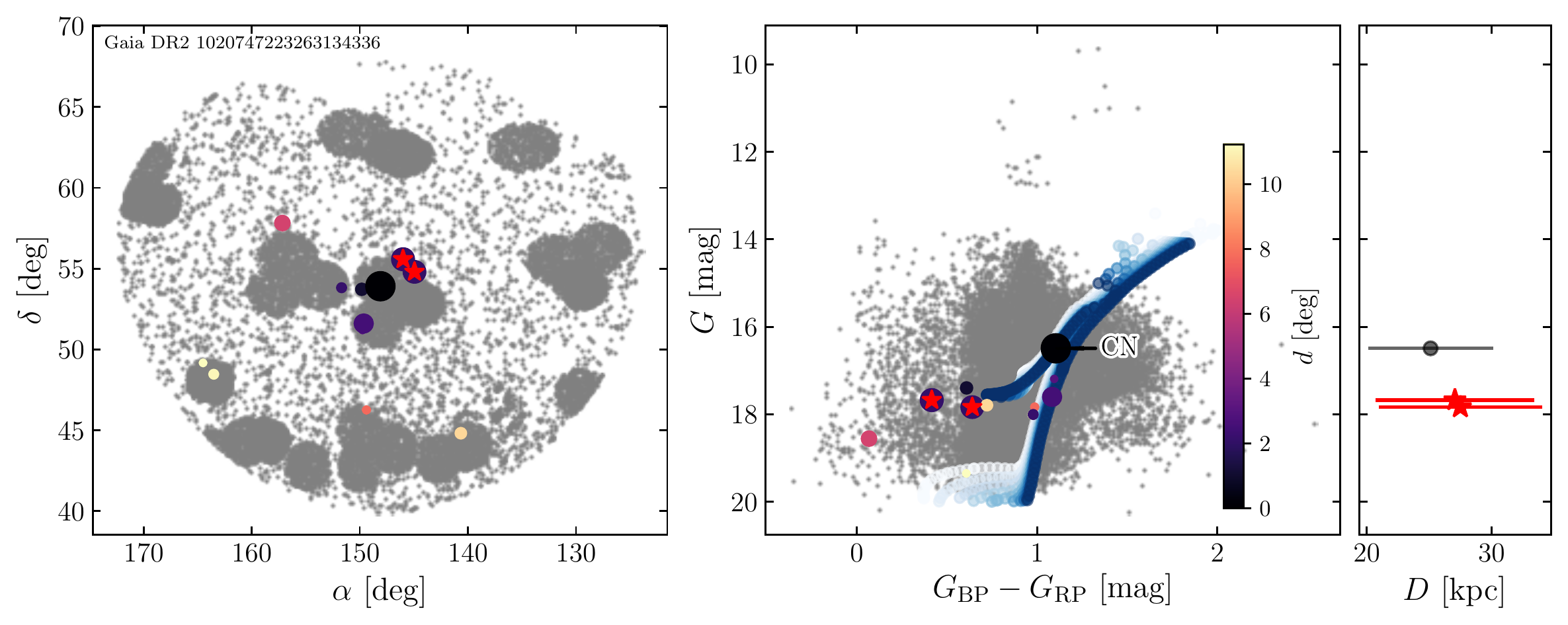}}
      \caption{Same as Fig. \ref{Fig:SgrCNs3_method3} but for associations to the CN-strong star \gaia DR2 1020747223263134336.
              }
\end{figure*}
\begin{figure*}
    \centering
    \resizebox{0.73\hsize}{!}{\includegraphics{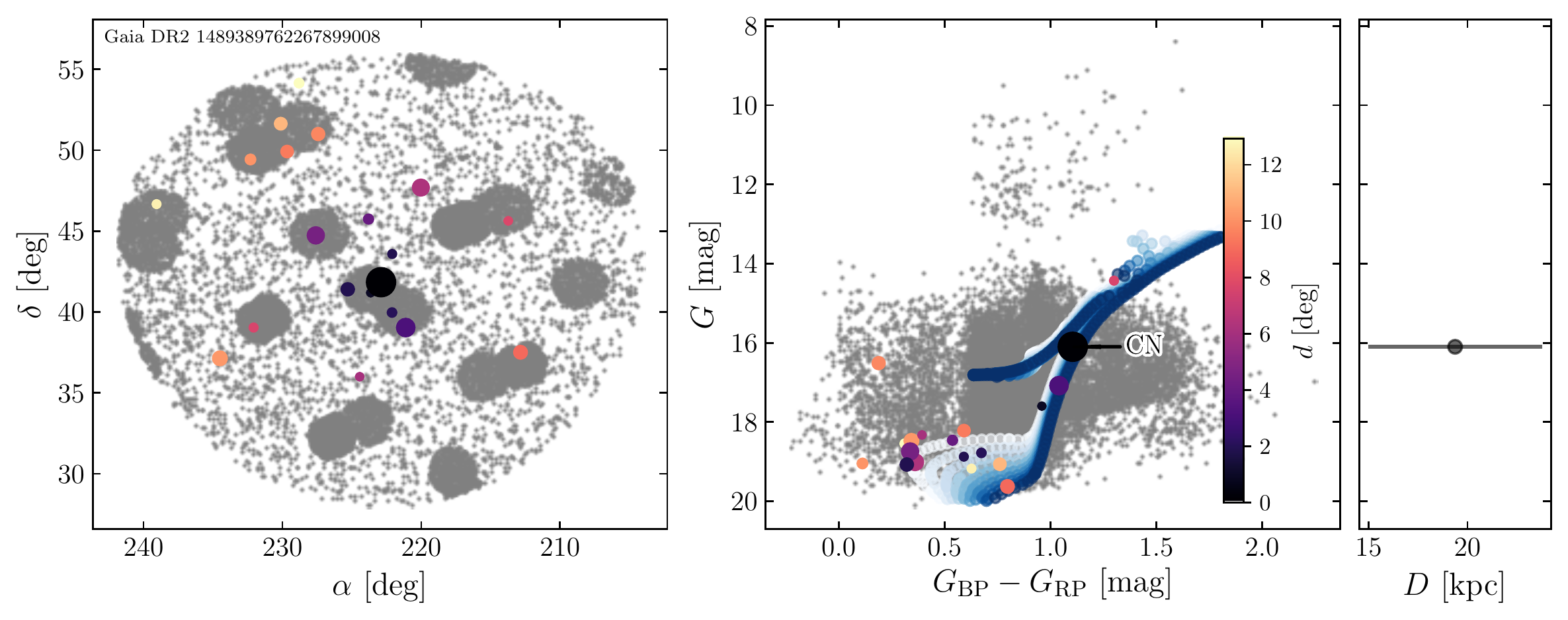}}
      \caption{Same as Fig. \ref{Fig:SgrCNs3_method3} but for associations to the CN-strong star \gaia DR2 1489389762267899008.
              }
\end{figure*}
\begin{figure*}
    \centering
    \resizebox{0.73\hsize}{!}{\includegraphics{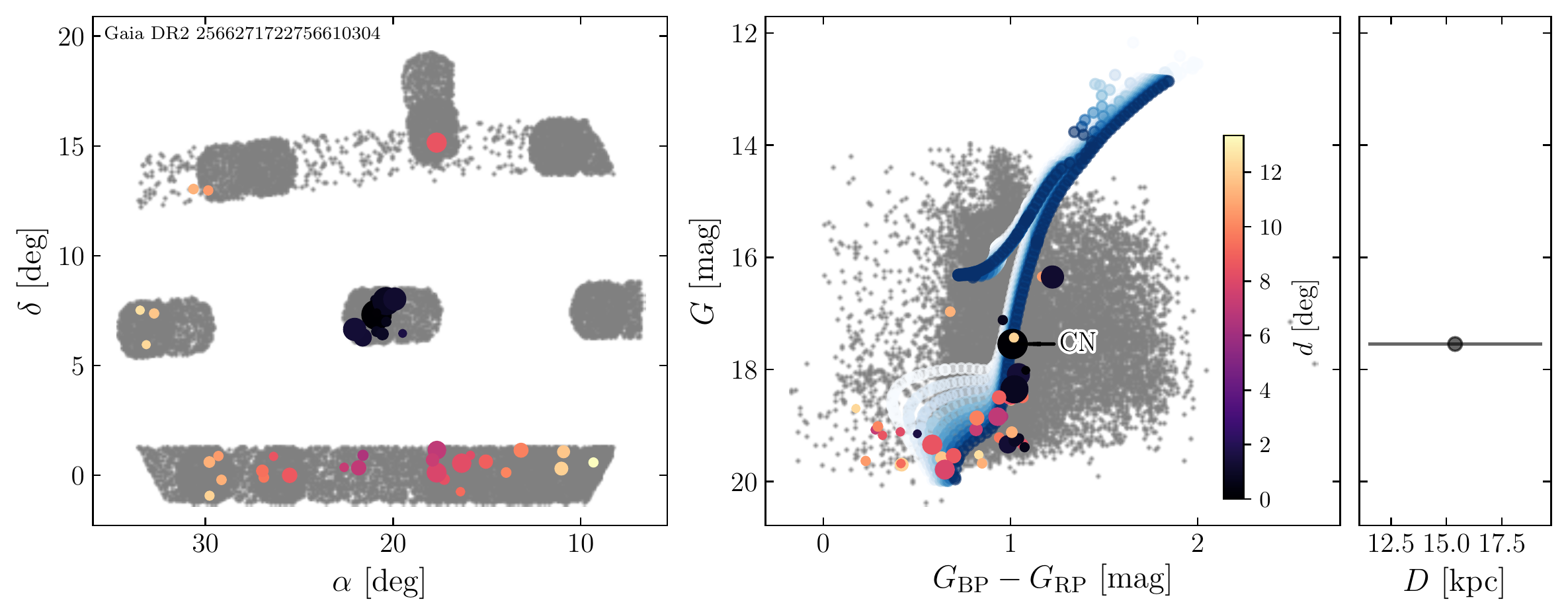}}
      \caption{Same as Fig. \ref{Fig:SgrCNs3_method3} but for associations to the CN-strong star \gaia DR2 2566271722756610304.
              }
\end{figure*}
\begin{figure*}
    \centering
    \resizebox{0.73\hsize}{!}{\includegraphics{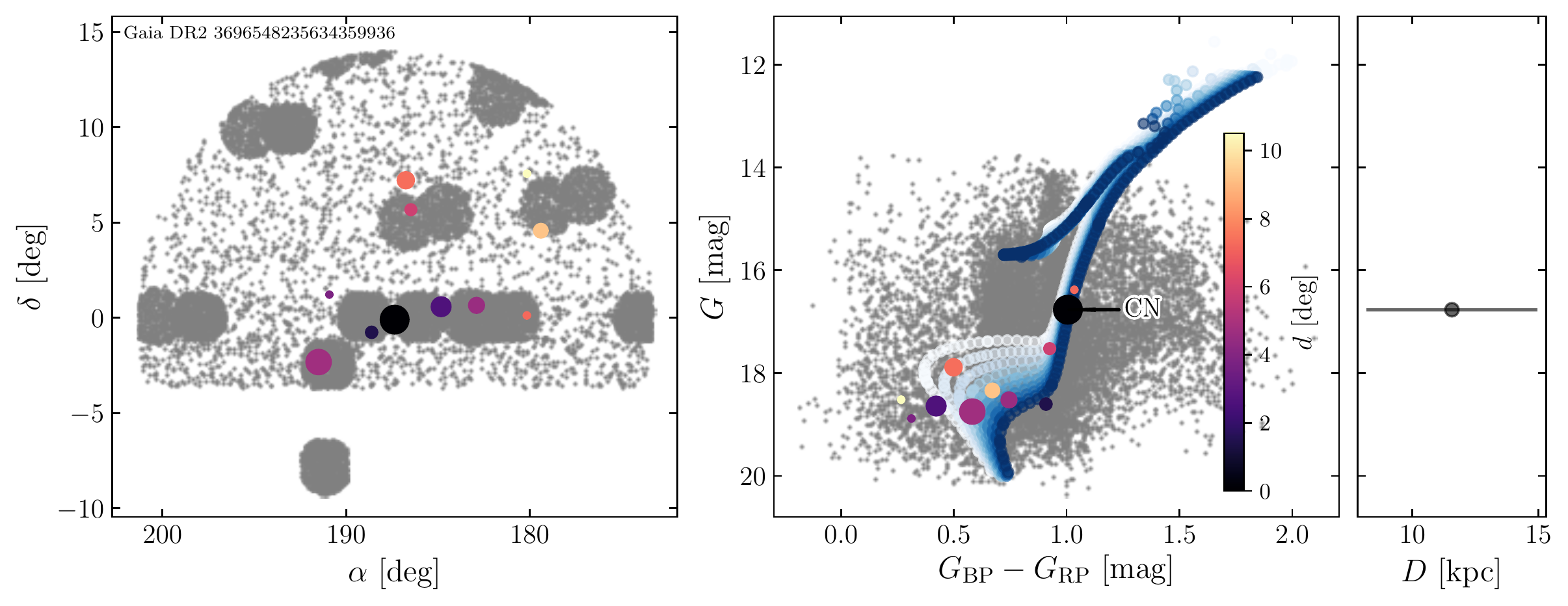}}
      \caption{Same as Fig. \ref{Fig:SgrCNs3_method3} but for associations to the CN-strong star \gaia DR2 3696548235634359936. We note that this group of stars may be associated with Gaia-Enceladus through its association with NGC\,1261 (Sects. \ref{Subsec: Associating CN-strong stars with clusters} and \ref{Subsec: Associations around CN-strong field stars}),
              }
\end{figure*}
\begin{figure*}
    \centering
    \resizebox{0.73\hsize}{!}{\includegraphics{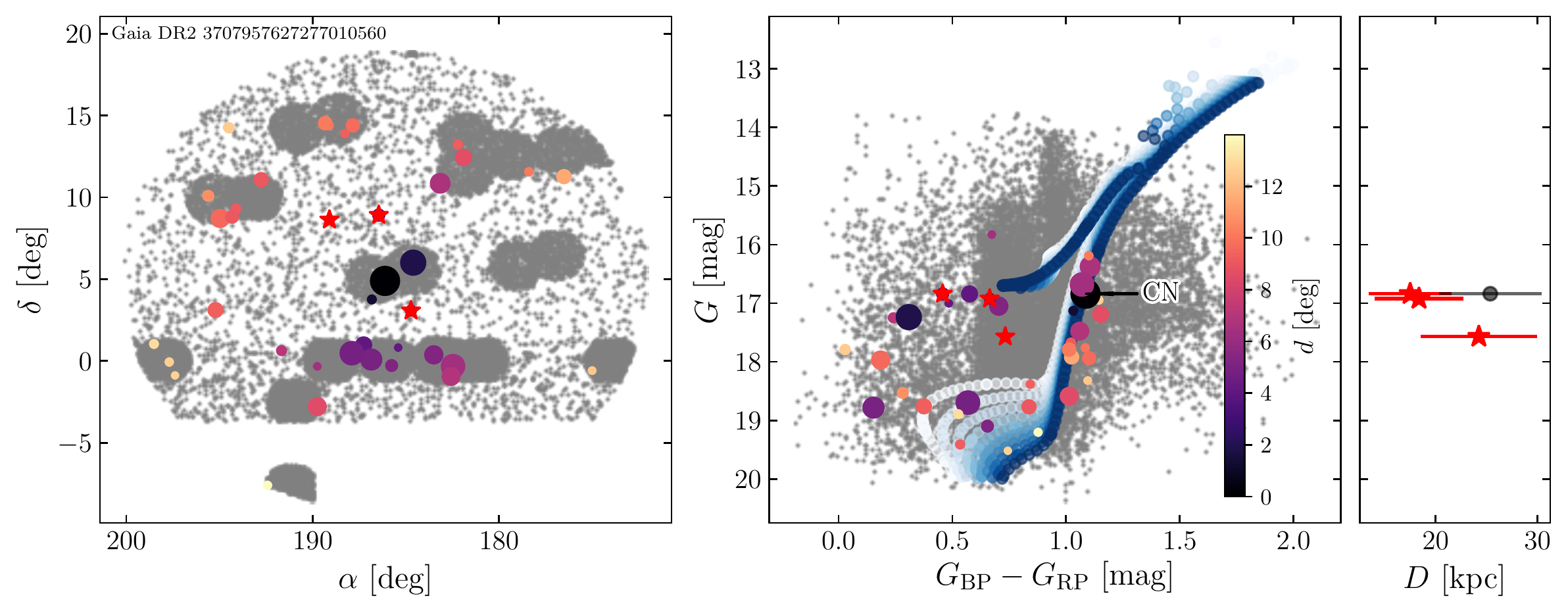}}
      \caption{Same as Fig. \ref{Fig:SgrCNs3_method3} but for associations to the CN-strong star \gaia DR2 3707957627277010560.
              }
\end{figure*}
\begin{figure*}
    \centering
    \resizebox{0.73\hsize}{!}{\includegraphics{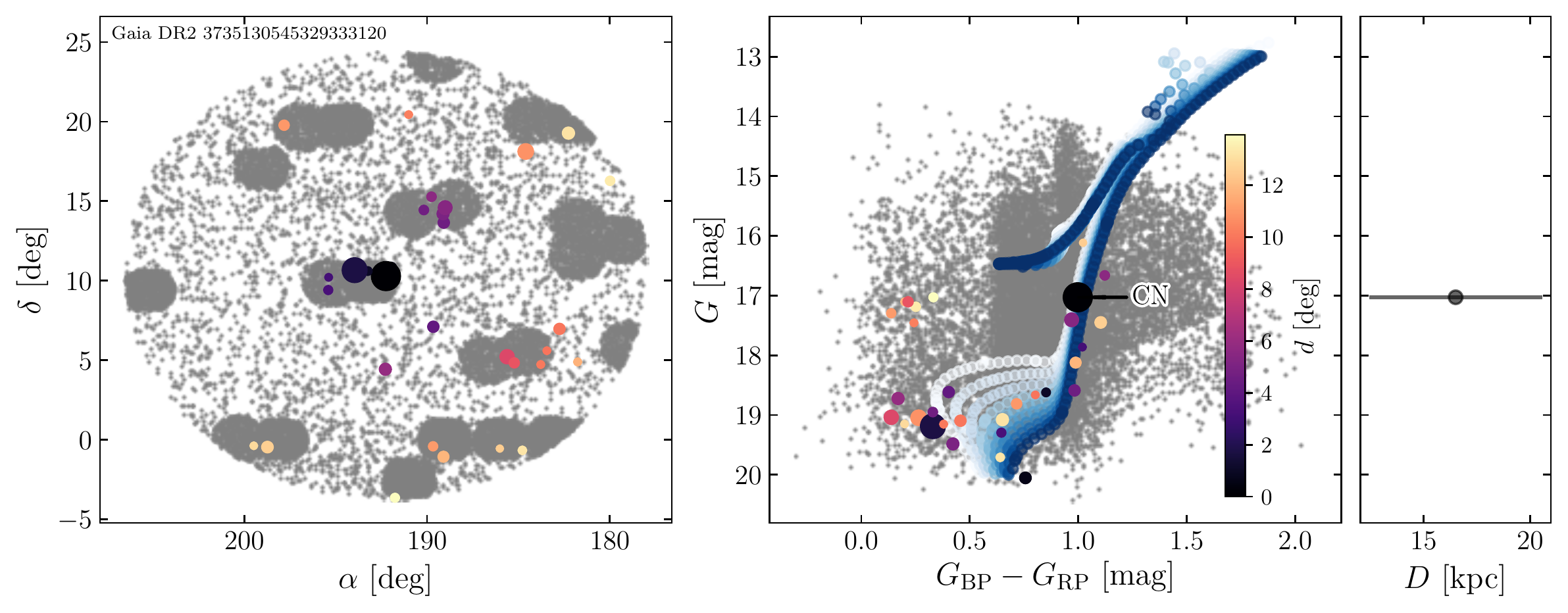}}
      \caption{Same as Fig. \ref{Fig:SgrCNs3_method3} but for associations to the CN-strong star \gaia DR2 3735130545329333120.
              }
\end{figure*}
\begin{figure*}
    \centering
    \resizebox{0.73\hsize}{!}{\includegraphics{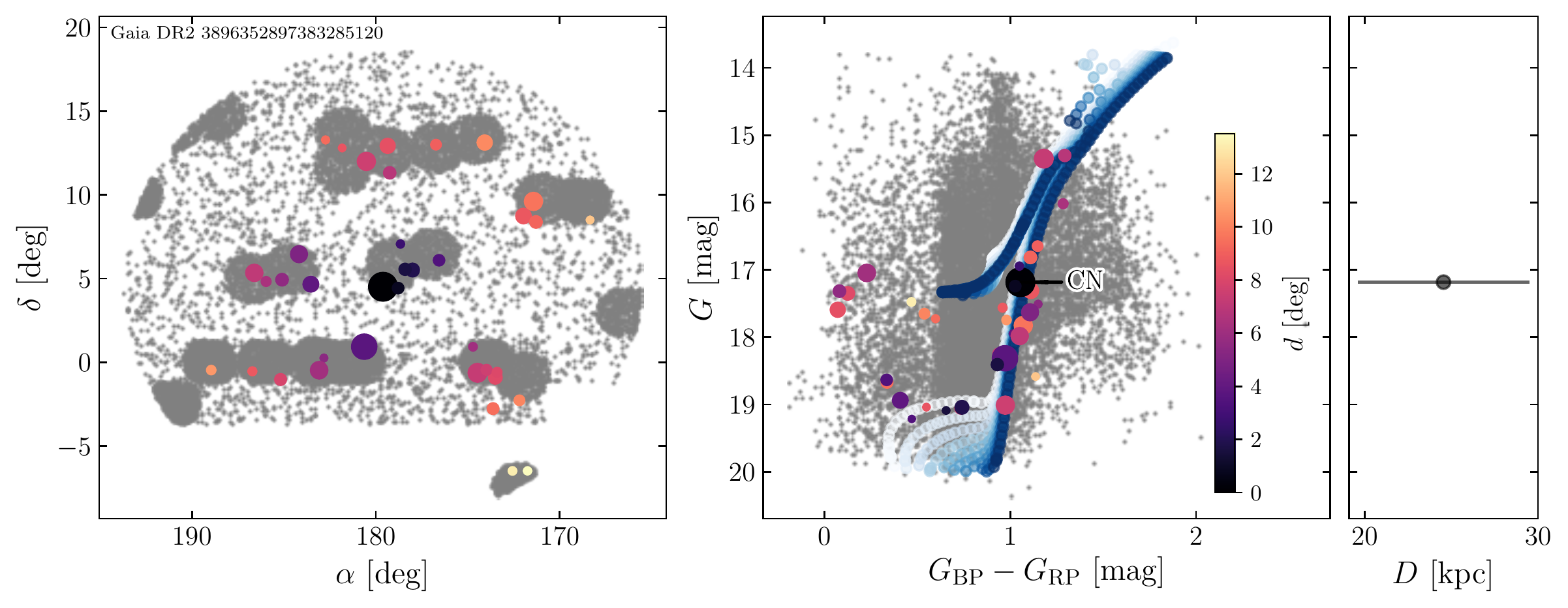}}
      \caption{Same as Fig. \ref{Fig:SgrCNs3_method3} but for associations to the CN-strong star \gaia DR2 3896352897383285120.
              }
\end{figure*}
\begin{figure*}
    \centering
    \resizebox{0.73\hsize}{!}{\includegraphics{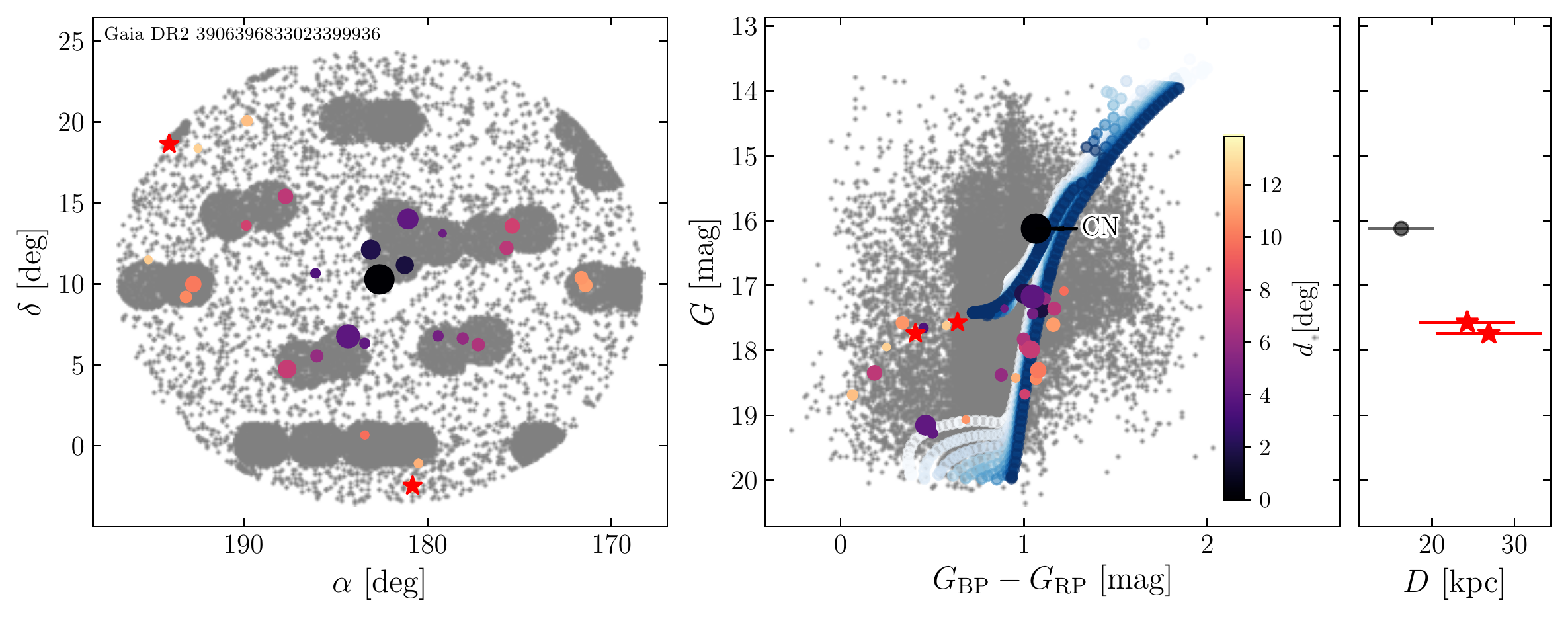}}
      \caption{Same as Fig. \ref{Fig:SgrCNs3_method3} but for associations to the CN-strong star \gaia DR2 3906396833023399936.
              }
\end{figure*}
\begin{figure*}
    \centering
    \resizebox{0.73\hsize}{!}{\includegraphics{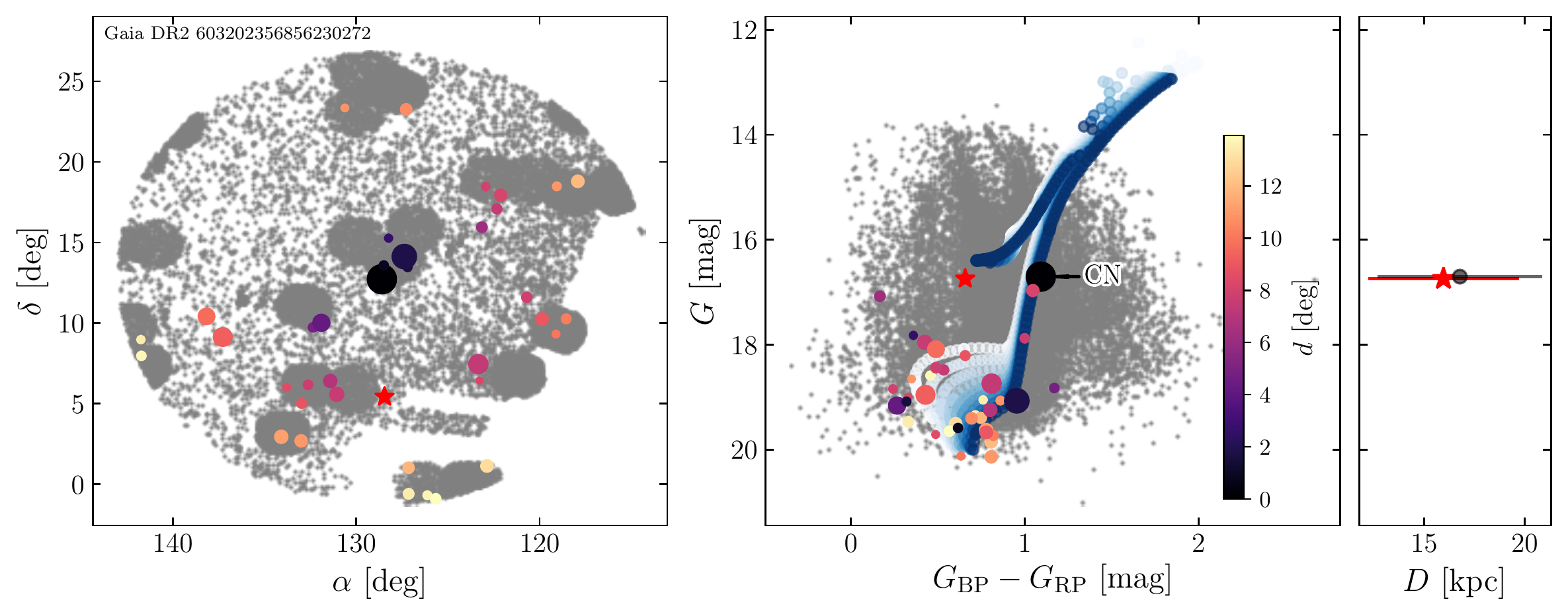}}
      \caption{Same as Fig. \ref{Fig:SgrCNs3_method3} but for associations to the CN-strong star \gaia DR2 603202356856230272.
              }
\end{figure*}
\begin{figure*}
    \centering
    \resizebox{0.73\hsize}{!}{\includegraphics{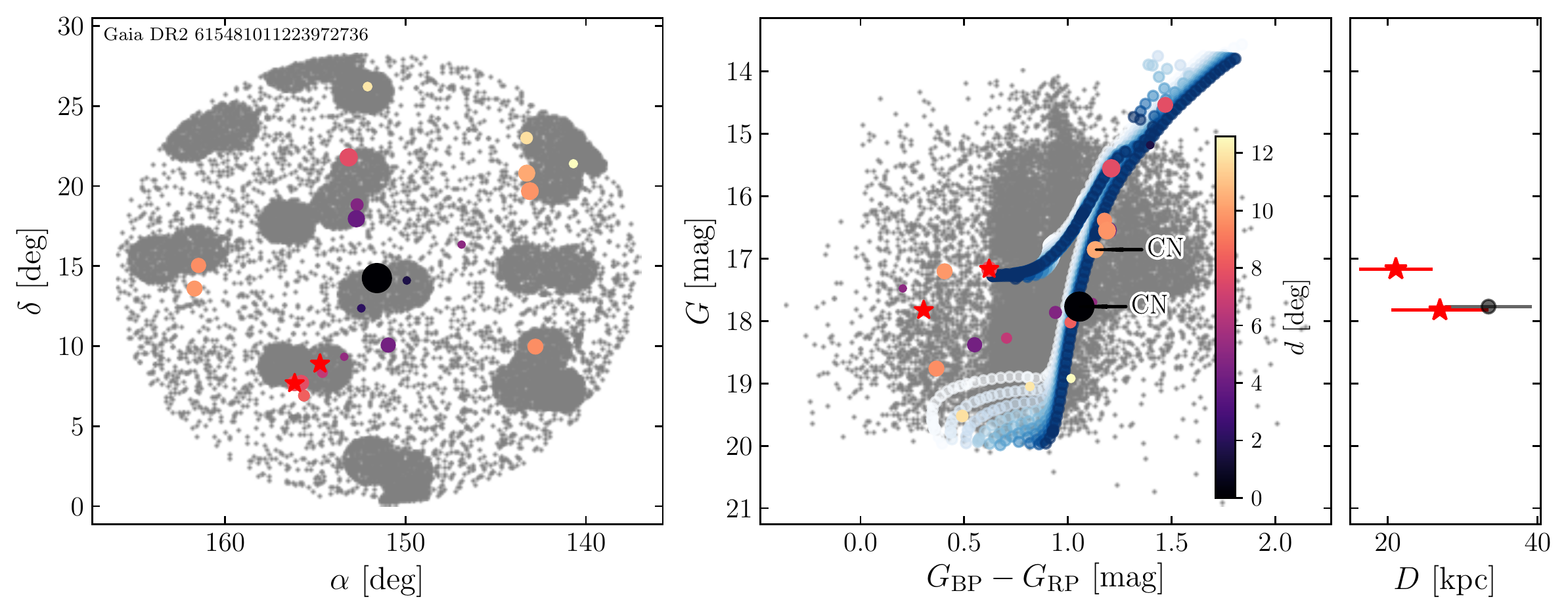}}
      \caption{Same as Fig. \ref{Fig:SgrCNs3_method3} but for associations to the CN-strong star \gaia DR2 615481011223972736.
              }
      \label{Fig: pair1}
\end{figure*}
\begin{figure*}
    \centering
    \resizebox{0.73\hsize}{!}{\includegraphics{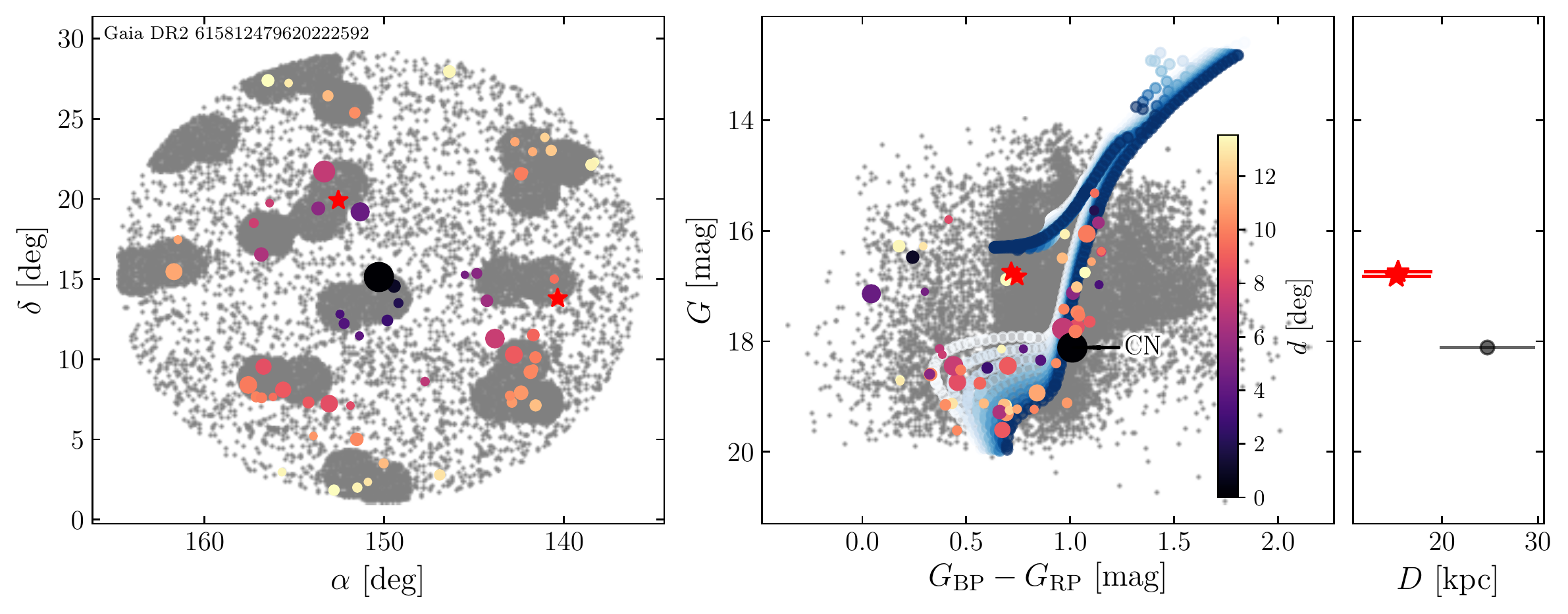}}
      \caption{Same as Fig. \ref{Fig:SgrCNs3_method3} but for associations to the CN-strong star \gaia DR2 615812479620222592.
              }
\end{figure*}
\begin{figure*}
    \centering
    \resizebox{0.73\hsize}{!}{\includegraphics{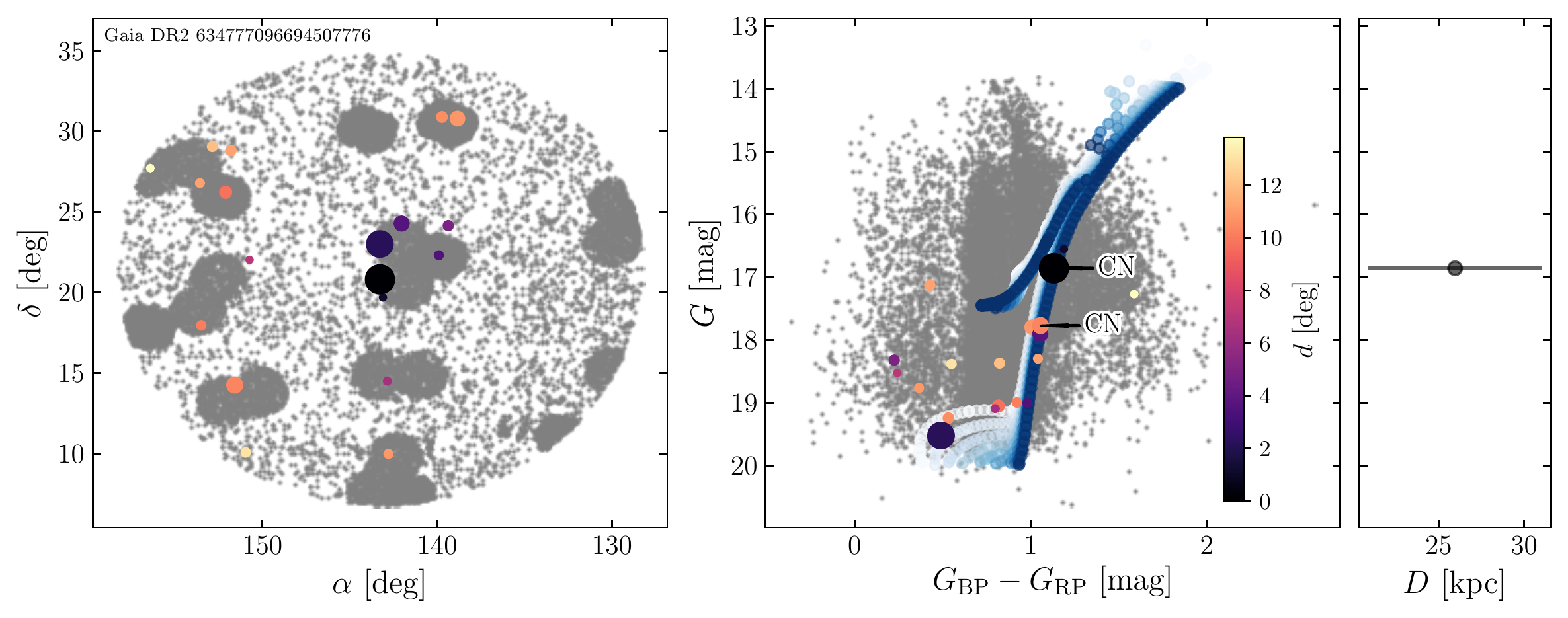}}
      \caption{Same as Fig. \ref{Fig:SgrCNs3_method3} but for associations to the CN-strong star \gaia DR2 634777096694507776.
              }
      \label{Fig: pair2}
\end{figure*}
\begin{figure*}
    \centering
    \resizebox{0.73\hsize}{!}{\includegraphics{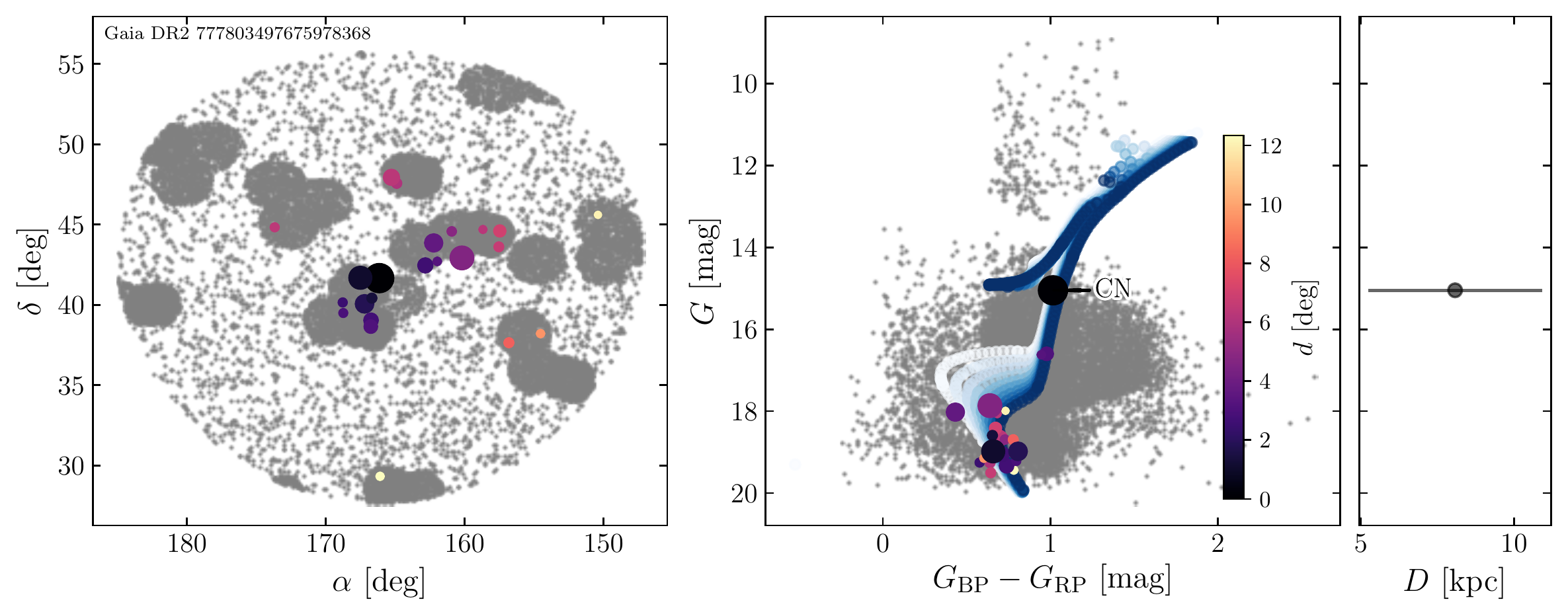}}
      \caption{Same as Fig. \ref{Fig:SgrCNs3_method3} but for associations to the CN-strong star \gaia DR2 777803497675978368.
              }
      \label{Fig: C16}
\end{figure*}

\section{Additional tables}
\begin{sidewaystable}[!h]
\caption{Candidate cluster escapees in the near-field of clusters (method I).}
\label{Table: Extratidal target information}
\centering%\small
\begin{tabular}{l@{\hspace{2ex}}c@{\hspace{2ex}}c@{\hspace{2ex}}c@{\hspace{2ex}}rrcr@{\hspace{0.1ex}}c@{\hspace{0.1ex}}lc@{\hspace{2ex}}c@{\hspace{2ex}}c@{\hspace{3ex}}cc}
\hline\hline
\\[-5pt]
\gaia DR2 source ID & $\alpha^{(a)}$ & $\delta^{(a)}$ & \multicolumn{1}{c}{$r/r_t$} & \multicolumn{1}{c}{$\varpi$} & $\mu_\alpha \cos{\delta}$ & $\mu_\delta$ & \multicolumn{3}{c}{$v_r$} & [Fe/H] & $P(A/B)$ & $G$ & $G_\mathrm{BP}-G_\mathrm{RP}$ & Rem.$^{(b)}$\\
 & [deg] & [deg] &  & \multicolumn{1}{c}{[mas]} & [mas yr$^{-1}$] & [mas yr$^{-1}$] & \multicolumn{3}{c}{[km s$^{-1}$]} & [dex] & & [mag] & [mag] & \\[5pt]
\hline\\[-5pt]
\\[-7pt]\multicolumn{15}{c}{NGC\,4147             }\\[3pt]
3950157700750904064 & 182.459 &  18.773 &  2.4 & $ 0.06\pm0.15$ & $-1.9\pm0.2$ & $-2.1\pm0.2$ & $ 184.8$ & $\pm$ & $ 5.7$ & $-1.53\pm0.16$ & 0.87 & 17.69 & 1.01 &\\
3950527823852135808 & 183.537 &  19.185 & 11.4 & $-0.57\pm0.38$ & $-3.8\pm1.0$ & $-1.7\pm0.4$ & $ 188.1$ & $\pm$ & $10.9$ & $-1.90\pm0.17$ & 0.06 & 19.22 & 0.77 &\\
3950992882911234944 & 182.416 &  19.691 & 11.4 & $-0.48\pm0.46$ & $-3.2\pm0.9$ & $-1.7\pm0.4$ & $ 160.6$ & $\pm$ & $10.5$ & $-1.91\pm0.15$ & 0.05 & 19.41 & 0.72 &\\
\hline
\end{tabular}
\tablefoot{Only a small portion of the data is shown to indicate the form and content of the table. The full table is available through the CDS. 
\tablefoottext{a}{Coordinates are in the \gaia DR2 reference epoch J2015.5.}
\tablefoottext{b}{Remarks: (1): Classified as being a CN-strong giant in Paper~I. (2): Identified as RR Lyrae star.}
}
\end{sidewaystable}

\begin{table*}
\caption{Star-cluster pairs from method II with $p_{ij}'\geq0.05$.}
\label{Table: CN-strong cluster pairs}
\centering
\resizebox{\textwidth}{!}{%
\begin{tabular}{@{\extracolsep{18pt}}rccccrcccc}
\hline\hline
\\[-5pt]
\gaia DR2 source ID & GC & $p_{ij}'$ & m.m.$^{(a)}$ & $p_\mathrm{ref}^{(b)}$ & \gaia DR2 source ID & GC & $p_{ij}'$ & m.m.$^{(a)}$ & $p_\mathrm{ref}^{(b)}$\\[5pt]
\cline{1-5}\cline{6-10}\\[-4pt]
1327315426141900672$^{(c)}$ & NGC 6205 (M 13)       & 0.95 & \dots & \dots & 4564058919924158208 & NGC 5904 (M 5)        & 0.13 & \dots & \dots\\
3906396833023399936 & NGC 5272 (M 3)        & 0.90 & \dots & \dots & 2732393700487077504 & NGC 7089 (M 2)        & 0.18 & Enc. & \dots\\
1314863834914856576$^{(c)}$ & NGC 6205 (M 13)       & 0.83 & \dots & \dots & 2749152495372410880 & NGC 5904 (M 5)        & 0.18 & \dots & \dots\\
1464049970616941952 & NGC 6864 (M 75)       & 0.63 & Enc. & \dots & 2749152495372410880 & NGC 6584              & 0.15 & \dots & \dots\\
2731803537620554112 & NGC 6981 (M 72)       & 0.56 & \dots & \dots & 1532601977988111872 & Pal 3                 & 0.17 & \dots & \dots\\
3696548235634359936 & NGC 1261              & 0.54 & Enc. & \dots & 1532601977988111872 & NGC 5824              & 0.09 & \dots & \dots\\
1889299458599695488 & IC 1257               & 0.51 & \dots & \dots & 1532601977988111872 & Eridanus              & 0.07 & \dots & \dots\\
1020747223263134336 & NGC 6229              & 0.48 & \dots & \dots & 2464872110448024960 & Pal 14 (Arp 1)        & 0.16 & \dots & \dots\\
1020747223263134336 & NGC 6981 (M 72)       & 0.11 & \dots & \dots & 2464872110448024960 & Pal 5                 & 0.12 & \dots & \dots\\
1020747223263134336 & NGC 6584              & 0.05 & \dots & \dots & 2464872110448024960 & Rup 106               & 0.09 & \dots & \dots\\
1266132414419596160 & NGC 6284              & 0.47 & \dots & \dots & 2464872110448024960 & Eridanus              & 0.06 & \dots & \dots\\
1266132414419596160 & NGC 288               & 0.13 & \dots & \dots & 2730283398370974208 & NGC 2298              & 0.15 & Enc. & \dots\\
3938161242913396352 & Eridanus              & 0.42 & \dots & \dots &  603202356856230272 & NGC 1904 (M 79)       & 0.15 & Enc. & \dots\\
3707957627277010560 & NGC 6981 (M 72)       & 0.39 & \dots & \dots &  603202356856230272 & NGC 3201              & 0.08 & Seq. & \dots\\
3707957627277010560 & NGC 6229              & 0.12 & \dots & \dots &  603202356856230272 & NGC 1851              & 0.07 & Enc. & 0.22\\
4425543651944958848 & NGC 6681 (M 70)       & 0.38 & \dots & \dots &  100528661660795648 & Eridanus              & 0.15 & \dots & \dots\\
4425543651944958848 & NGC 6205 (M 13)       & 0.35 & \dots & \dots &  777803497675978368 & NGC 6779 (M 56)       & 0.15 & Enc. & \dots\\
 658003150255826048 & NGC 5634              & 0.36 & \dots & \dots &  777803497675978368 & NGC 6656 (M 22)       & 0.09 & \dots & \dots\\
 658003150255826048 & NGC 6584              & 0.31 & \dots & \dots & 2732599171722435968 & NGC 6584              & 0.14 & \dots & \dots\\
1505296706224813824 & NGC 1261              & 0.32 & Enc. & \dots & 2732599171722435968 & ESO 280-06            & 0.10 & \dots & \dots\\
1505296706224813824 & IC 4499               & 0.09 & \dots & \dots & 2732599171722435968 & NGC 7089 (M 2)        & 0.05 & Enc. & \dots\\
1352831345811056384 & NGC 6584              & 0.31 & \dots & \dots & 3065507927991629824 & NGC 6981 (M 72)       & 0.14 & \dots & \dots\\
3793673590978231168 & Eridanus              & 0.31 & \dots & \dots & 3065507927991629824 & NGC 6229              & 0.13 & \dots & \dots\\
1336539538425553280 & NGC 1904 (M 79)       & 0.30 & Enc. & \dots & 3065507927991629824 & NGC 5904 (M 5)        & 0.06 & \dots & \dots\\
1336539538425553280 & NGC 7089 (M 2)        & 0.08 & Enc. & \dots & 3065507927991629824 & IC 4499               & 0.05 & \dots & \dots\\
1336539538425553280 & NGC 6981 (M 72)       & 0.05 & \dots & \dots & 2507434583516170240 & Pal 3                 & 0.14 & \dots & \dots\\
2619708262744328064 & NGC 1261              & 0.30 & Enc. & \dots & 2507434583516170240 & Eridanus              & 0.11 & \dots & \dots\\
2619708262744328064 & NGC 7089 (M 2)        & 0.17 & Enc. & \dots & 2507434583516170240 & NGC 6715 (M 54)       & 0.08 & \dots & \dots\\
1319293939061613312 & NGC 362               & 0.29 & \dots & \dots & 2507434583516170240 & Pal 5                 & 0.06 & \dots & \dots\\
2686927211051932032$^{(c)}$ & NGC 6656 (M 22)       & 0.28 & \dots & \dots &    2910342854781696 & Pal 3                 & 0.13 & \dots & \dots\\
2686927211051932032$^{(c)}$ & NGC 7089 (M 2)        & 0.20 & Enc. & \dots &    2910342854781696 & Eridanus              & 0.08 & \dots & \dots\\
2686927211051932032$^{(c)}$ & ESO 280-06            & 0.10 & \dots & \dots & 3896352897383285120 & NGC 7089 (M 2)        & 0.13 & Enc. & 0.22\\
 634777096694507776 & NGC 5904 (M 5)        & 0.27 & \dots & \dots & 3896352897383285120 & IC 4499               & 0.09 & \dots & \dots\\
 634777096694507776 & NGC 6229              & 0.09 & \dots & \dots & 3896352897383285120 & NGC 1904 (M 79)       & 0.08 & Enc. & 0.14\\
 634777096694507776 & NGC 6981 (M 72)       & 0.07 & \dots & \dots & 4141123766486595200 & NGC 6139              & 0.13 & \dots & \dots\\
 634777096694507776 & NGC 6584              & 0.05 & \dots & \dots & 4448177751636921088 & NGC 6544              & 0.11 & \dots & \dots\\
 860782815590943488 & Rup 106               & 0.25 & \dots & \dots & 4448177751636921088 & NGC 6273 (M 19)       & 0.07 & \dots & \dots\\
 860782815590943488 & NGC 7006              & 0.09 & \dots & \dots & 2470086991019454848 & Eridanus              & 0.10 & \dots & \dots\\
 860782815590943488 & NGC 6934              & 0.05 & \dots & \dots & 2470086991019454848 & Pal 5                 & 0.10 & \dots & \dots\\
4602060554337732992$^{(c)}$ & NGC 6205 (M 13)       & 0.25 & \dots & 0.35 & 2470086991019454848 & Rup 106               & 0.09 & \dots & \dots\\
 764888767240012928 & NGC 4147              & 0.25 & \dots & \dots & 2470086991019454848 & Pal 14 (Arp 1)        & 0.08 & \dots & \dots\\
 764888767240012928 & NGC 6981 (M 72)       & 0.16 & \dots & \dots & 2720484958765839104 & IC 4499               & 0.10 & \dots & \dots\\
3563020451180259456 & IC 1257               & 0.24 & \dots & \dots & 2720484958765839104 & NGC 7492              & 0.06 & \dots & \dots\\
3563020451180259456 & NGC 7089 (M 2)        & 0.11 & Enc. & \dots & 1592645066735097088 & NGC 1904 (M 79)       & 0.10 & Enc. & \dots\\
2688138220030361600 & NGC 6752              & 0.24 & \dots & \dots & 1592645066735097088 & NGC 6981 (M 72)       & 0.08 & \dots & \dots\\
2688138220030361600 & NGC 6656 (M 22)       & 0.14 & \dots & \dots & 1592645066735097088 & NGC 4147              & 0.06 & \dots & \dots\\
2688138220030361600 & NGC 6254 (M 10)       & 0.10 & \dots & \dots & 3591471723298342400 & IC 1257               & 0.10 & \dots & \dots\\
2688138220030361600 & NGC 6397              & 0.07 & \dots & \dots & 3932598122098282496 & Eridanus              & 0.10 & \dots & \dots\\
3735130545329333120 & ESO 280-06            & 0.23 & \dots & \dots & 3932598122098282496 & Pal 14 (Arp 1)        & 0.08 & \dots & \dots\\
3735130545329333120 & NGC 7089 (M 2)        & 0.11 & Enc. & \dots & 3932598122098282496 & NGC 5272 (M 3)        & 0.08 & \dots & \dots\\
3735130545329333120 & NGC 7492              & 0.10 & \dots & \dots & 3932598122098282496 & NGC 7492              & 0.07 & \dots & \dots\\
4425538154386975616 & NGC 6273 (M 19)       & 0.22 & \dots & \dots & 2686860763615798912 & NGC 6284              & 0.10 & \dots & \dots\\
4425538154386975616 & NGC 6287              & 0.08 & \dots & \dots & 1325256040863360128 & NGC 6205 (M 13)       & 0.09 & \dots & 0.19\\
1546577767213867136 & NGC 6229              & 0.21 & \dots & \dots & 1325256040863360128 & NGC 5139 (omega Cen)  & 0.08 & Seq. & \dots\\
1546577767213867136 & NGC 4147              & 0.15 & \dots & \dots & 1325256040863360128 & NGC 6681 (M 70)       & 0.05 & \dots & \dots\\
1546577767213867136 & NGC 6981 (M 72)       & 0.12 & \dots & \dots & 1666212668195069184 & NGC 5824              & 0.09 & \dots & \dots\\
1546577767213867136 & NGC 6584              & 0.08 & \dots & \dots & 1666212668195069184 & Rup 106               & 0.08 & \dots & \dots\\
 864027195230709504 & NGC 6981 (M 72)       & 0.21 & \dots & \dots & 1307939316841792512 & Pal 3                 & 0.09 & \dots & \dots\\
 864027195230709504 & NGC 1904 (M 79)       & 0.07 & Enc. & \dots & 1307939316841792512 & Eridanus              & 0.06 & \dots & \dots\\
 864027195230709504 & NGC 1851              & 0.06 & Enc. & \dots & 3651209835007571456 & Eridanus              & 0.09 & \dots & \dots\\
2566271722756610304 & Pal 14 (Arp 1)        & 0.21 & \dots & \dots & 3651209835007571456 & IC 4499               & 0.06 & \dots & \dots\\
2566271722756610304 & Rup 106               & 0.11 & \dots & \dots & 3559140961841845888 & NGC 6981 (M 72)       & 0.09 & \dots & \dots\\
2566271722756610304 & Eridanus              & 0.10 & \dots & \dots & 4018168336083916800 & Pal 3                 & 0.08 & \dots & \dots\\
4141127962669782400 & NGC 6093 (M 80)       & 0.20 & \dots & \dots & 4021877298042177920 & NGC 6715 (M 54)       & 0.08 & \dots & \dots\\
4034859648443488896 & NGC 4147              & 0.20 & \dots & \dots & 4021877298042177920 & Eridanus              & 0.07 & \dots & \dots\\
4034859648443488896 & NGC 5634              & 0.12 & \dots & \dots & 3697011301828434816 & Eridanus              & 0.07 & \dots & \dots\\
4034859648443488896 & NGC 5272 (M 3)        & 0.07 & \dots & \dots & 3697011301828434816 & NGC 7089 (M 2)        & 0.05 & Enc. & \dots\\
4034859648443488896 & NGC 6981 (M 72)       & 0.07 & \dots & \dots & 3563015262859711744 & NGC 7089 (M 2)        & 0.06 & Enc. & \dots\\
1678398697300158208 & NGC 7006              & 0.19 & \dots & \dots & 1395074303376247296 & Pal 5                 & 0.06 & \dots & \dots\\
1678398697300158208 & Eridanus              & 0.10 & \dots & \dots & 3200603517242311808 & NGC 6981 (M 72)       & 0.06 & \dots & \dots\\
1678398697300158208 & NGC 288               & 0.06 & \dots & \dots & 3303140501315921152 & NGC 6535              & 0.05 & Seq. & \dots\\
4564058919924158208 & NGC 6229              & 0.18 & \dots & \dots & 4437207133854765056 & Pal 3                 & 0.05 & \dots & \dots\\
4564058919924158208 & NGC 6981 (M 72)       & 0.16 & \dots & \dots &  &  & & \\

\hline
\end{tabular}}
\tablefoot{
\tablefoottext{a}{Cluster might be associated with the major merger (m.m.) events Gaia-Enceladus \citep[Enc., ][]{Myeong18} or Sequoia \citep[Seq., ][]{Myeong19}.}
\tablefoottext{b}{Confidence attributed to the same pair by \citet{Savino19}. We note that the values are inverted -- that is unity minus reported rejection confidence -- with respect to the original study.}
\tablefoottext{c}{Associations already reported in Sect. \ref{Subsec: Chemodynamical matches in the fields surrounding CN-strong stars} and listed in Table \ref{Table: Extratidal target information}.}
}
\end{table*}

\begin{sidewaystable*}
\caption{Stars that were chemodynamically associated with CN-strong giants (method III).}
\label{Table: Associations to CN-strong stars}
\centering
\begin{tabular}{l@{\hspace{2ex}}c@{\hspace{2ex}}c@{\hspace{2ex}}c@{\hspace{2ex}}r@{\hspace{2ex}}c@{\hspace{2ex}}c@{\hspace{2ex}}r@{\hspace{0.1ex}}c@{\hspace{0.1ex}}l@{\hspace{2ex}}c@{\hspace{2ex}}c@{\hspace{2ex}}c@{\hspace{3ex}}c@{\hspace{3ex}}c@{\hspace{0.1ex}}c}
\hline\hline
\\[-5pt]
\gaia DR2 source ID & $\alpha^{(a)}$ & $\delta^{(a)}$ & \multicolumn{1}{c}{$r$} & \multicolumn{1}{c}{$\varpi$} & $\mu_\alpha \cos{\delta}$ & $\mu_\delta$ & \multicolumn{3}{c}{$v_r$} & [Fe/H] & $P(A/B)$ & $G$ & $G_\mathrm{BP}-G_\mathrm{RP}$ & RRx$^{(b)}$ & $D_\mathrm{RR}$\\
 & [deg] & [deg] & [deg] & \multicolumn{1}{c}{[mas]} & [mas yr$^{-1}$] & [mas yr$^{-1}$] & \multicolumn{3}{c}{[km s$^{-1}$]} & [dex] & & [mag] & [mag] & & [kpc]\\[5pt]
\hline\\[-5pt]  
\\[-7pt]\multicolumn{16}{c}{\gaia DR2 4018168336083916800 (Sgr CNs 1)}\\
4018168336083916800$^{(c)}$ & 175.083 &  27.533 &  0.0 & $ 0.05\pm0.12$ & $-1.4\pm0.1$ & $-1.0\pm0.2$ & $ -63.0$ & $\pm$ & $ 5.2$ & $-1.40\pm0.17$ & 1.00 & 16.67 & 1.18 & &\\
4017874216723852672 & 175.438 &  26.772 &  0.8 & $ 0.00\pm0.18$ & $-1.3\pm0.3$ & $-1.3\pm0.2$ & $ -67.2$ & $\pm$ & $ 5.4$ & $-1.13\pm0.18$ & 0.47 & 17.91 & 1.07 & &\\
4004075586313197312 & 175.879 &  23.347 &  4.2 & $-0.22\pm0.34$ & $-1.7\pm0.4$ & $-1.2\pm0.3$ & $ -53.5$ & $\pm$ & $ 7.4$ & $-1.41\pm0.16$ & 0.43 & 18.14 & 0.90 & &\\
3951337858979970560 & 181.428 &  20.513 &  9.1 & $ 0.01\pm0.05$ & $-1.3\pm0.1$ & $-1.0\pm0.1$ & $ -56.2$ & $\pm$ & $ 5.1$ & $-1.33\pm0.16$ & 0.41 & 15.68 & 1.47 & &\\
4017707606352394880 & 174.097 &  26.485 &  1.4 & $ 0.04\pm0.17$ & $-1.4\pm0.2$ & $-1.2\pm0.3$ & $ -77.0$ & $\pm$ & $ 5.7$ & $-1.61\pm0.17$ & 0.38 & 17.60 & 1.12 & &\\
3951642591203760384 & 182.275 &  20.433 &  9.7 & $-0.00\pm0.18$ & $-1.3\pm0.3$ & $-0.9\pm0.2$ & $ -49.9$ & $\pm$ & $ 5.4$ & $-1.39\pm0.17$ & 0.26 & 17.94 & 1.12 & &\\
3950619942311040000 & 183.795 &  19.849 & 11.1 & $ 0.18\pm0.18$ & $-1.6\pm0.3$ & $-1.1\pm0.2$ & $ -71.6$ & $\pm$ & $ 6.2$ & $-1.32\pm0.16$ & 0.23 & 18.04 & 0.85 & &\\
3950348916990151296 & 183.925 &  19.179 & 11.6 & $ 0.12\pm0.11$ & $-1.6\pm0.2$ & $-1.0\pm0.2$ & $ -63.5$ & $\pm$ & $ 5.2$ & $-1.26\pm0.17$ & 0.23 & 16.55 & 1.33 & &\\
3999380469799933952 & 180.989 &  20.676 &  8.7 & $ 0.08\pm0.18$ & $-1.7\pm0.3$ & $-0.9\pm0.4$ & $ -60.2$ & $\pm$ & $ 5.4$ & $-0.94\pm0.16$ & 0.19 & 17.87 & 1.16 & &\\
3960517711624629632 & 188.877 &  25.658 & 12.5 & $-0.07\pm0.29$ & $-1.7\pm0.3$ & $-1.0\pm0.2$ & $ -53.0$ & $\pm$ & $15.6$ & $-1.45\pm0.19$ & 0.18 & 18.40 & 0.47 & RRab$^1$ & $34.2\pm8.3$\\
\hline
\end{tabular}
\tablefoot{Only a small portion of the data is shown to indicate the form and content of the table. The full table is available through the CDS.
\tablefoottext{a}{Coordinates are in the \gaia DR2 reference epoch J2015.5.}
\tablefoottext{b}{RR Lyrae references: (1): \citet{Drake13}. (2): \citet{Drake13b}. (3): \citet{Drake14}. (4): \citet{Abbas14}, (5): This study.}
\tablefoottext{c}{Classified CN-strong in Paper I.}
}
\end{sidewaystable*}
 
\end{appendix}

\end{document}